\documentclass[prper,aps,twocolumn,groupedaddress,floats,showpacs,final,superscriptaddress]{revtex4-1}
\usepackage[latin3]{inputenc}
\usepackage[makeroom]{cancel}
\usepackage{graphicx}
\usepackage{amsmath}
\usepackage{amsfonts}
\usepackage{amssymb}
\usepackage{color}
\usepackage[left=2cm,right=2cm,top=2cm,bottom=2cm]{geometry}
\usepackage[export]{adjustbox}
\usepackage{graphicx} 
\usepackage{dcolumn}
\usepackage{bm}
\usepackage{simplewick}
\usepackage{array}
\usepackage{appendix}
\RequirePackage[
   hyperindex,colorlinks,bookmarksnumbered,
   plainpages=true,pdfstartview=FitH]{hyperref}
\hypersetup{linkcolor=blue,urlcolor=blue,citecolor=blue}
\usepackage{hyperref}

\definecolor{purple}{rgb}{0.5,0,0.6}

\usepackage{ulem}
\renewcommand{\emph}[1]{\textit{#1}}
\definecolor{darkblue}{rgb}{0,0,0.5}
\definecolor{darkgreen}{rgb}{0,0.5,0}
\definecolor{darkred}{rgb}{.7,0,0}
\definecolor{purple}{rgb}{0.5,0,0.6}
\definecolor{orange}{rgb}{1,0.5,0}
\definecolor{grey}{rgb}{.6,.6,.6}
\definecolor{lightpink}{rgb}{1,0.7,0.75}
\definecolor{pink}{rgb}{1,0.4,0.58}
\definecolor{deeppink}{rgb}{1,0.08,0.58}

\newcommand{\DK}[1]{{\color{black} #1}}

\newcommand{\pdag}{{\phantom{\dagger}}}
 
\renewcommand{\emph}[1]{\textit{#1}}

\begin{document}
\author{D. B. Karki}
\affiliation{The  Abdus  Salam  International  Centre  for  Theoretical  Physics  (ICTP),
Strada  Costiera 11, I-34151  Trieste,  Italy}
\affiliation{International School for Advanced Studies (SISSA), Via Bonomea 265, 34136 Trieste, Italy}
\author{Mikhail N. Kiselev}
\affiliation{The  Abdus  Salam  International  Centre  for  Theoretical  Physics  (ICTP),
Strada  Costiera 11, I-34151  Trieste,  Italy}

\title{Full counting statistics of two-stage Kondo effect }

\date{\today}

\begin{abstract}
We developed a theoretical framework which extends the method of \textit{full counting statistics} (FCS) from conventional single channel Kondo screening schemes to multi-channel Kondo paradigm. The developed idea of FCS has been demonstrated considering an example of two-stage Kondo (2SK) model. We analyzed the charge transferred statistics in the strong-coupling regime of a 2SK model using non-equilibrium Keldysh formulation. A bounded value of Fano factor, $1\leq F\leq 5/3$, confirmed the cross-over regimes of charge transfered statistics in 2SK effect, from Poissonian to super-Poissonian. An innovative way of measuring transport properties of 2SK effect, by the independent measurements of charge current and noise, has been proposed.
\end{abstract}


\maketitle

\section{Introduction}
\vspace*{-5mm}
Quantized charge in nanoscale systems results in large current fluctuations \cite{Butt_Blt}. Besides, thermal fluctuations are ubiquitous at finite temperature. These fluctuations are prevalently measured by charge current and its noise, the first and second cumulant of fluctuating current~\cite{Blanter_Nazarov}. The study of noise in a generic nanodevice provides underlying transport informations that cannot be inferred from the average current measurements
 \cite{Levitov_JETP, Levitov_2, Levitov_1,Butt_Blt, Levitov, Nazarov_1,Levitov_PRB(2004), Blanter_Nazarov}. In particular, noise measurement imparts an effective way of probing the dynamics of charge transfer \cite{Levitov_PRB(2004), Blanter_Nazarov}. Moreover, noise has revealed the nature of quasi-particle interactions and different types of entanglements associated with the system \cite{Gustavsson, Belzig, clerk}. In addition to first and second cumulant of the fluctuating current, the fundamental relevance of higher order cumulants to describe the transport processes in nanostructure has been also demonstrated~\cite{m1, m2, Fujisawa1634,Gustavsson, m4, m5, m6, m7, m8}.

The method of full counting statistics (FCS) furnishes an elegant way to scrutinize an \textit{arbitrary} ($n$-th) order cumulant of current through a nanodevice~\cite{Levitov_JETP, Levitov_2, Levitov_1}. The probabilistic interpretation of charge transport is at the core of FCS theory. The primary object of FCS is the moment generating function (MGF) for the probability distribution function (PDF) of transferred charges within a given time interval~\cite{Levitov_JETP, Levitov_2, Levitov_1}. The moments of PDF of order $n{\geq}2$ characterize the current fluctuations. The FCS scheme permits in this way a transparent study of the quantum transport in various nanostructures. Notably, FCS of the normal metal-superconductor hybrid structures, superconducting weak links, tunnel junctions, chaotic cavities,
entangled electrons, spin-correlated systems, charge shuttles, nanoelectromechanical systems are most striking examples~\cite{sita1, sita2, sita4, sita5, sita6, sita7, sita8, sita9}.

In nanoscale transport studies, an archetype of electronic device consists of an impurity sandwiched between two reservoirs of conduction electrons \cite{Butt_Blt,Blanter_Nazarov}. The artificial atom, molecule, quantum dot (QD), carbon nano tube (CNT) etc., plays the role of an impurity. Given their low tunneling rate, the QDs represents archetypal setups for the study of a highly accurate FCS~\cite{gora}, the main concern of present work. The transport through the QD depends strongly on the associated number of electronic levels, while the orbitals of the impurity play the major role to define the underlying transport characteristics~\cite{Hewson}. Out of all the impurities-mediated transport processes, those with intrinsic magnetic moment; hence magnetic in nature, have attracted an ever increasing interests~\cite{Hewson,Coleman_book}. One can expect variant transport fingerprints when such magnetic impurities exchange coupled to conduction electrons (for review see Ref.~\cite{Cox_Adv_Phys(47)_1998}). 

In the low energy regime of transport measurements, the correlation between the localized spin of impurity and the spin of conduction electrons results in the well known many-body phenomenon, the Kondo screening effect~\cite{dk1}. The fundamental role of Kondo effect in enhancing and controlling the transport through a nanostructure is the acknowledged evidence~\cite{Nozieres, Nozieres_Blandin_JPhys_1980, Affleck_Lud_PRB(48)_1993, Wilkins_PRL(54)_1985,Goldhaber_nat(391)_1998,  Cronewett_SCI(281)_1998, Jesper_NAT(408)_2000,revival, Pierre_SCI(342)_2013, Keller_Goldhaber_nat_phys(10)_2014, Pierre_NAT(526)_2015,  Pierre_NAT(536)_2016}. In a transport setup with two reservoirs (leads), the Kondo screening of the localized spin is caused by at most two conduction channels, the symmetric and anti-symmetric combination of electron states in the leads. The interplay between the number of conduction channels ($\mathcal{K}{=}1, 2$) and the effective spin of magnetic impurity ($\mathcal{S}{\geq}1/2$) boosts up further the richness of Kondo physics. Specifically, $\mathcal{K}{=}2\mathcal{S}$ put forward the controllable 
comprehension of Kondo effects in nanodevices~\cite{GP_Review_2005}. In this particular case, the effective spin of impurity get completely screened by the spin of conduction electrons. Such fully-screened Kondo effects are of immense interest given their low energy behavior described by a local Fermi-Liquid (FL) theory~\cite{Nozieres, Nozieres_Blandin_JPhys_1980, Affleck_Lud_PRB(48)_1993}.

The Kondo screening involving only a single channel of conduction electrons ($\mathcal{K}{=}1$) and a spin half impurity ($\mathcal{S}{=}1/2$) forms a prototypical example of fully-screened Kondo effect. The magnetic impurities with only one orbital manifest the single channel Kondo (1CK) effect. Tremendous perseverance efforts~\cite{GP_Review_2005,Mora_Leyronas_Regnault_PRL_(100)_2008, Mora_Clerk_Hur_PRB(80)_2009,HWDK_PRB_(89)_2014} has been devoted in understanding the transport behavior in paradigmatic 1CK schemes. Moreover, various seminal works~\cite{Gogolin_Komnik_PRL(2006), Gogolin_Komnik_PRB(2006),golub,tschmidt_2, tschmidt_1,ahes} paved the way to access the associated FCS in 1CK realm. Unlike the 1CK, the transport characteristics of a multi-orbital impurity has been less explored. In this facet, many orbitals of the conduction channels are involved in screening the impurity spin (multi-channel screening), which make the problem more obscure~\cite{Glazman_PRL_2001}. Nonetheless, several cogent evidences~\cite{intspin,KMaddy,Glazman_PRL_2001,lla, bdaama}, theoretical and experimental, are available to demonstrate the relevance of multi-channel screening effect in a generic transport setup. The simplest multi-channel screening involves two conduction channels ($\mathcal{K}{=}2$) and $\mathcal{S}{\geq}1/2$; general manifestation of a two leads geometry. In the present work we focus only on the particular case of multi-channel screening such that $\mathcal{K}{=}2\mathcal{S}$ in a two leads setup. Thus the $\mathcal{S}{=}1$ impurity interacting with two channels of conduction electrons forms the minimal description of multi-channel screening in FL regime~\cite{Coleman_PRL(94)_2005,KMDK_2018}.

Multi-orbital quantum impurity with effective spin $\mathcal{S}{=}1$ connected
to two terminals can leads to a Kondo
effect exhibiting two-stage screening~\cite{KMDK_2018}. The first-stage screening process
constitutes in an under-screened Kondo effect where the impurity spin is
effectively reduced from $\mathcal{S}{=}1$ to $\mathcal{S}{=}1/2$. Subsequently, second-stage
screening leads to complete screening of the impurity spin and the
formation of a Kondo singlet. This feature of screening is called two-stage Kondo (2SK) effect~\cite{Glazman_PRL_2001,Coleman_PRL(94)_2005}. The low energy description of such 2SK effects is still governed by a local FL theory. Nonetheless, transport properties of such FL get modified in dramatic ways compared to 1CK~\cite{KMDK_2018}. The strong interplay between two conduction channels, both close to resonance scattering, causes aforesaid different transport features over 1CK. The lack of compatible cure of the two Kondo resonances makes the 2SK paradigm far from being trivial for several years \cite{Glazman_PRL_2001,Coleman_PRL(94)_2005,Coleman_PRB(75)_2007}. To analyze the equilibrium and non-equilibrium transport properties of a generic 2SK effect, a two-color local FL theory has been recently developed~\cite{KMDK_2018}. Here, the absence of zero-bias anomaly and non-monotonicity of FL transport coefficients are demonstrated as the hallmarks of 2SK effect. 

These two traits of the 2SK effect, which are contrasting over 1CK, have raised many  fascinating concerns. For instance, how these fingerprints can affect the higher cumulants of charge current, particularly the noise to signal ratio. This ratio is commonly known as the Fano factor ($F$). The zero temperature limit of $F$ is of extreme experimental interest~\cite{Sela_Oreg_Oppen_Koch_PRL_(97)_2006, Mora_Clerk_Hur_PRB(80)_2009}. In theoretical perspective, the method of FCS pertaining to the two resonance channels of conduction electrons has not been developed yet. 

In this work we \DK{take an important} step of revealing the FCS for 2SK effect. The structure of this paper is as follows. The basis description of 2SK effect setup and model Hamiltonian is given in Sec.~\ref{setup}. We present the theory of FCS for two resonance channels of a local FL in Sec.~\ref{fcs}. The section~\ref{rd} is devoted to discussing the results of applying many-body method of FCS developed in Sec.~\ref{fcs} to the 2SK model of Sec.~\ref{setup}. Conclusions and future perspectives are presented in Sec.~\ref{con}. Details of mathematical calculations are given in Appendices. 
\vspace*{-5mm}
\section{Setup and model Hamiltonian}\label{setup}
\vspace*{-2mm}
\setlength\belowcaptionskip{-5ex}
\begin{figure}[t]
\begin{center}
\includegraphics[scale=0.6]{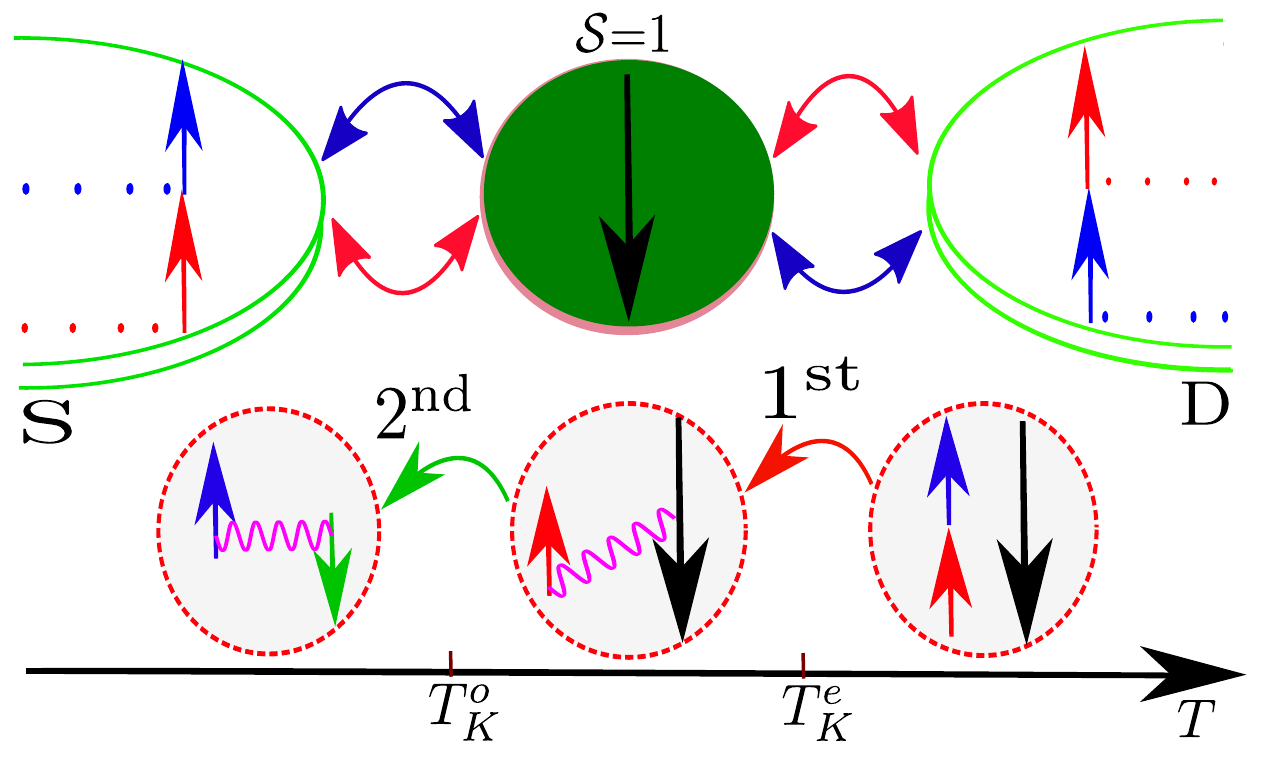}
\caption{Upper panel: Schematic representation of a generic 2SK effect setup. The effective spin $\mathcal{S}{=}1$ impurity is tunnel-coupled with two external leads, the source S and the drain D. Lower panel: Flow diagram of 2SK model from weak to strong coupling. For the entries in the figure and their explanations see Sec.~\ref{setup}.}\label{setupfig}
\end{center}
\end{figure}
The cartoon representing the generic 2SK effect is as shown in upper panel of Fig.~\ref{setupfig}. \DK{The generic quantum impurity sandwiched between two conducting leads (the source S and the drain D) is described by the Anderson model with the Hamiltonian 
\begin{align}\label{model}
H_{\rm A }&=\sum_{k\alpha\sigma}\xi_k c^\dagger_{\alpha k \sigma}c_{\alpha k \sigma}+\sum_{\alpha k i \sigma} t_{\alpha i} c^\dagger_{\alpha k \sigma} d_{i \sigma} + \text{H.c.}\nonumber\\
&+\sum_{i\sigma}
\varepsilon_i d^\dagger_{i \sigma}d_{i \sigma}
+E_c \hat{\cal N}^2- {\cal J} \hat{\mathbf{S}}^2.
\end{align}
The conducting leads are described by the first term of Eq.~\eqref{model} such that the operator $c^{\dagger}_{\alpha k\sigma}$ creates an electron with momentum $k$ and spin $\sigma=\uparrow{\color{black}(+)},\downarrow{\color{black}(-)}$ in the $\alpha$ ($\alpha=$S, D) lead. Here $\xi_k$
is the energy of conduction electrons with respect to the chemical potential $\mu$. The operator $d_{i\sigma}$ describes electrons with energy $\varepsilon_i$ and spin $\sigma$ in the $i$-th
orbital state of the quantum impurity. The tunneling matrix elements are represented by 
$t_{\alpha i}$, the charging energy of the impurity (dot) is $E_c$ and
${\cal J}\ll E_c$ is an exchange integral accounting for the Hund's rule \cite{Coleman_PRB(75)_2007}. The total number of electrons in the dot is given by an operator $\hat{\cal N}=\sum_{i\sigma} d^{\dagger}_{i\sigma}d_{i\sigma}$. The two electrons in the dot ensures the expectation value of $\hat{\cal N}$ to be
$\bar n_d=2$  and the total spin $S=1$. Application of the Schrieffer-Wolff transformation~\cite{Schrieffer_Wolf_PR(149)_1966} to the Hamiltonian Eq.~\eqref{model} {\color{black} results in the effective Kondo Hamiltonian for the spin-1 quantum impurity~\cite{GP_Review_2005,Coleman_PRB(75)_2007}}.

}

\DK{To proceed with the calculation of FCS relevant to the setup in Fig.~\ref{setupfig}, we assume that the} thermal equilibrium is maintain in source and drain, separately, at temperature $T$. The chemical potentials of source and drain electrodes are $\mu_{S}$ and $\mu_D$ respectively. The applied voltage bias across source and drain drives the impurity-leads system out-of-equilibrium. For the sake of simplicity, we consider symmetrically applied bias voltage such that $\mu_S{-}\mu_D{=}eV$, where $e$ is the electronic charge. In this frame, the symmetrical (even, $e$) and anti-symmetrical (odd, $o$) combinations of electron operators in the two leads interact with the impurity. Assuming $c_{S/D}$ as an operator that annihilates an electron in the source/drain, the even/odd combinations of electron operators are $b_{e/o}{=}\left(c_S\pm c_D\right)/\sqrt{2}$. These states are also known as conduction channels. In Fig.~\ref{setupfig}, we used arrows with different color to show that the electrons form even and odd channels (${\color{blue}\uparrow}$, electron forms channel-$e$ and ${\color{red}\uparrow}$, electron forms channel-$o$). Likewise, the interaction between even (odd) channel and impurity is represented by two-headed arrow with blue (red) color. In our convention, blue (red) color is generic for even (odd) channel. 

In conventional 1CK effect the odd channel is completely decoupled from the impurity~\cite{Nozieres}. The interacting channel, the even one, is characterized by the Kondo temperature $T^e_K$. Depending upon the applied bias $eV$ and the temperature $T$ in particular setup, different coupling regimes come into play. Namely, $(eV, T){\leq}T^e_K$, the strong-coupling regime and $(eV, T){\geq}T^e_K$, the weak-coupling regime. Immense efforts, experimental inclusive of theory, have been paid for the transport description of 1CK effect in both of the above regimes (see Refs.~\cite{Hewson,GP_Review_2005} for review).

In two-leads setup with a generic quantum impurity having more than one orbitals, nonetheless, neither of the electron combinations remain decoupled from the impurity~\cite{Glazman_PRL_2001,Coleman_PRL(94)_2005}. Consequently, both of the conduction channels take part for the screening of localized spin of the impurity. In addition to $T^e_K$ the another energy scale characterizing the Kondo temperature of the odd channel, $T^o_K$, engage to the problem of 2SK effect. The interplay between two Kondo temperatures ($T^a_K,a{=}e,o$) makes the 2SK problem far away richer then 1CK, but, at the same time notoriously difficult. Perturbation treatments of weak, $(eV, T) {>} \text{max}(T^e_K, T^o_K)$ and intermediate, $T^o_K{\leq} (eV, T) {\leq} T^e_K$, coupling regimes have been formulated \cite{Coleman_PRL(94)_2005,Coleman_PRB(75)_2007}. In the intermediate regime the impurity spin get partially screened via first (${\rm 1^{st}}$)-stage of screening. Still swapping the temperature and bias voltage down, to satisfy the condition $(eV, T)\ll\text{min}(T^e_K, T^o_K)$, results in strong-coupling regime of 2SK effect. In this second (${\rm 2^{nd}}$)-stage the complete screening of impurity spin is achieved. These three coupling regimes are shown in lower panel of Fig.~\ref{setupfig}. 

Furthermore, it has been argued \cite{Glazman_PRL_2001,Coleman_PRL(94)_2005} that the most nontrivial part of 2SK effect is the strong-coupling regime, where both of the interacting channels are close to the resonance scattering. Since, the 2SK effect satisfies the identity $\mathcal{K}{=}2\mathcal{S}$, it offers the transport description in terms of a local FL. From now on, we focus only on the strong-coupling regime of 2SK effect. Owing to its low energy behavior as a local FL, we describes the strong-coupling regime of 2SK effect in the spirit of Nozieres FL theory~\cite{Nozieres, Affleck_Lud_PRB(48)_1993}. Accordingly, the Kondo singlet (Kondo cloud) acts as the scatterer for the incoming electrons from the leads. Outgoing and incoming electrons then differ from each other by the phase shifts $\delta^a_\sigma(\varepsilon)$. At low energy, $\varepsilon\ll\text{min}(T^e_K,T^o_K)$, we expand the phase shifts in terms of phenomenological parameters to write~\cite{Nozieres}
\begin{equation}\label{pss}
\delta^a_\sigma(\varepsilon)=\delta^a_0+\alpha_a\varepsilon-\phi_a\delta N^a_{\bar{\sigma}}
+\Phi\sigma\left(\delta N^{\bar{a}}_{\uparrow}-\delta N^{\bar{a}}_{\downarrow}\right).
\end{equation}
Here, $\delta^a_0{=}\pi/2$ are the resonance phase shifts considered to be the same for both channels and both spin components.
Writing Eq.~\eqref{pss} we explicitly consider the particle-hole (PH) symmetric limit; $\sigma{=}\uparrow,\downarrow$ ($\bar{\sigma}{=}\downarrow,\uparrow$). First two terms of Eq.~\eqref{pss} represents the purely elastic effects associated with two channels. These are, equivalently, known as the scattering terms. The parameters $\alpha_a$ are the Nozieres FL coefficients characterizing the scattering. Although, for $\varepsilon{=}T{=}eV{=}0$ both channels are at resonance, the way phase shifts changes with energy is different in two channels. This consequence can be accounted for by defining the Kondo temperatures as~\cite{HWDK_PRB_(89)_2014,Karki_MK_TE}
\begin{equation}\label{kot}
T^a_K=1/\alpha_a.
\end{equation}
For definiteness, we consider $T^o_K\leq T^e_K$ throughout the paper.

Besides, the third and fourth terms of Eq.~\eqref{pss} are due to the finite inelastic effects. These are known as interaction terms. The parameters $\phi_a$ quantify the intra-channel interactions, and the inter-channel interaction is accounted for by $\Phi$. The notation $\delta N^a_{\sigma}$ is defined by,
$$ \delta N^a_{\sigma}{=}\int^{\infty}_{-\infty}  \left[\langle b^{\dagger}_{a\varepsilon\sigma} b_{a\varepsilon\sigma}\rangle_0-\Theta(\varepsilon_{F}-\varepsilon)\right] d\varepsilon.$$
Here, $\varepsilon_F$, in the argument of step function $\Theta$, is the Fermi energy. The average $\langle . .\rangle_0$ is taken with respect to non-interacting Hamiltonian describing the free electrons in two channels,
\begin{equation}\label{nih}
H_0=\nu\sum_{a\sigma}\int_\varepsilon\varepsilon\;
b^\dagger_{a\varepsilon\sigma}b^\pdag_{a\varepsilon\sigma},
\end{equation} 
where, $\nu$ is the density of states per species for a one-dimensional channel. We see that the phase shifts expression, Eq.~\eqref{pss}, consist of five FL parameters ($\alpha_e$, $\alpha_o$, $\phi_e$, $\phi_o$ and $\Phi$). \DK{However, the invariance of phase shifts under the shift of reference energy (the floating of the Kondo resonance~\cite{Mora_Clerk_Hur_PRB(80)_2009}) recovers the FL identity $\alpha_a{=}\phi_a$.} Thereupon, three independent FL parameters ($\alpha_e$, $\alpha_o$ and $\Phi$) completely describes the low energy sector of 2SK problem. With the specification of $T^a_K$ in terms of $\alpha_a$ as in Eq.~\eqref{kot}, we have only one FL parameter ($\Phi)$ to relate with the physical observables. 
The response functions measurements could provide the way to access the parameter $\Phi$~\cite{KMDK_2018}. Therefore, all the phenomenological parameters in Eq.~\eqref{pss} are under control in an experiment. 

In the facet of theoretical ground, the finding of seminal work~\cite{Affleck_Lud_PRB(48)_1993} paved the way to formulate the Hamiltonian describing the scattering and interaction processes in Eq.~\eqref{pss}. The PH symmetry of the problem demands the scattering terms to be represented by the Hamiltonian,
\begin{equation}\label{halpha}
 H_{\rm el} =-\frac{\alpha_{a}}{2 \pi} \sum_{a\sigma}
  \int_{\varepsilon_{1-2}}  
\left(\varepsilon_1+\varepsilon_2\right)\;
 \! b^\dagger_{a\varepsilon_1\sigma}b^\pdag_{a\varepsilon_2\sigma}.
\end{equation}
Similarly, the intra-channel and inter-channel quasi-particles interactions are designed by the Hamiltonians $H_{\phi}$ and $H_{\Phi}$ respectively, $H_{\rm in}=H_{\phi}+H_{\Phi}$ represents the total interactions associated with 2SK effect. Here,
\begin{align}
H_\phi &=\phantom{-}\frac{\phi_{a}}{2\pi\nu}\sum_{a\sigma} \int_{\varepsilon_{1-4}}  :\rho^a_{\varepsilon_1\varepsilon_2\sigma}\;\rho^a_{\varepsilon_3\varepsilon_4\bar{\sigma}}:,\label{hphi}\\
H_\Phi &=-\frac{\Phi}{{2}{\pi}{\nu}}{\sum_{\sigma_{{1{-}4}}}}
{\int_{\varepsilon_{{1{{-}}4}}}}
:S^o_{\varepsilon_1\varepsilon_2\sigma_1\sigma_2}
S^e_{\varepsilon_3\varepsilon_4\sigma_3\sigma_4}:.\label{hPhi}
\end{align}
The colon, $:\cdot\cdot\cdot:$, denotes the normal ordering. In Eqs.~\eqref{hphi} and~\eqref{hPhi} we used the short-hand notations
\begin{equation}\nonumber
\rho^a_{\varepsilon_1\varepsilon_2\sigma}\equiv
b^{\dagger}_{a\varepsilon_1\sigma}b_{a\varepsilon_2\sigma},\; S^a_{\varepsilon_1\varepsilon_2\sigma_1\sigma_2}\equiv b^{{\dagger}}_{{a\varepsilon_{1}\sigma_{1}}}{\boldsymbol{\tau}_{{{\sigma_{1}\sigma_2}}}}
{b_{{a\varepsilon_{2}\sigma_{2}}}},
\end{equation}
with $\boldsymbol{\tau}_{\sigma_{i}\sigma_j}$ the elements of Pauli-matrices. The scattering and interaction parts of Hamiltonian given in Eqs.~\eqref{halpha}$-$\eqref{hPhi} are first-order in $1/T^a_K$. The two-leg vertex $\alpha_a$ and the four-leg vertices $\phi_a$ and $\Phi$ are shown in Fig.~\ref{tablecodex}. In general the symmetry of the problem also allows one to construct the Hamiltonian with eight-leg vertex, for instance  $\tilde{\mathcal{H}}\propto\tilde{\phi}\left(\rho_\sigma \rho_{\bar{\sigma}}\right)^a\left(\rho_\sigma \rho_{\bar{\sigma}}\right)^{\bar{a}}$. Note that in the present work we restrict ourselves to the second order correction to the CGF, thus the relevant terms are upto the $\mathcal{O}(T/T^a_K)^2$. Since the vertex $\tilde{\phi}$ is already second order in $1/T^a_K$, it does not contribute to the cumulants of charge current within second order perturbative calculation, hence has been neglected. Thus the Hamiltonian $H\equiv H_0+H_{\rm el}+H_{\rm in}$ constitutes the minimal model Hamiltonian of a generic 2SK effect. This particular model has the channel symmetry at the point $\alpha_e=\alpha_o$ and $\alpha_a=3/2\Phi$, where the conductance vanishes due to the destructive interference between two interacting channels~\cite{KMDK_2018}. \DK{It is worth noting that the effects of breaking PH symmetry can be accounted for by introducing extra first and second generation of FL coefficients into Eq.~\eqref{pss} in the spirit of Ref.~\cite{Mora_Clerk_Hur_PRB(80)_2009}. The $n$-th generation of FL coefficients refers to the $n$-th order coefficients in the Taylor expansion of the scattering phase shifts with respect to the energy. Moreover for the description of FCS beyond PH symmetric point, the density-density inter-channel interaction should be added. The finite potential scattering amounts to renormalizes the resonance phase shifts in such a way that $\delta^a_0\to \delta^a_0+\delta^a_P$, $\delta^a_P\ll\delta^a_0$
\cite{Karki_MK_TE}.} 
\vspace*{-2mm}
\section{Full counting statistics}\label{fcs}
\vspace*{-2mm}
The randomness of transferred charge ($q$) through a nanodevice during a measurement time ($\mathcal{T}$) is specified by the PDF, $\mathcal{P}(q)$. Then, the central object of FCS, the MGF is given by
\begin{equation}
\chi(\lambda){=}\sum_q \mathcal{P}(q)e^{i\lambda q}.
\end{equation}
Here, $\lambda$ is the charge counting field.
Following the spirit of pioneering works~\cite{Gogolin_Komnik_PRL(2006), Gogolin_Komnik_PRB(2006)}, we conceal the 2SK many-body Hamiltonian into the MGF (see text below). The complete charge transferred statistics of 2SK effect is, then, obtained via cumulant generating function (CGF) $\ln \chi(\lambda)$. The $n$-th order differentiation of CGF with respect to the counting field, delivers the arbitrary moment (central) of charge current. Besides, the counting field, $\lambda$, is explicitly time dependent which takes different value in forward ($\mathcal{C}_-$) and backward ($\mathcal{C}_+$) Keldysh contour 
\begin{align}\label{cfield}
\lambda(t)&=\begin{cases}
\phantom{-}\lambda ,& \text{if } 0<t<\mathcal{T}\; \text{and}\; t\in \mathcal{C}_- \\
-\lambda,&  \text{if } 0<t<\mathcal{T}\; \text{and}\; t\in \mathcal{C}_+\\
\phantom{-}0,& \text{else}
\end{cases} 
\end{align}
Here the Keldysh contour extends from $-\infty$ to $\mathcal{T}$ and back to $\infty$. \DK{Note that, in order to calculate the FCS, the current measurement device has
to be included in the Hamiltonian description. Such terms in the Hamiltonian due to the measuring device can be eliminated by means of unitary transformation of the form $U\sim e^{-i\lambda(t)\hat{N}_{\alpha}}$, $\hat{N}_{\alpha}$ being the number operator of the electrons in $\alpha$ reservoir~\cite{Gogolin_Komnik_PRB(2006)}. This transformation changes only the tunneling part of the Hamiltonian Eq.~\eqref{model}. Analogously, in the strong-coupling regime the charge
measuring field causes the rotation of the even and odd electron states in the reservoirs such that~\cite{Gogolin_Komnik_PRL(2006)}}
\begin{align}
b^{\lambda}_{a}&=\cos\left(\lambda/4\right)b_a-i\sin\left(\lambda/4\right)b_{\bar{a}}.
\end{align}
Where, for simplicity, we omit the spin degrees of freedom. Under this transformation the free part of Hamiltonian, $H_0$, remains unchanged. Nevertheless, the Hamiltonian corresponding to the sum of scattering and interaction effects , $\mathcal{H}$ ($\equiv H_{\rm el}+H_{\rm in}$), transforms to $ \mathcal{H}^{\lambda}=\mathcal{H}+\lambda/4\; \hat{I}_{\rm bs}$. Here, we considered only the lowest order terms in counting field. The backscattering current, $\hat{I}_{\rm bs}$, is given by the commutator $\hat{I}_{\rm bs}=i\left[Q, H\right]$, where $Q$ is the charge transferred operator across the junction
$Q=1/2\sum_{k\sigma}(b^{\dagger}_{ek\sigma}b_{ok\sigma}+\text{H.c.})$. Since there are no zeroth order transmission processes in 2SK \cite{KMDK_2018}, the MGF is given by
\begin{equation}\label{cdf1}
\chi(\lambda)=\Big\langle T_C \exp\left[-i\int_C \mathcal{H}^{\lambda}( t)dt\right]\Big\rangle_0.
\end{equation}
Where $T_C$ is time ordering operator in Keldysh contour, $C$. The expansion of Eq.~\eqref{cdf1} in $\mathcal{H}^{\lambda}$ and use of Wick's theorem paved the way to proceed with the perturbative study of MGF, $\chi(\lambda)$. Then the $n$-th order (arbitrary) moment of charge current is given by
\begin{equation}\label{mgfx}
\mathcal{C}_n=\left.\frac{1}{\mathcal{T}}(-i)^{n}\frac{d^{n}}{d\lambda^{n}}\ln \chi(\lambda)\right|_{\lambda=0}.
\end{equation}
\setlength\belowcaptionskip{-3ex}
\begin{figure}
\begin{tabular}{|c|}
\hline 
\;\;\;\;\;\;\;\;\;\;\;\;\;\;\;\;\;\;\\
\includegraphics[width=2.0cm, valign=c]{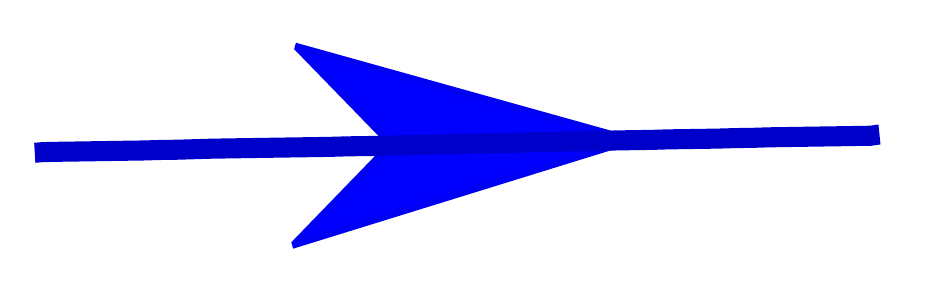}  $\equiv $ Green's function of channel-e ($\check{\mathcal{G}}_{ee}$)\\
\includegraphics[width=2.0cm, valign=c]{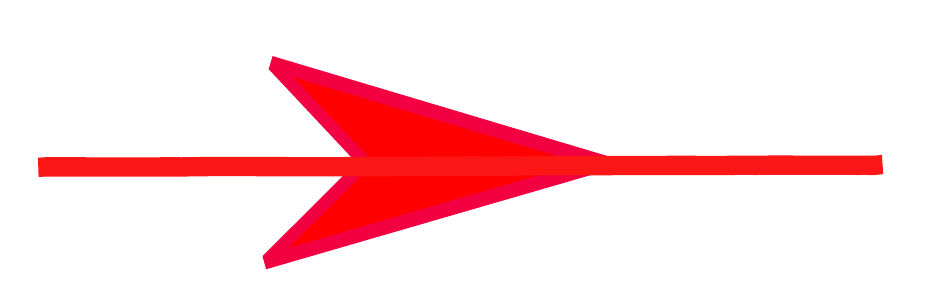}  $\equiv $ Green's function of channel-o ($\check{\mathcal{G}}_{oo}$)\\
\includegraphics[width=2.0cm, valign=c]{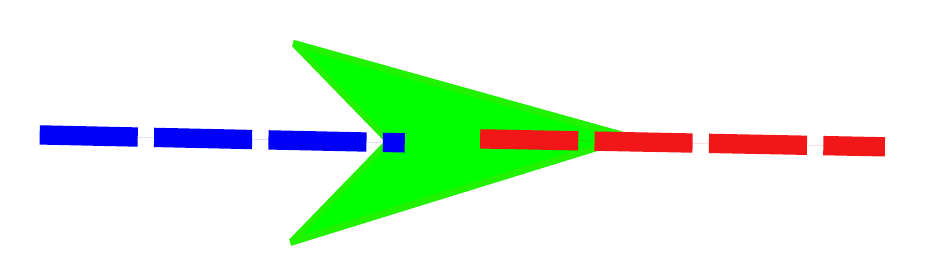} $\equiv $  Mixed Green's function ($\check{\mathcal{G}}_{eo}/\check{\mathcal{G}}_{oe}$)\;\;\\
\includegraphics[width=2.0cm, valign=c]{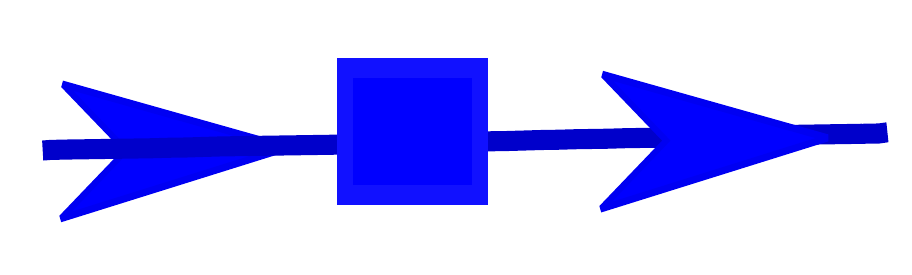} $\equiv $  Elastic scattering in channel-e ($\alpha_e$)\\
\includegraphics[width=2.0cm, valign=c]{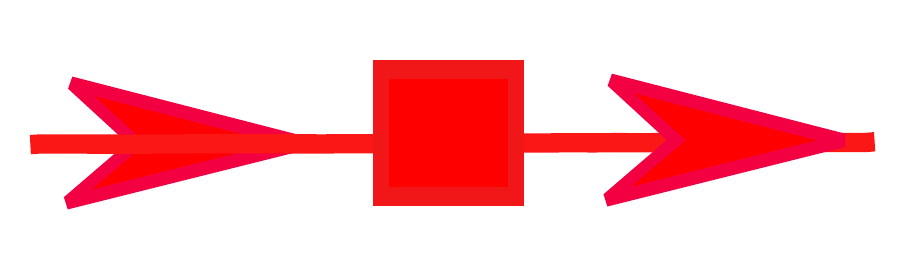} $\equiv $ Elastic scattering in channel-o ($\alpha_o$)\\
\includegraphics[width=2.0cm, valign=c]{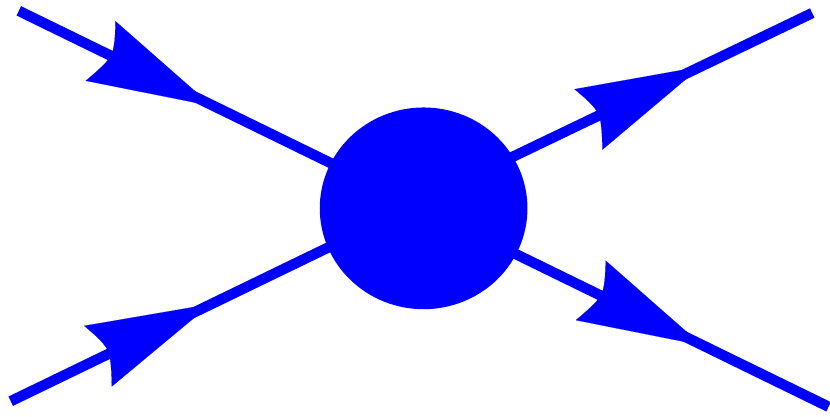} $\equiv $  Inelastic scattering in channel-e ($\phi_e$)\\
\includegraphics[width=2.0cm, valign=c]{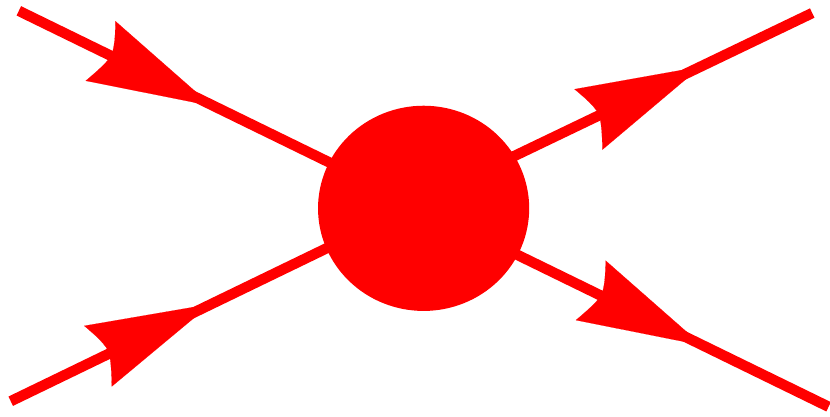} $\equiv$  Inelastic scattering in channel-o ($\phi_o$)\\
\includegraphics[width=2.0cm, valign=c]{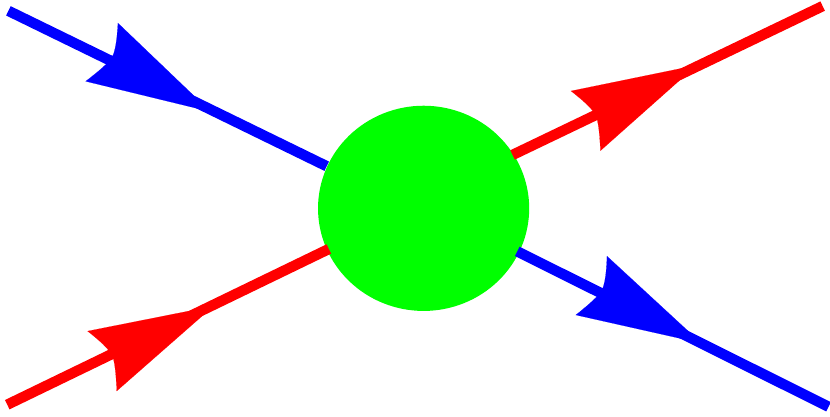} $\equiv$ Inter-channel inelastic scattering ($\Phi$)
 \\ 
\hline 
\end{tabular} 
\caption{Feynman diagrammatic codex used for the calculation of FCS in the presence of two conduction mode.}\label{tablecodex}
\end{figure}

To proceed with the calculation for the higher cumulants of charge current, we require Keldysh Green's functions (GFs) in $\lambda$-rotated basis. As the odd conduction channel remains completely decoupled, the Keldysh GFs of even channel ($\check{\mathcal{G}}_{ee}$) suffice to characterize the transport of 1CK schemes. However, the persistent treatment of 2SK effect requires two additional Keldysh GFs, the Keldysh GFs of odd channel ($\check{\mathcal{G}}_{oo}$)  and that of mixed channel ($\check{\mathcal{G}}_{eo/oe}$).
Note that, the spin index in these GFs is implied. Besides, we prefer the renaming of GFs
$\check{\mathcal{G}}_{ee}$ and $\check{\mathcal{G}}_{oo}$ as the channel-diagonal GFs, and $\check{\mathcal{G}}_{eo/oe}$ as mixed GFs, whenever necessary.
The energy representation of these Keldysh GFs is
\begin{equation}
\check{\mathcal{G}}_{aa/a\bar{a}}(\varepsilon)=
\begin{bmatrix}
\mathcal{G}^{--}_{aa/a\bar{a}}(\varepsilon)\;\;\;\;\; & \mathcal{G}^{-+}_{aa/a\bar{a}}(\varepsilon)\\
\mathcal{G}^{+-}_{aa/a\bar{a}}(\varepsilon)\;\;\;\;\; & \mathcal{G}^{++}_{aa/a\bar{a}}(\varepsilon)
\end{bmatrix}.
\end{equation}
Where, the diagonal GFs,
\begin{equation}\label{matrixgf}
\mathcal{G}^{--}_{aa/a\bar{a}}(\varepsilon)=\mathcal{G}^{++}_{aa/a\bar{a}}(\varepsilon)=i\pi\nu\left[\left(f_S-1/2\right){\pm} \left(f_D-1/2\right)\right],
\end{equation}
are independent of counting field $\lambda$. Here, $f_{S/D}\equiv f_{S/D}(\varepsilon)$ is the free-electron Fermi distribution function of source/drain reservoir.
The off-diagonal GFs, explicitly depend on $\lambda$, are given by
\begin{align}
\mathcal{G}^{+-}_{aa/a\bar{a}}(\varepsilon)=e^{i\lambda/2}\left(f_S-1\right)&\pm e^{-i\lambda/2}\left(f_D-1\right).\label{man1}\\
\mathcal{G}^{-+}_{aa/a\bar{a}}(\varepsilon)=e^{-i\lambda/2}f_S&\pm e^{i\lambda/2}f_D.\label{man2}
\end{align}
The pictorial representation of these GFs are shown in Fig.~\ref{tablecodex}. Neither of the above GFs included the principal parts, since they does not contribute to the local quantities in the flat band model~\cite{Gogolin_Komnik_PRL(2006)}. The Fourier transformation (FT) of Eq.~\eqref{man2} into the real time permits,
\begin{align}\label{twogfs}
\mathcal{G}^{-+}_{aa/a\bar{a}}(t)=\mp\pi\nu T\; \frac{e^{i\left(\frac{\lambda}{2}{+}\frac{eV}{2}t\right)}\pm e^{-i\left(\frac{\lambda}{2}{+}\frac{eV}{2}t\right)}}{2\sinh\left(\pi T t\right)}.
\end{align}
The singularity in Eq.~\eqref{twogfs} is removed by shifting the contour of integration upward from the origin such that $ t\rightarrow t+i\eta$ for $\eta\rightarrow 0$. The GFs $\mathcal{G}^{+-}(t)$ has the analogous expression as $\mathcal{G}^{-+}(t)$~\cite{tschmidt_2}. 

\begin{figure}[t]
\centering
\begin{align}\nonumber
&\;\;\;\;\;\;\;\underbrace{\includegraphics[scale=0.1]{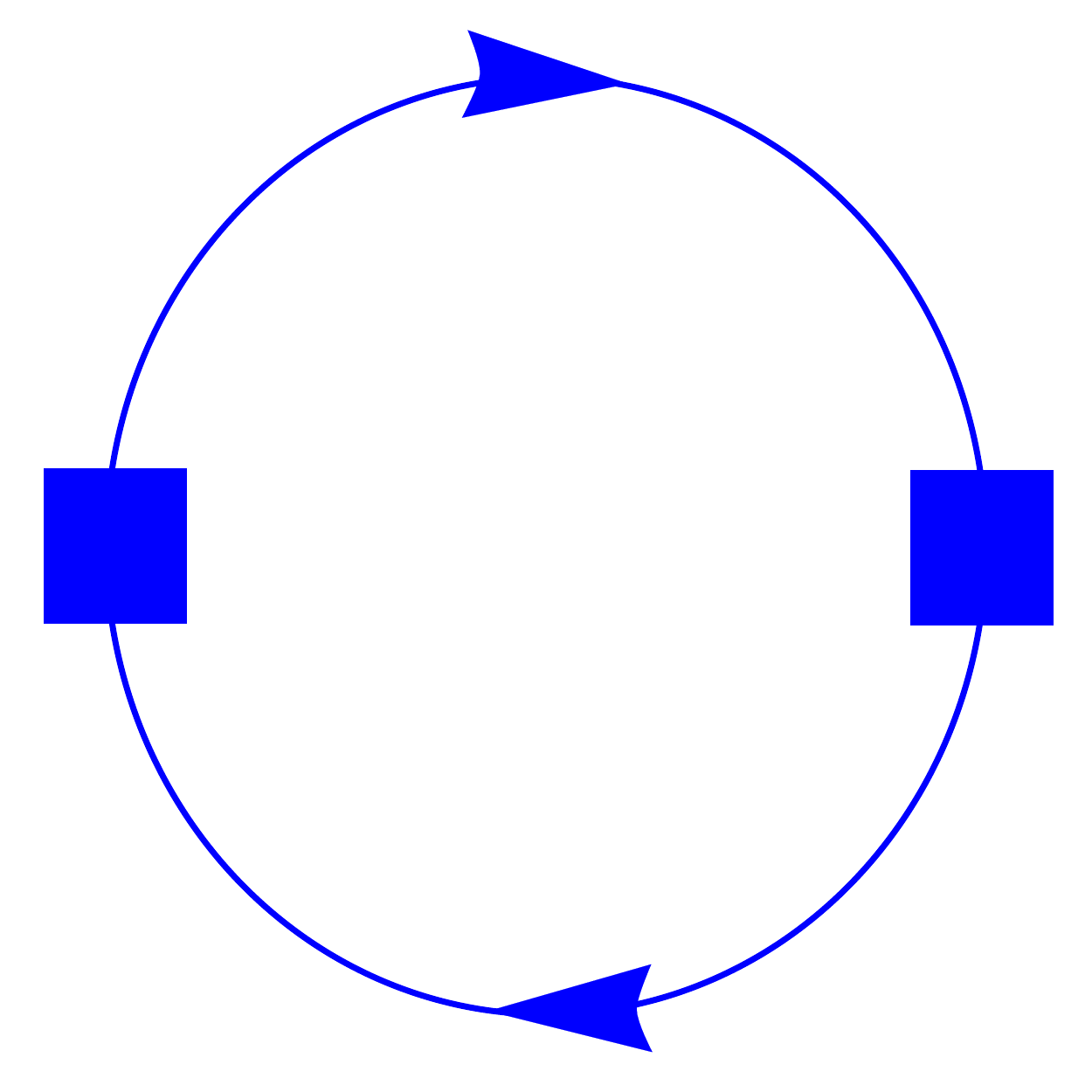}\;\;
\includegraphics[scale=0.1]{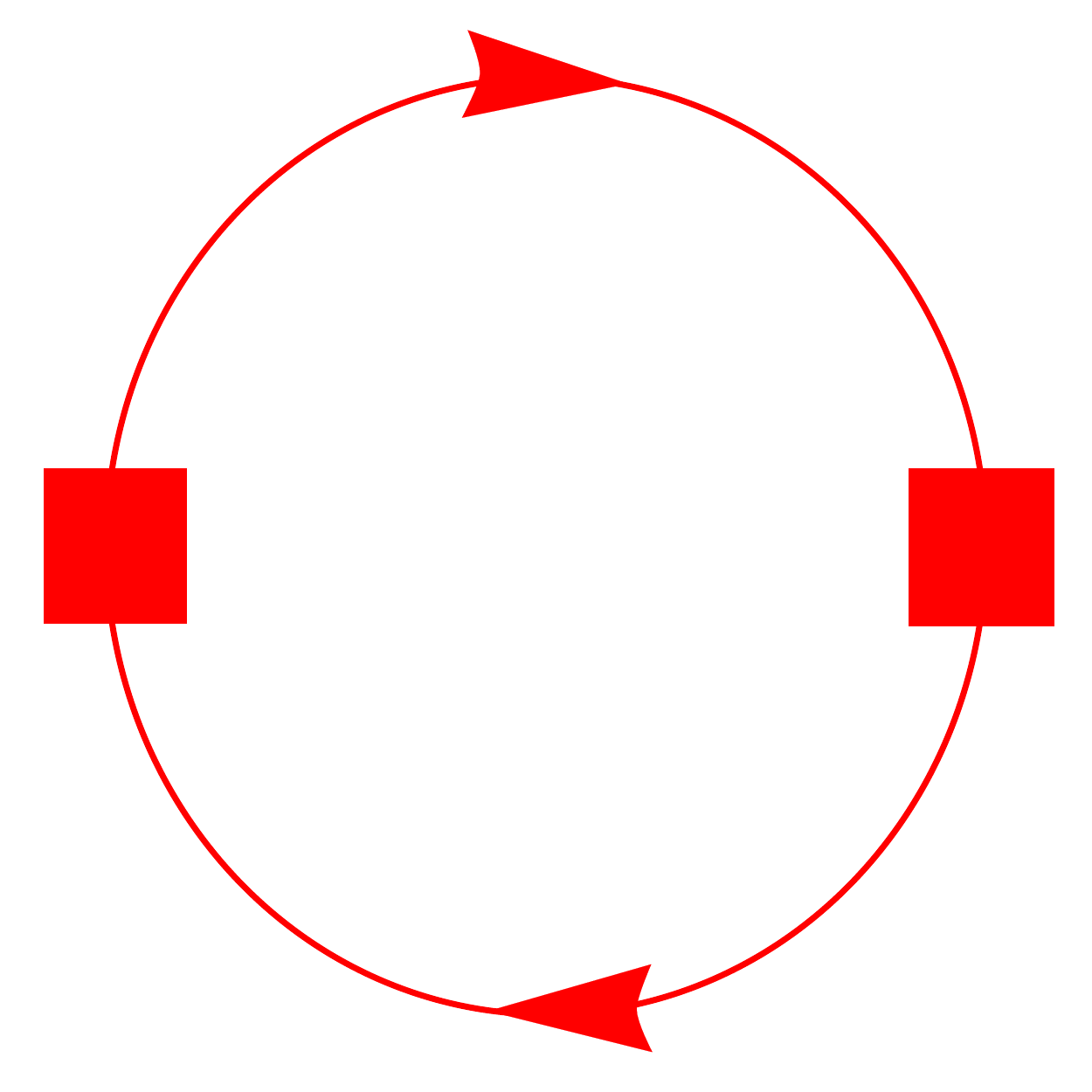}}_{\rm type-E1}\;\;\;\;\;
\underbrace{\includegraphics[scale=0.1]{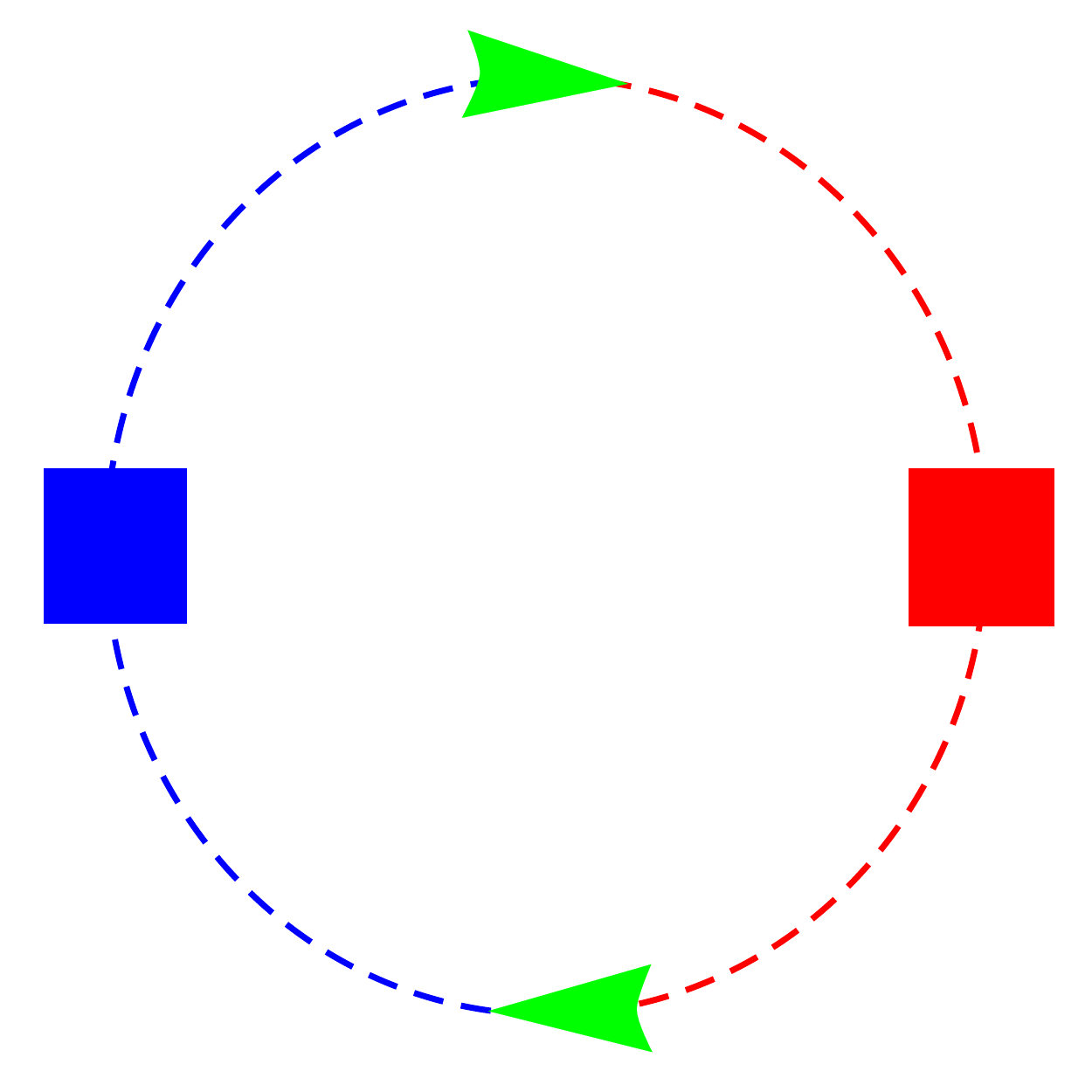}\;\;
\includegraphics[scale=0.1]{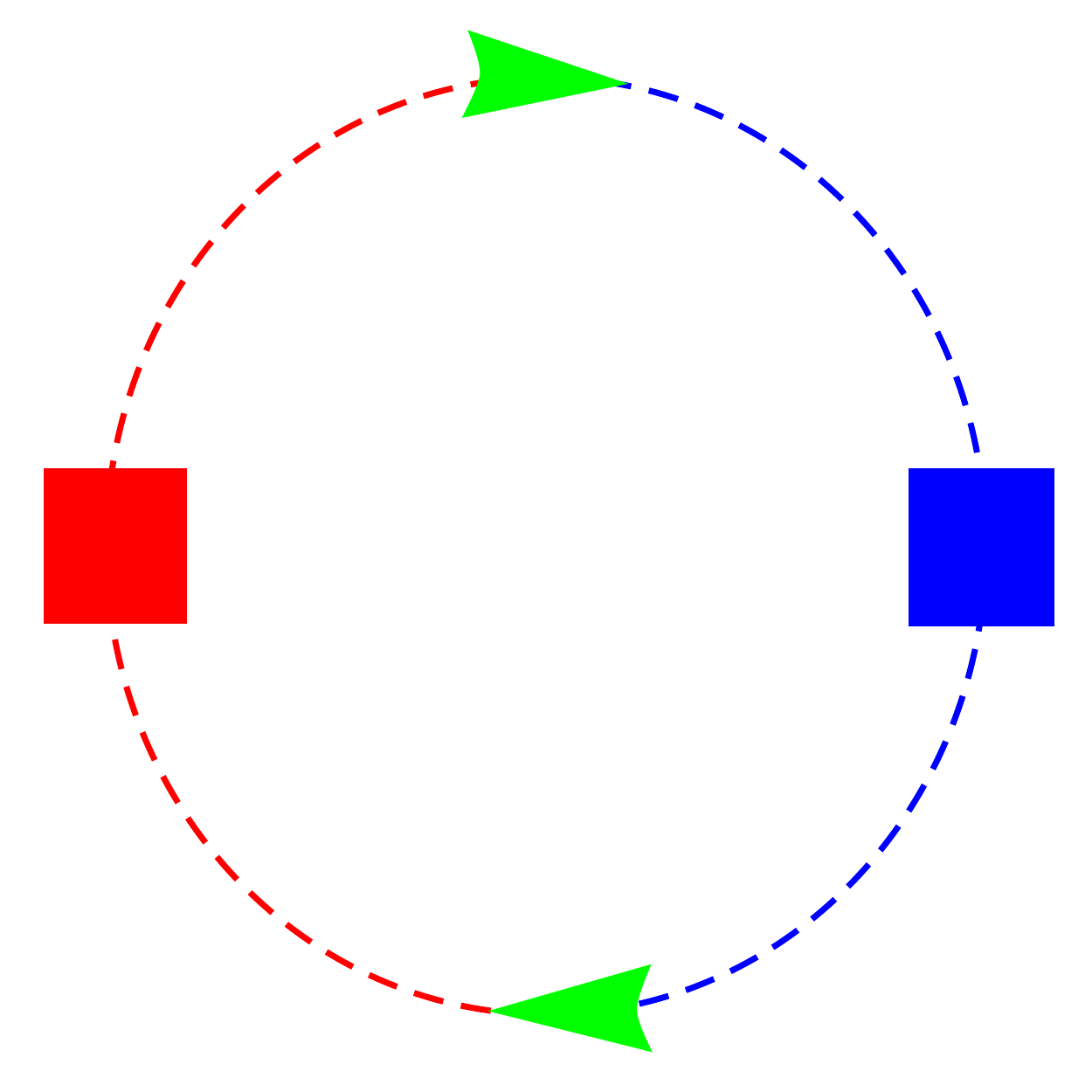}}_{\rm type-E2}\nonumber\\
&\;\;\;\;\;\;\;\underbrace{\includegraphics[width=1.8cm, valign=c]{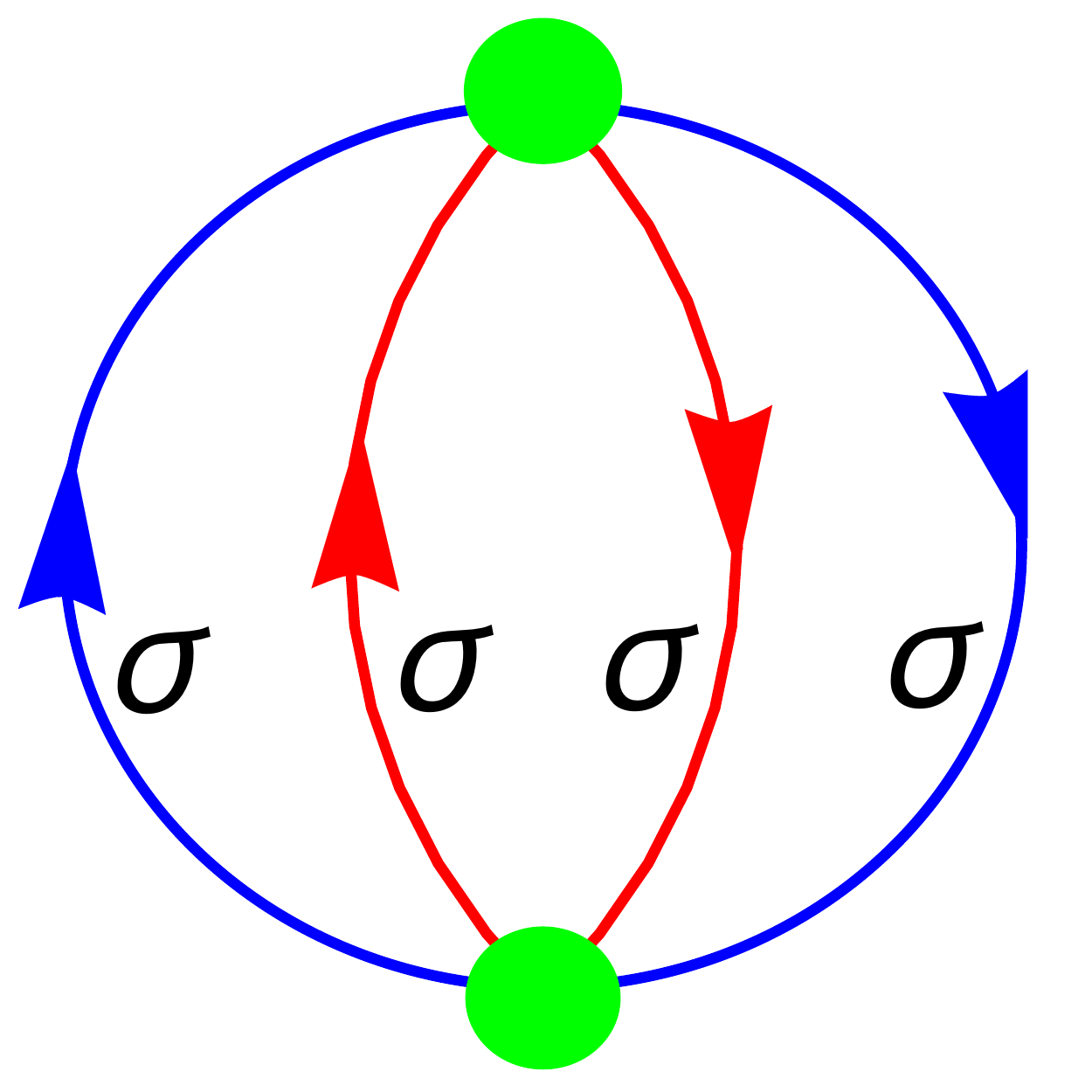}}_{\rm type-I1}\;\;\;
\underbrace{\includegraphics[width=1.8cm, valign=c]{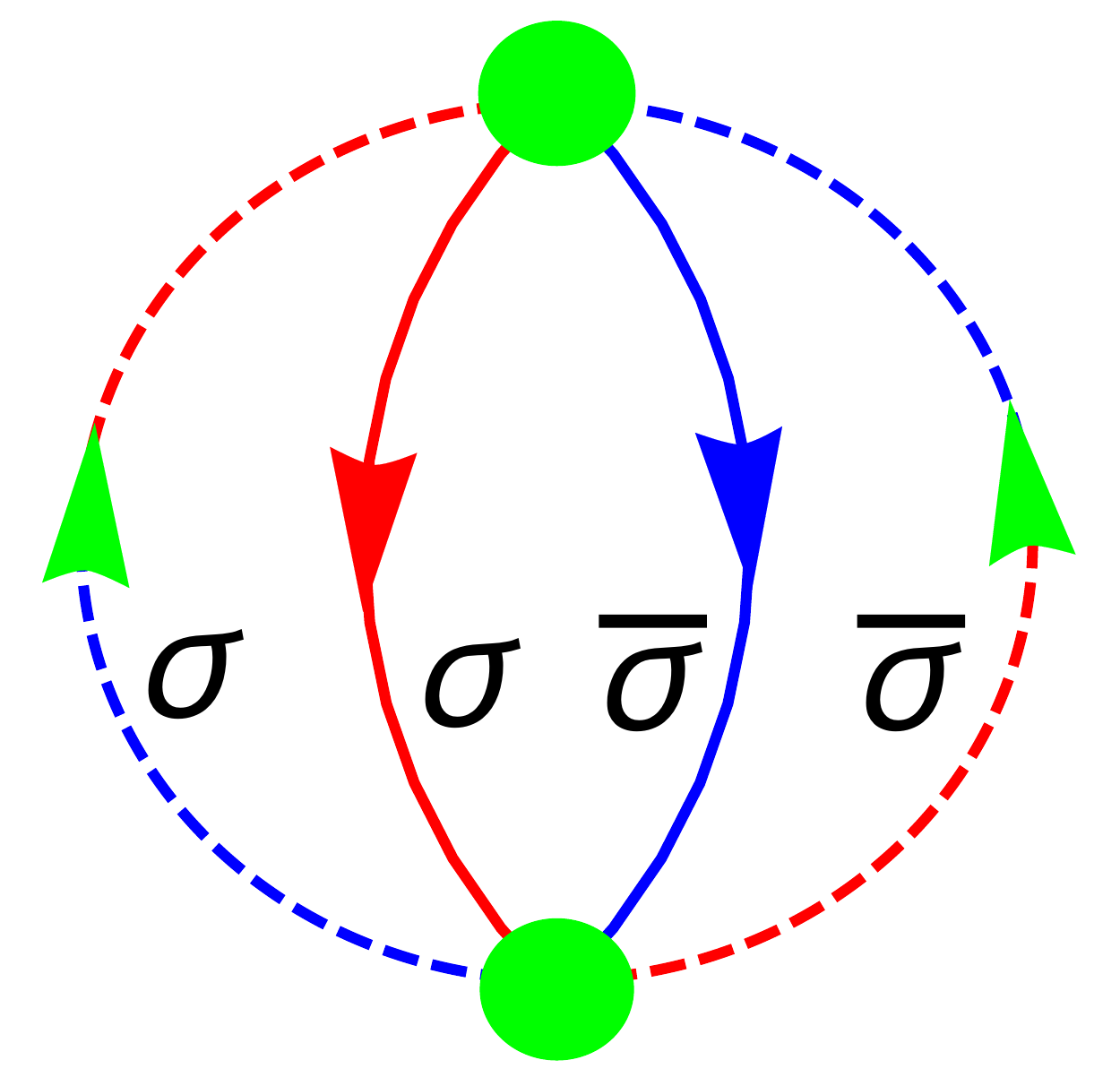}}_{\rm type-I2}\;\;\;
\underbrace{\includegraphics[width=1.8cm, valign=c]{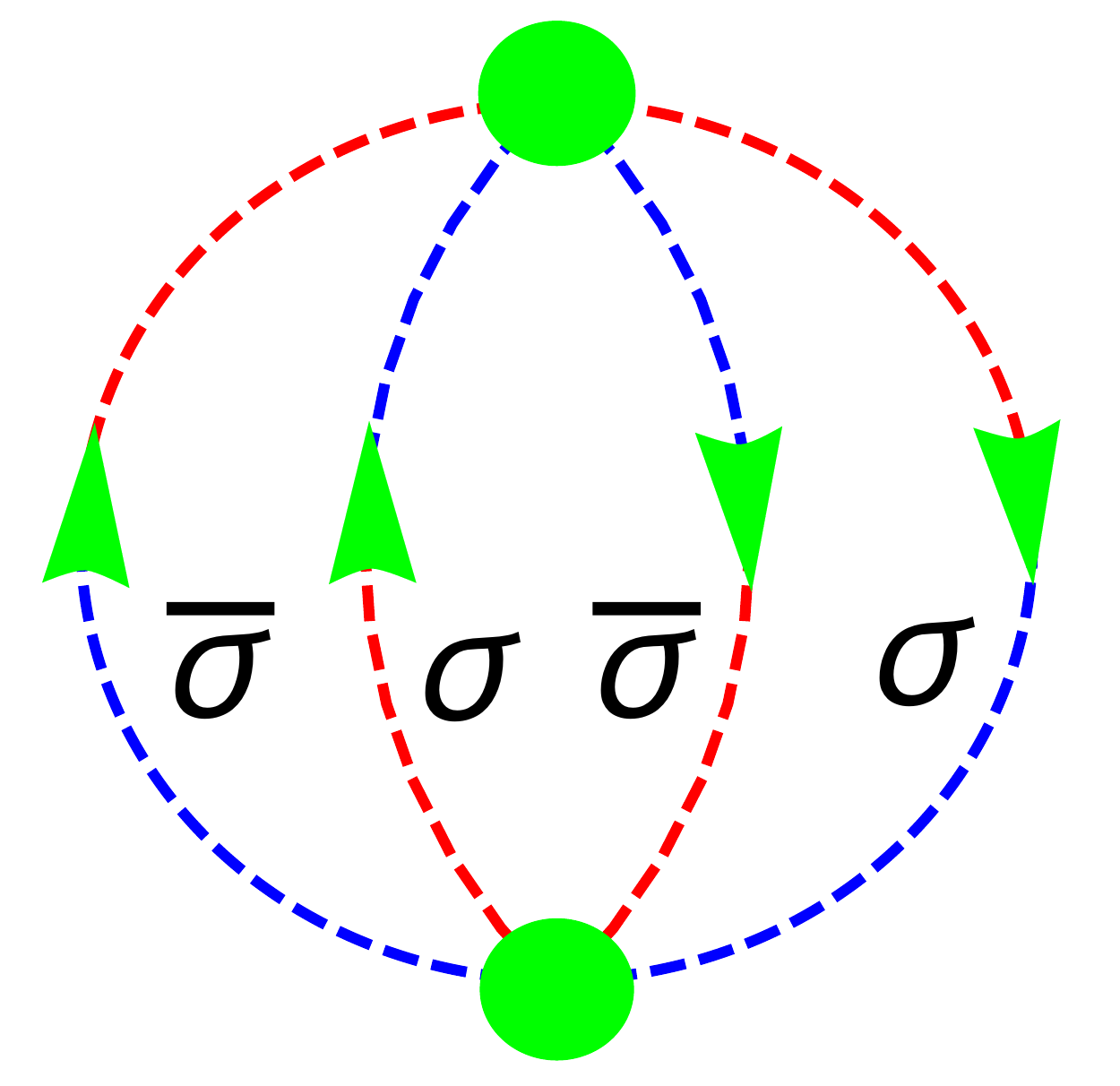}}_{\rm type-I3}\nonumber
\end{align}
\caption{Upper panel: Schematic representation of the CGF contribution of scattering effects, $\ln \chi_{\rm el}(\lambda)$. Lower panel: Topologically different diagrams accounting for the interaction contribution to the CGF, $\ln \chi_{\rm in}(\lambda)$.}\label{codex1}
\end{figure}

We substitute the scattering (elastic) part of the Hamiltonain, $H_{\rm el} $, into Eq.~\eqref{cdf1} and use the Wick's theorem to get the elastic contribution to CGF,  $\ln \chi_{\rm el}(\lambda)$. Following the diagrammatic-codex of Fig.~\ref{tablecodex}, we succeed to re-express $\ln \chi_{\rm el}(\lambda)$ in terms of two topologically different Feynman diagrams. These diagrams are classified as type-E1 and type-E2 (see upper panel of Fig.~\ref{codex1}). Following the standard technique of Feynman diagrammatic calculation with the GFs given in Eqs.~\eqref{matrixgf}, ~\eqref{man1} and~\eqref{man2}, we obtained the CGF contribution of type-E1 and type-E2 diagram. As detailed in Appendix~\ref{elastic_app}, the CGF for 2SK effect contributed by the scattering effects is
\begin{equation}\label{cgf1}
\frac{\ln \chi_{\rm el}}{(\alpha_e-\alpha_o)^2}=\frac{\mathcal{T}V}{24\pi}\frac{V^2+4(\pi T)^2}{\sinh(V/2T)}\sum_{x=\pm} \left(e^{-i\lambda x}{-}1\right)e^{xV/2T}.
\end{equation}
We have used the generalized notation $e{=}\hbar{=}k_B{=}1$ to write Eq.~\eqref{cgf1} and for the rest of discussion. Plugging in Eq.~\eqref{cgf1} into Eq.~\eqref{mgfx} and then taking the limit $T\to 0$, we bring the zero temperature contribution of scattering effects to the $n$-th moment of charge current,
\begin{equation}\label{m1}
\mathcal{C}^{\rm el}_n=\frac{V^3}{12\pi}(-1)^n\left(\alpha_e-\alpha_o\right)^2.
\end{equation}

\begin{figure}[h!]
\begin{equation}\nonumber
\begin{split}
\mathcal{C}^{\phi^2}_{n}&=\phantom{-}\phantom{-}\left[\underbrace{
\includegraphics[width=1.6cm, valign=c]{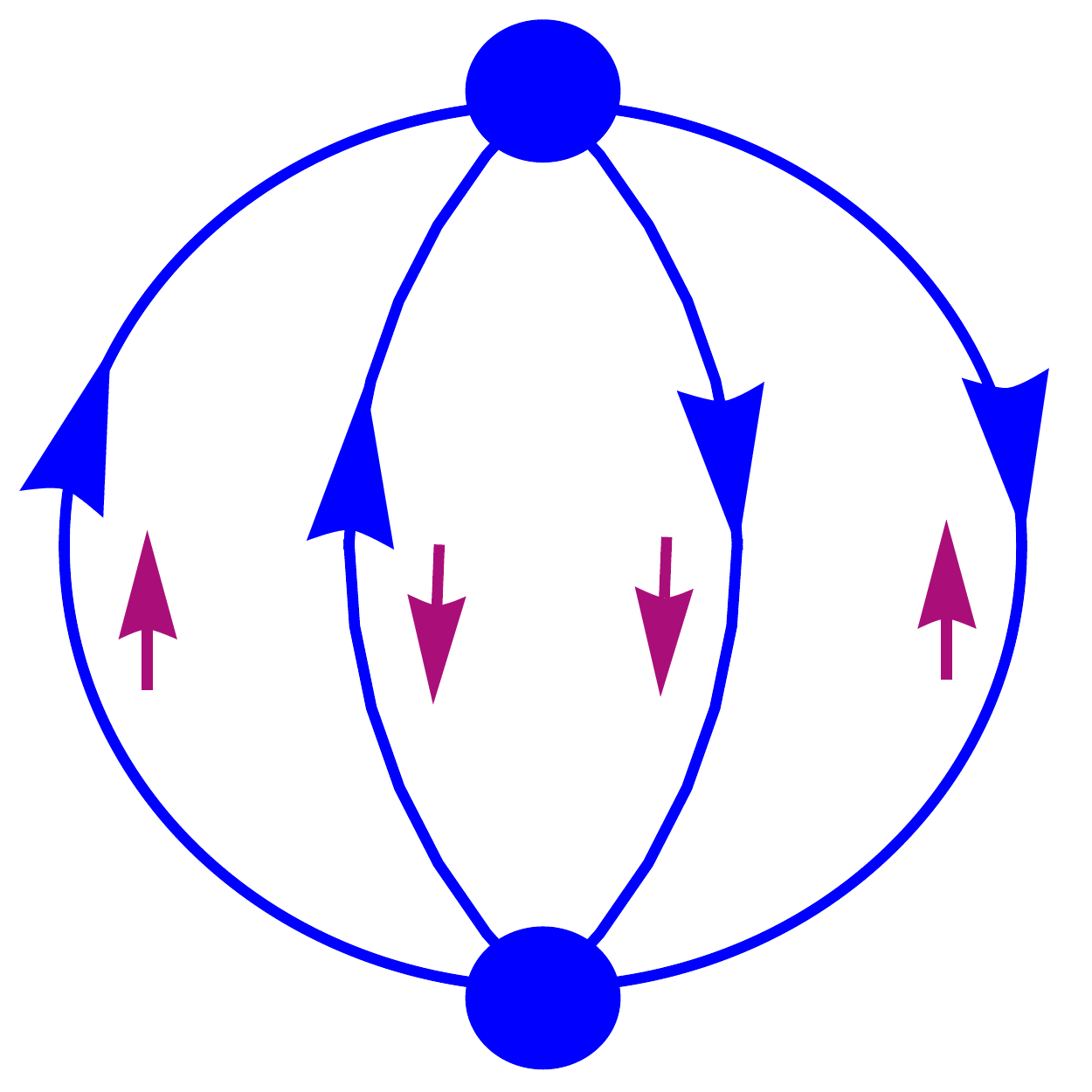}+\includegraphics[width=1.6cm, valign=c]{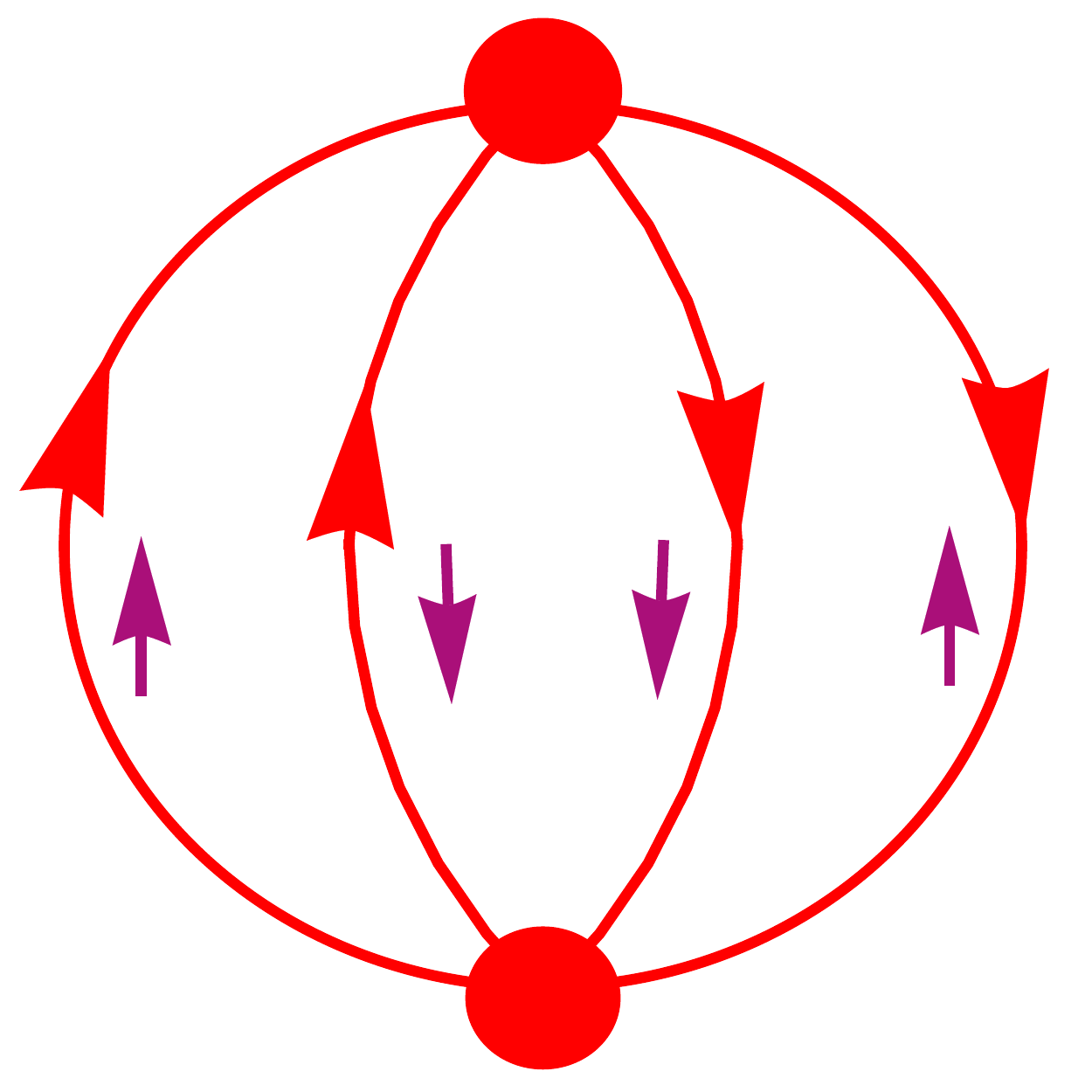}
}_{\rm type-I1}\right]\\
\mathcal{C}^{\Phi^2}_{n}&=+\frac{1}{4}\left[\underbrace{\phantom{-}
\includegraphics[width=1.6cm, valign=c]{Phisqt1_a.pdf}
+\includegraphics[width=1.6cm, valign=c]{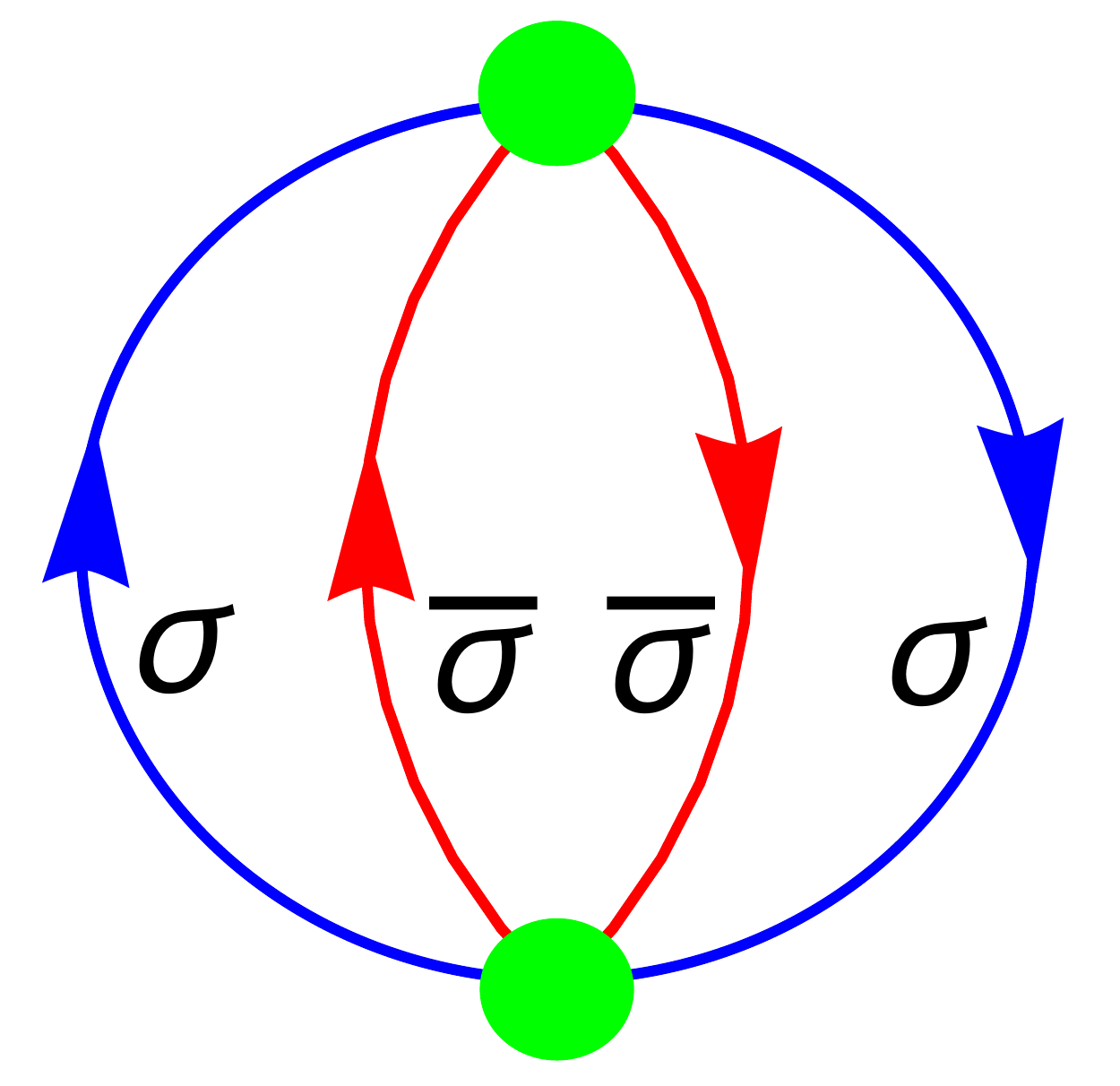}
+4\;\includegraphics[width=1.6cm, valign=c]{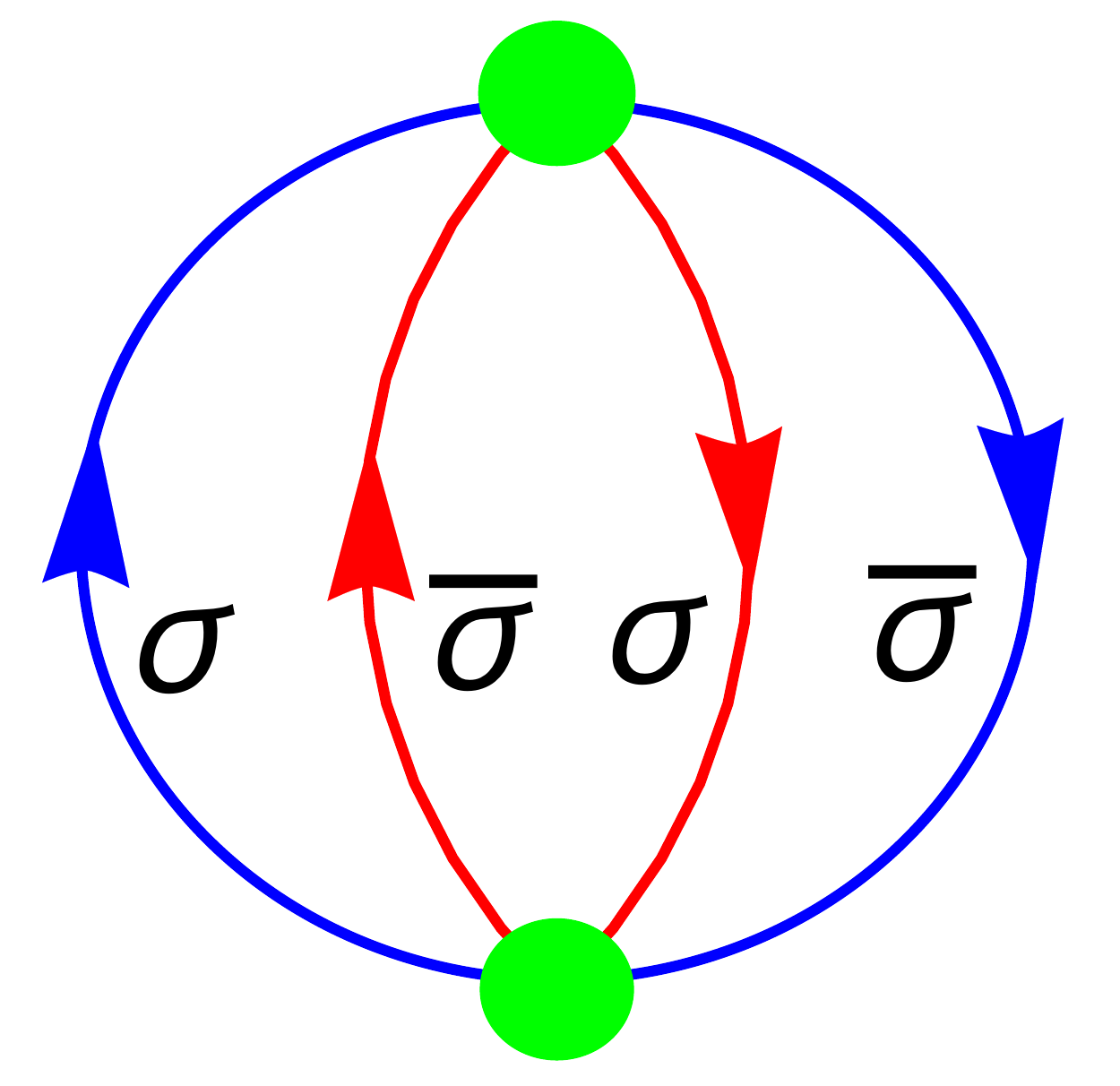}}_{\rm \text{type-I1}}
\right]\\
&
\phantom{-}+\frac{1}{4}\left[\underbrace{-
\includegraphics[width=1.6cm, valign=c]{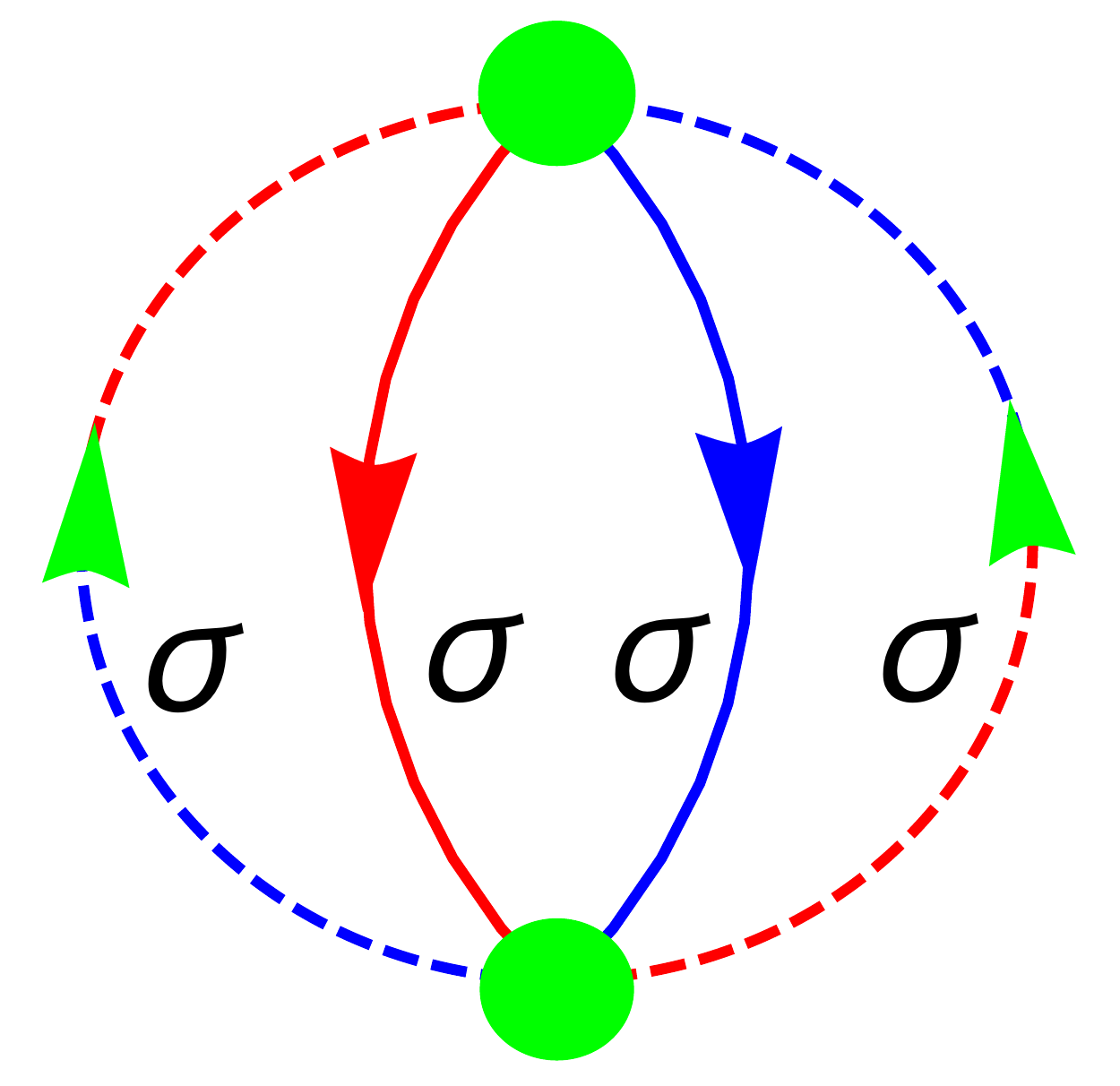}
-\includegraphics[width=1.6cm, valign=c]{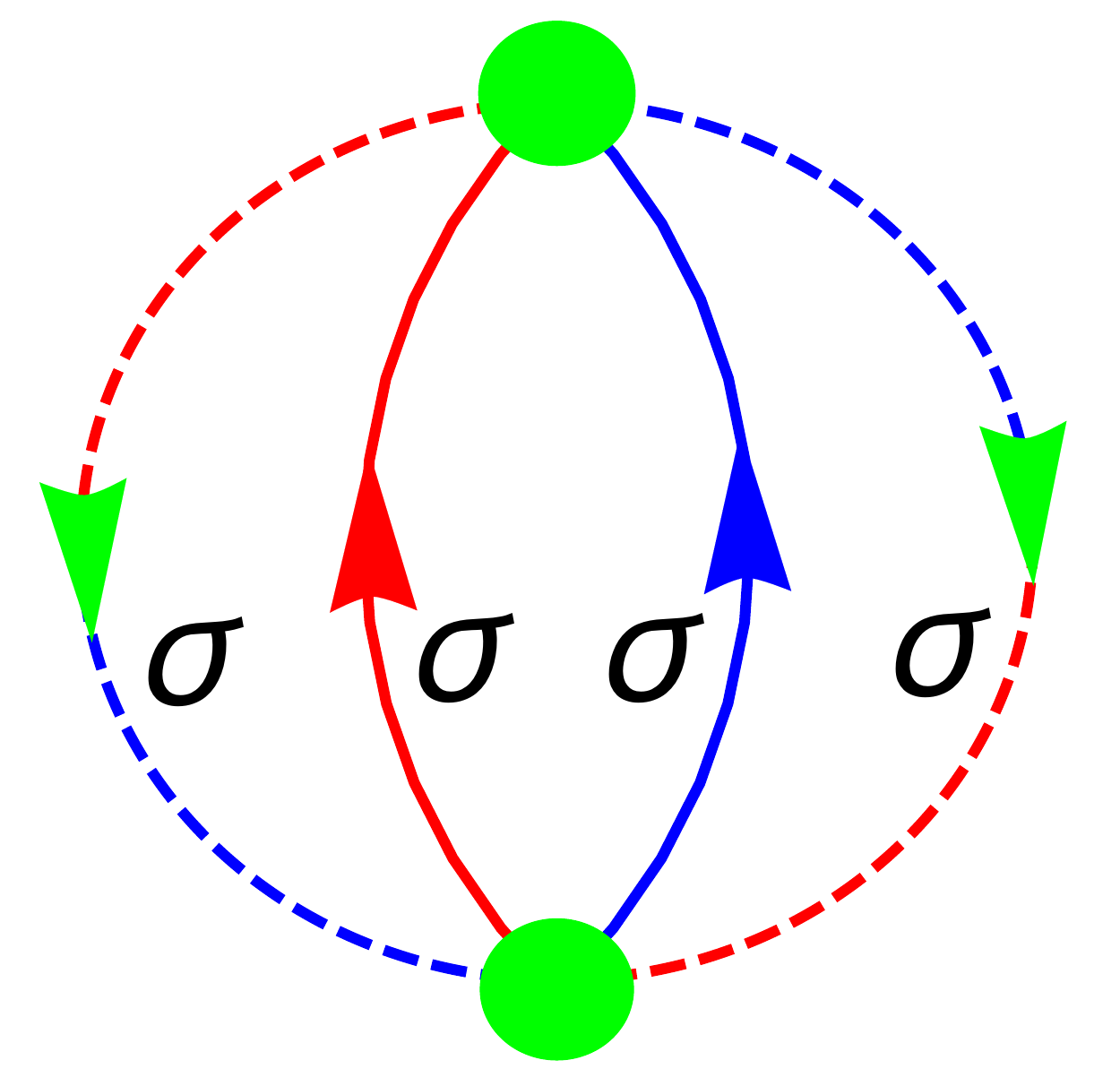}
+2\;\includegraphics[width=1.6cm, valign=c]{Phisqt2_c.pdf}}_{\rm \text{type-I2}}
\right]\\
&\phantom{-}+\frac{1}{4}\left[\underbrace{{+2}\includegraphics[width=1.55cm, valign=c]{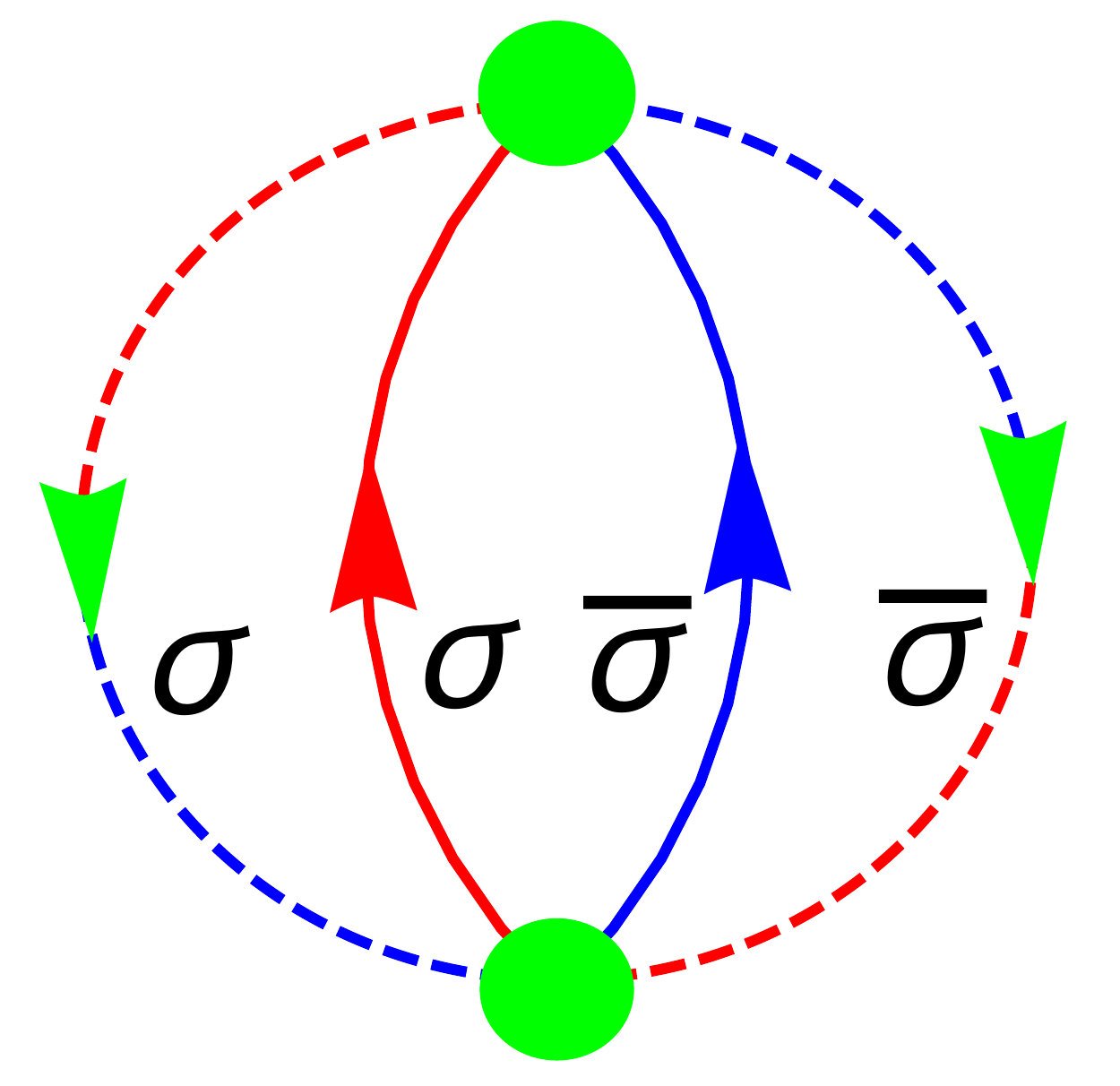}
{+2}\;\includegraphics[width=1.55cm, valign=c]{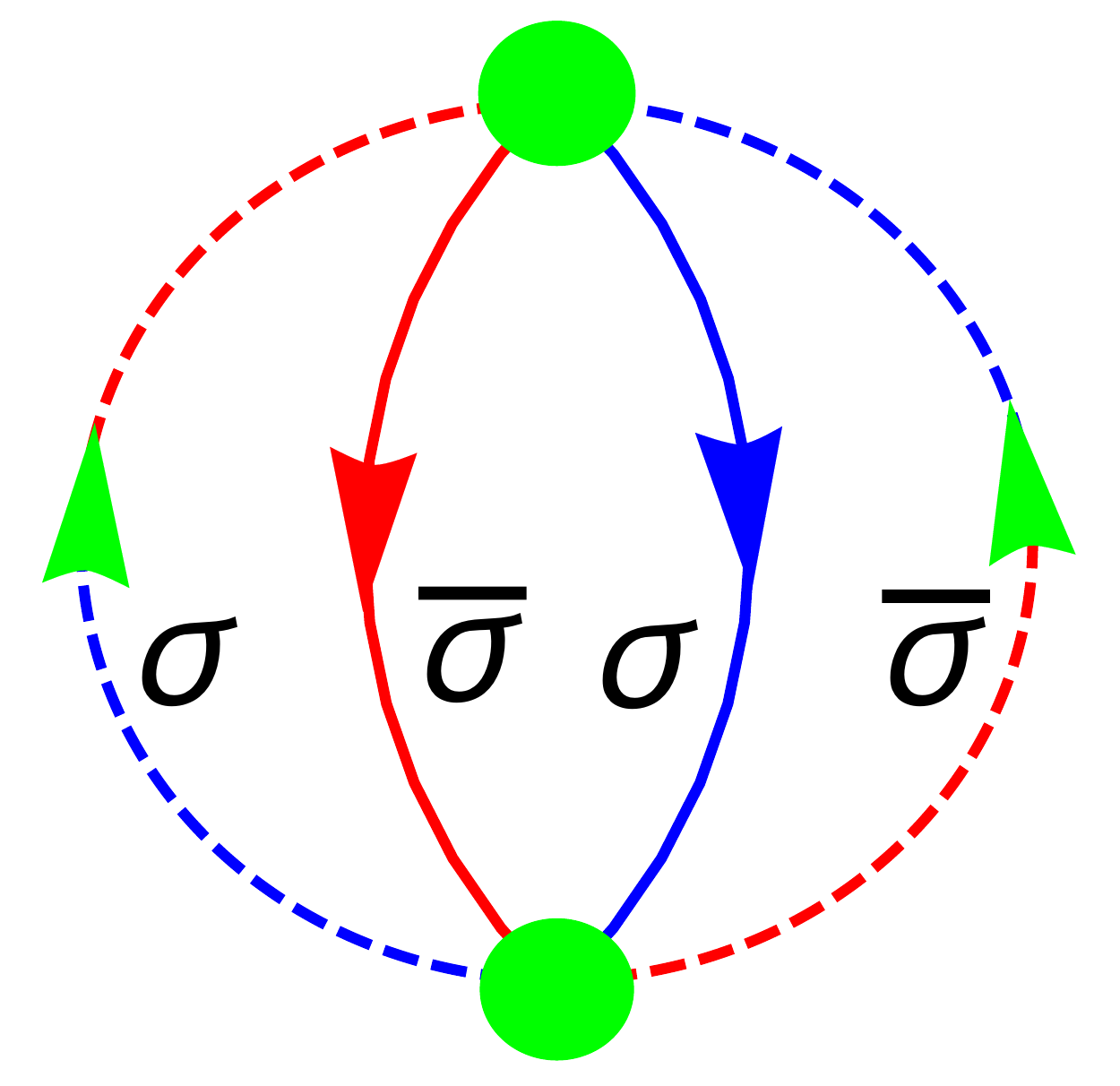}
{+2}\;\includegraphics[width=1.55cm, valign=c]{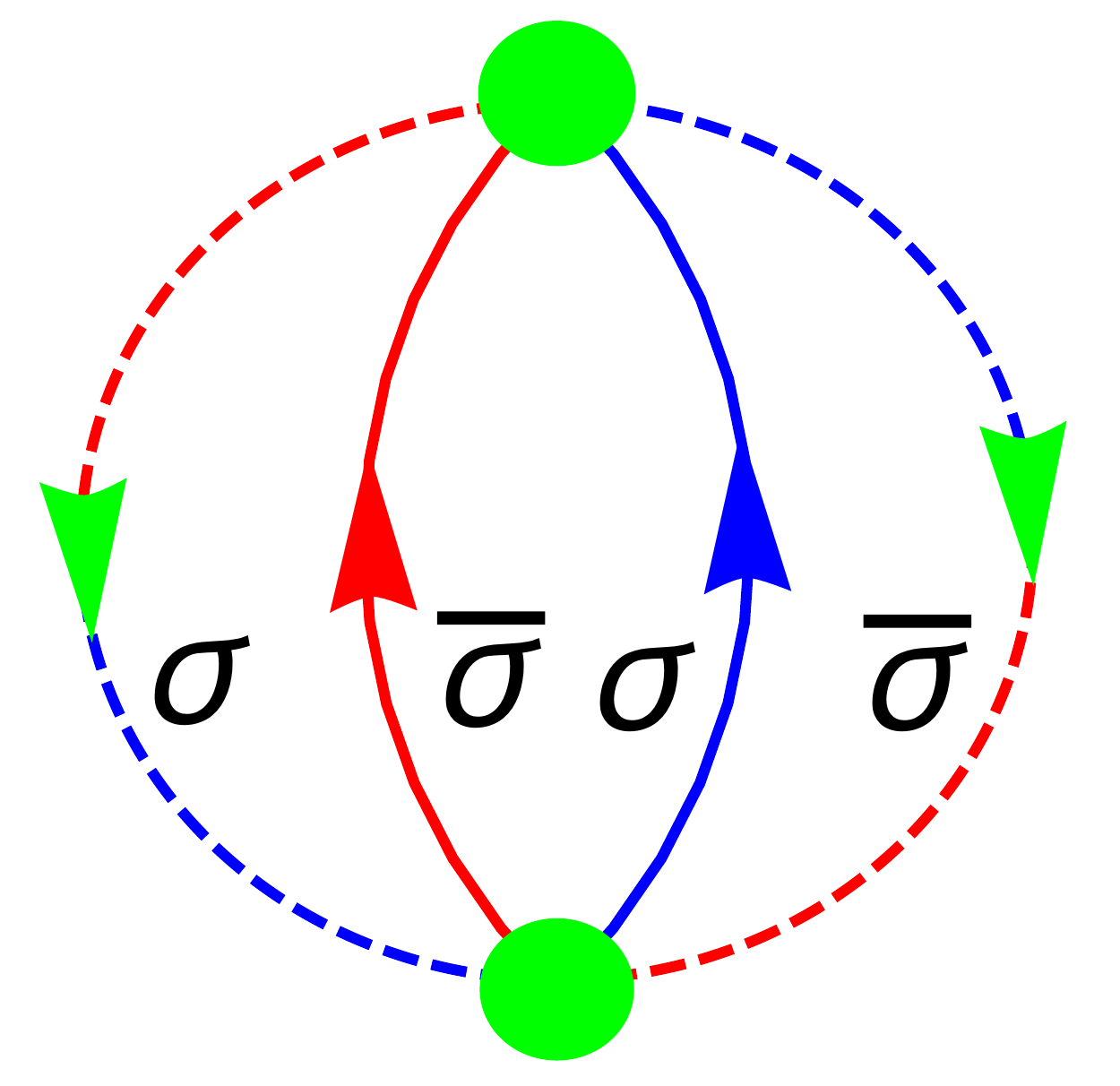}}_{\rm \text{type-I2}}\right]\\
&\phantom{-}+\frac{1}{4}\left[\underbrace{
\includegraphics[width=1.6cm, valign=c]{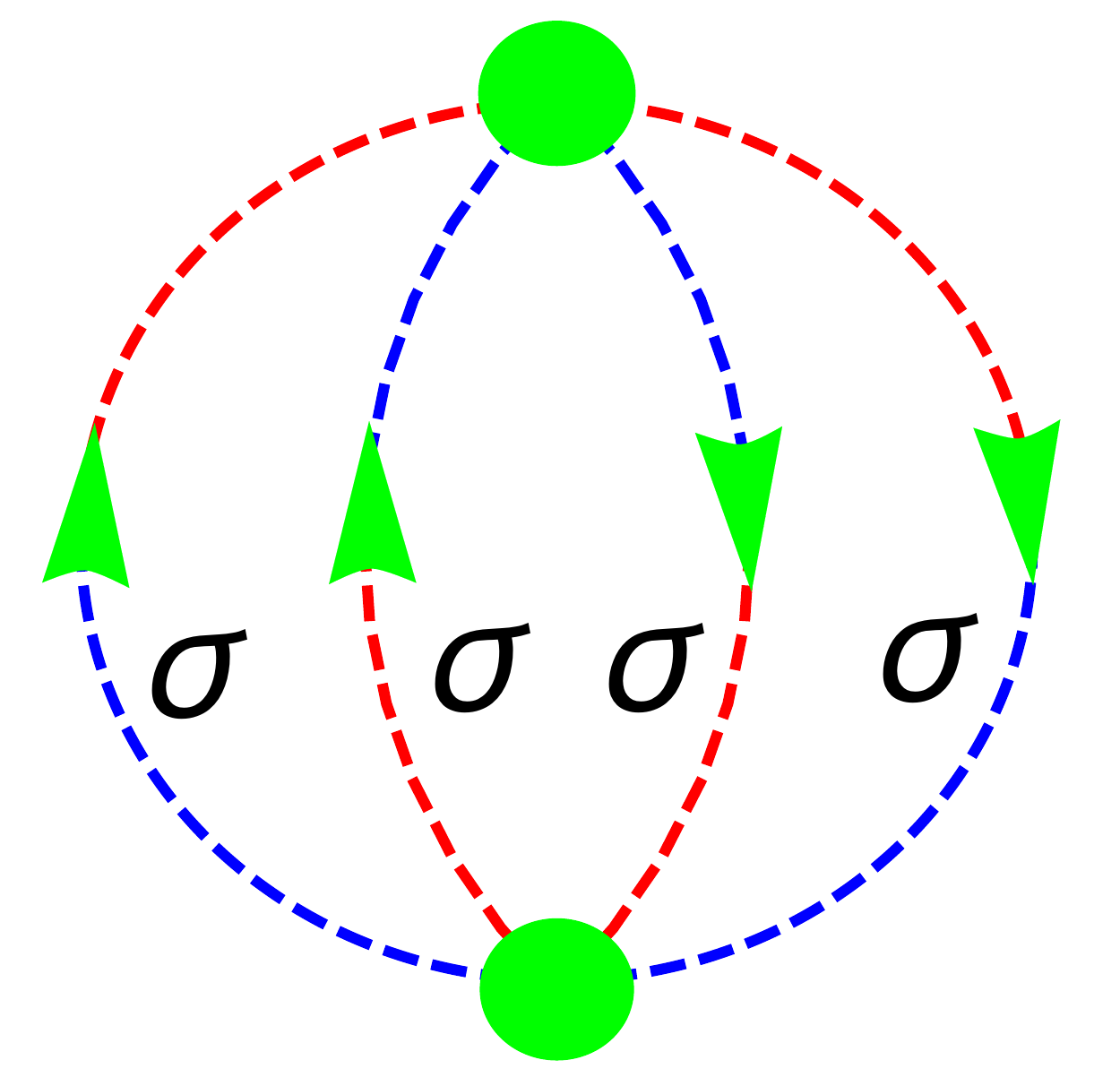}
+\;\includegraphics[width=1.6cm, valign=c]{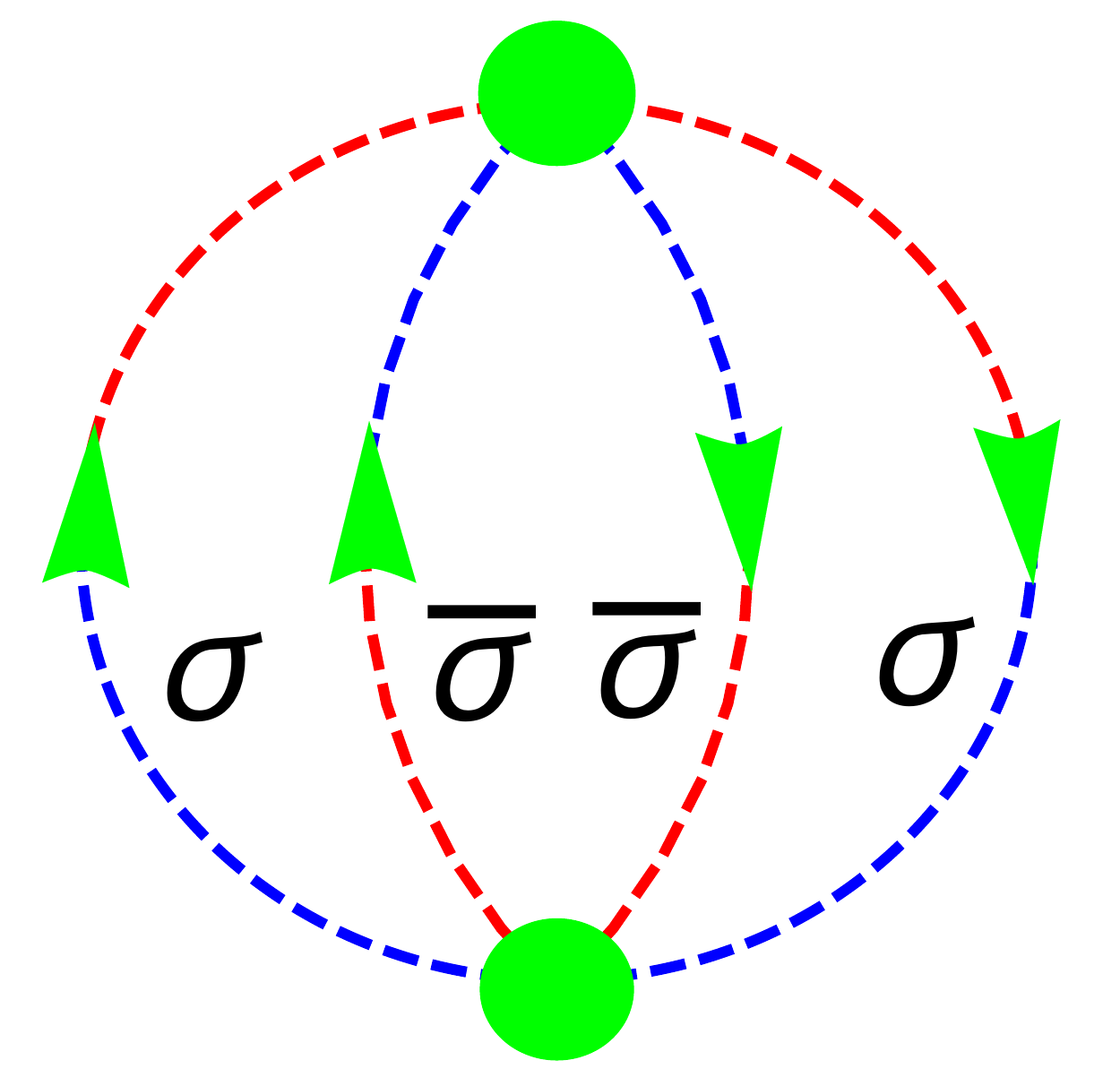}
+4\;\;\includegraphics[width=1.6cm, valign=c]{Phisqt3_c.pdf}}_{\rm \text{type-I3}}\right]\\
\mathcal{C}^{\phi_a\Phi}_{n}&={-}\frac{1}{2}\left[\underbrace{
+\includegraphics[width=1.7cm, valign=c]{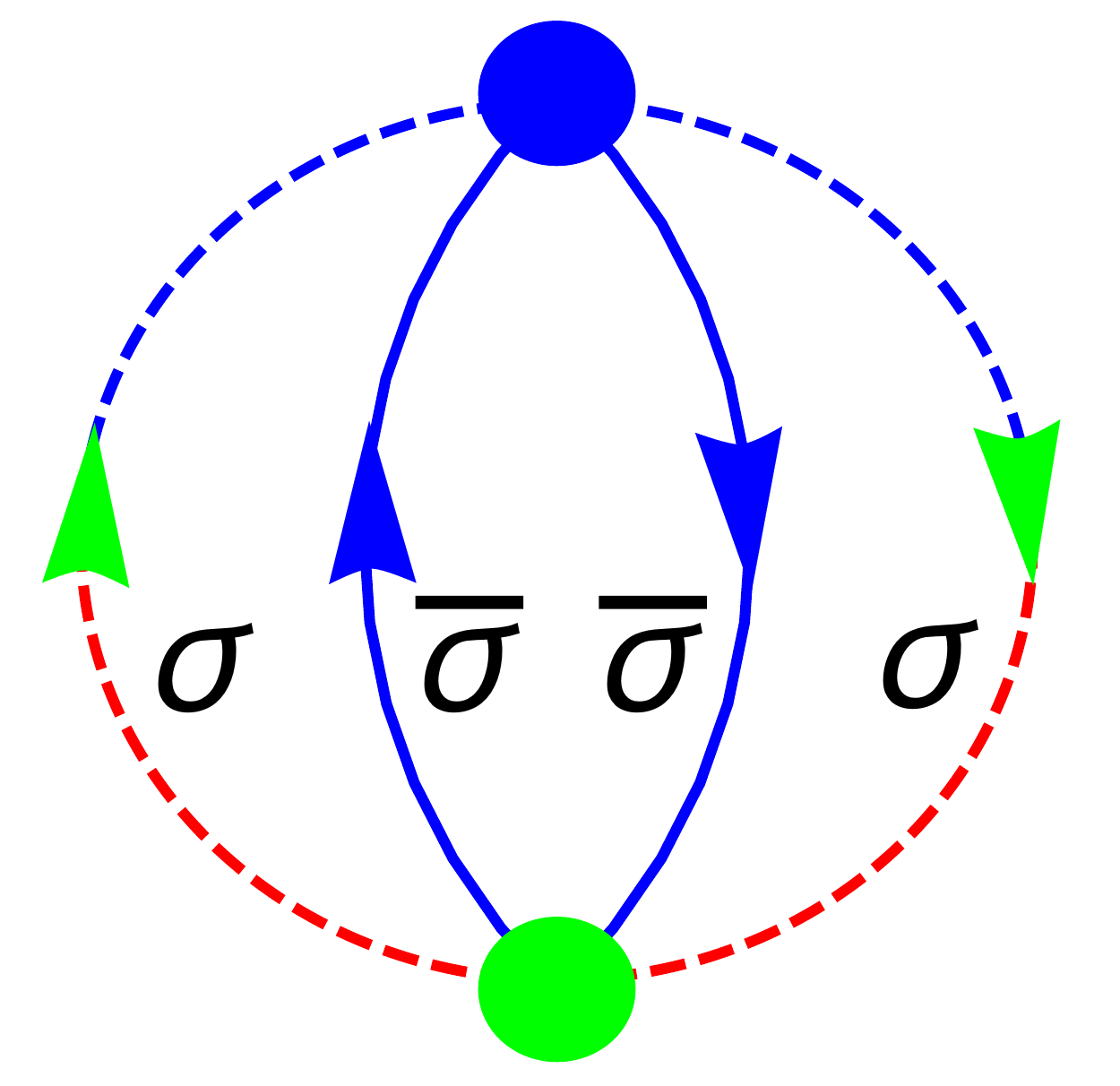}
+2\includegraphics[width=1.7cm, valign=c]{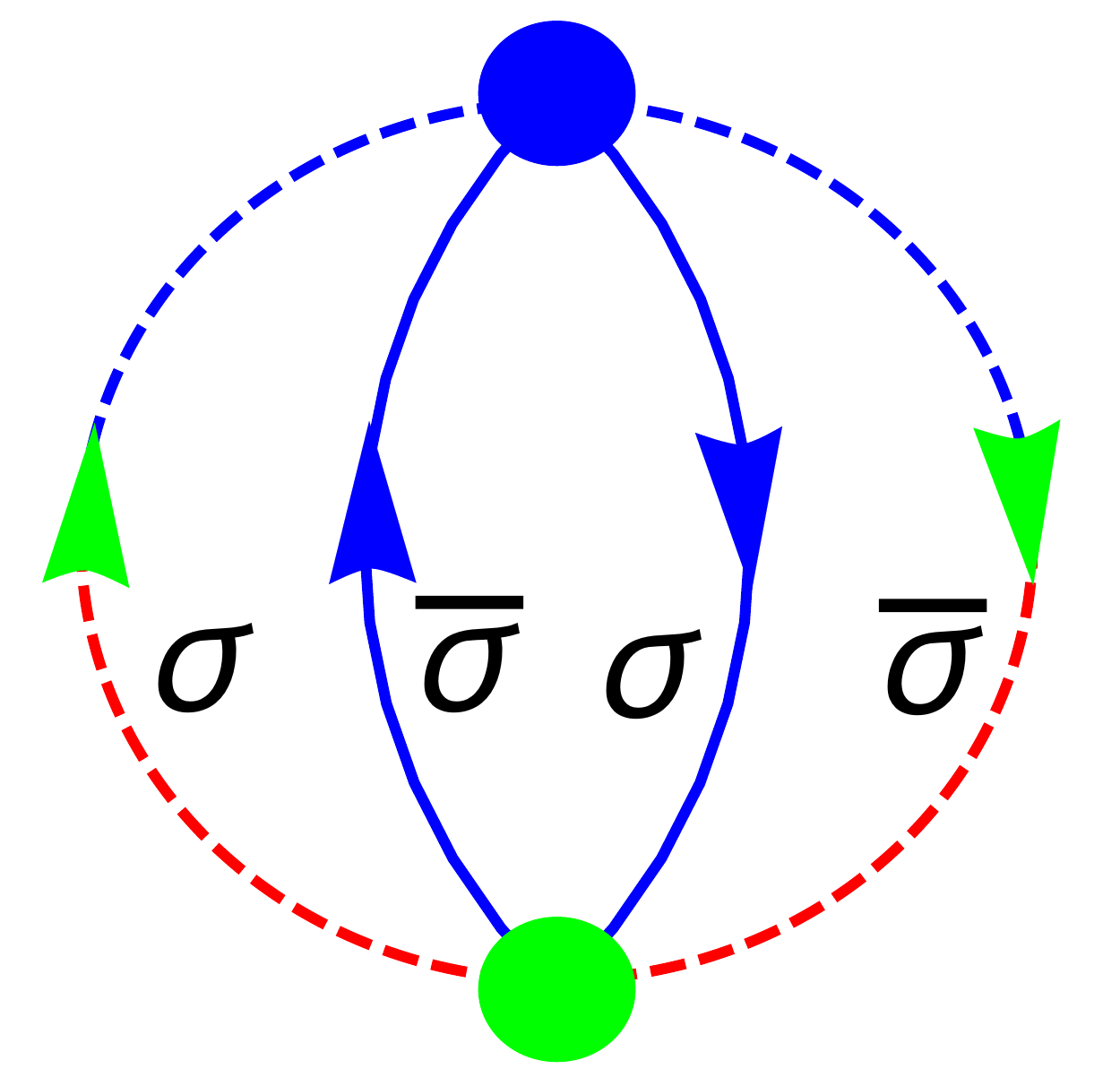}}_{\rm type-I2}
\right]\times 2+\mathcal{C}^{\phi_o\Phi}_{n}\\
\mathcal{C}^{\phi_a\phi_{\bar{a}}}_{n}&=\phantom{-}\phantom{-}\left[\underbrace{
\includegraphics[width=1.7cm, valign=c]{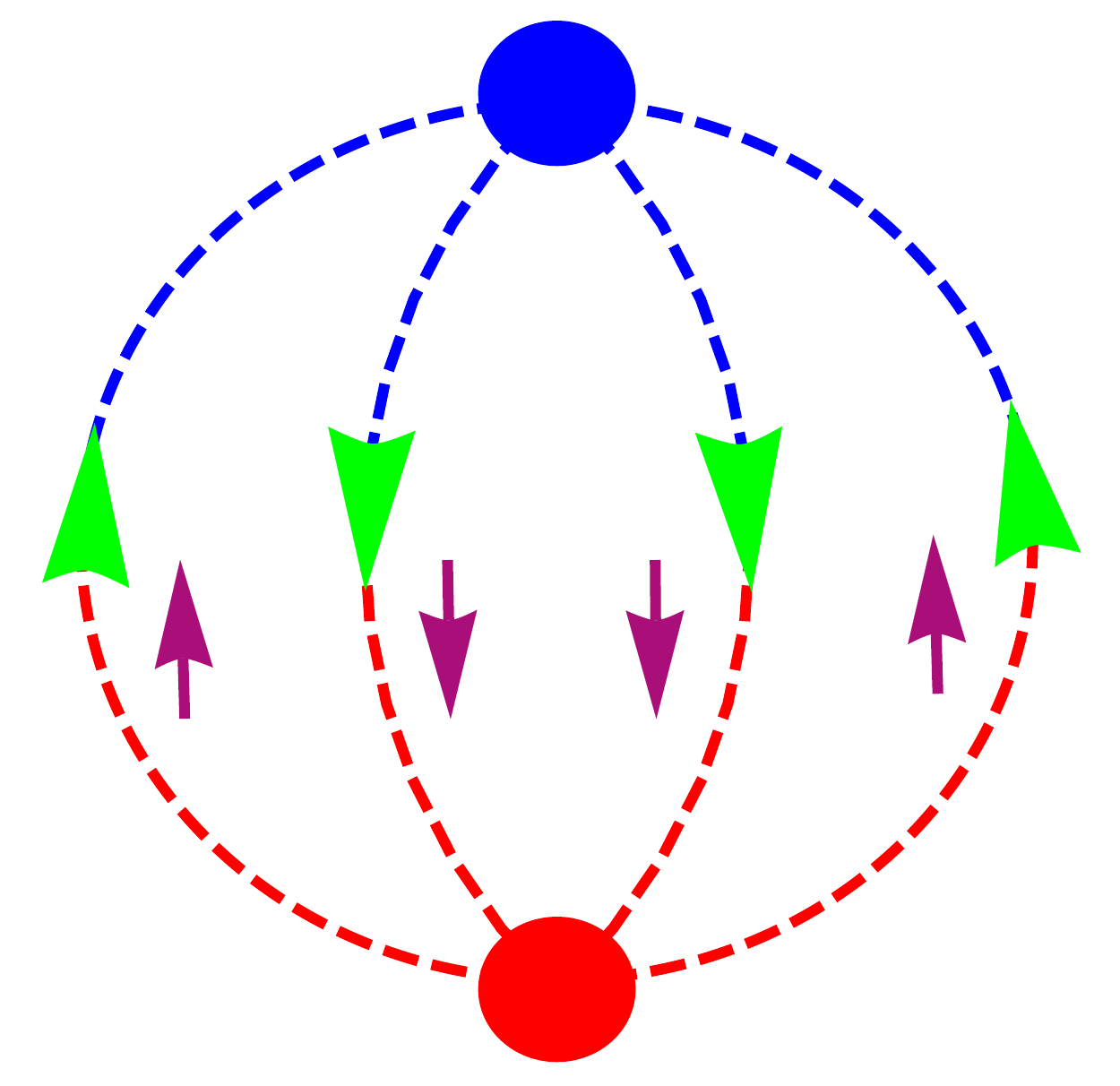}+\includegraphics[width=1.7cm, valign=c]{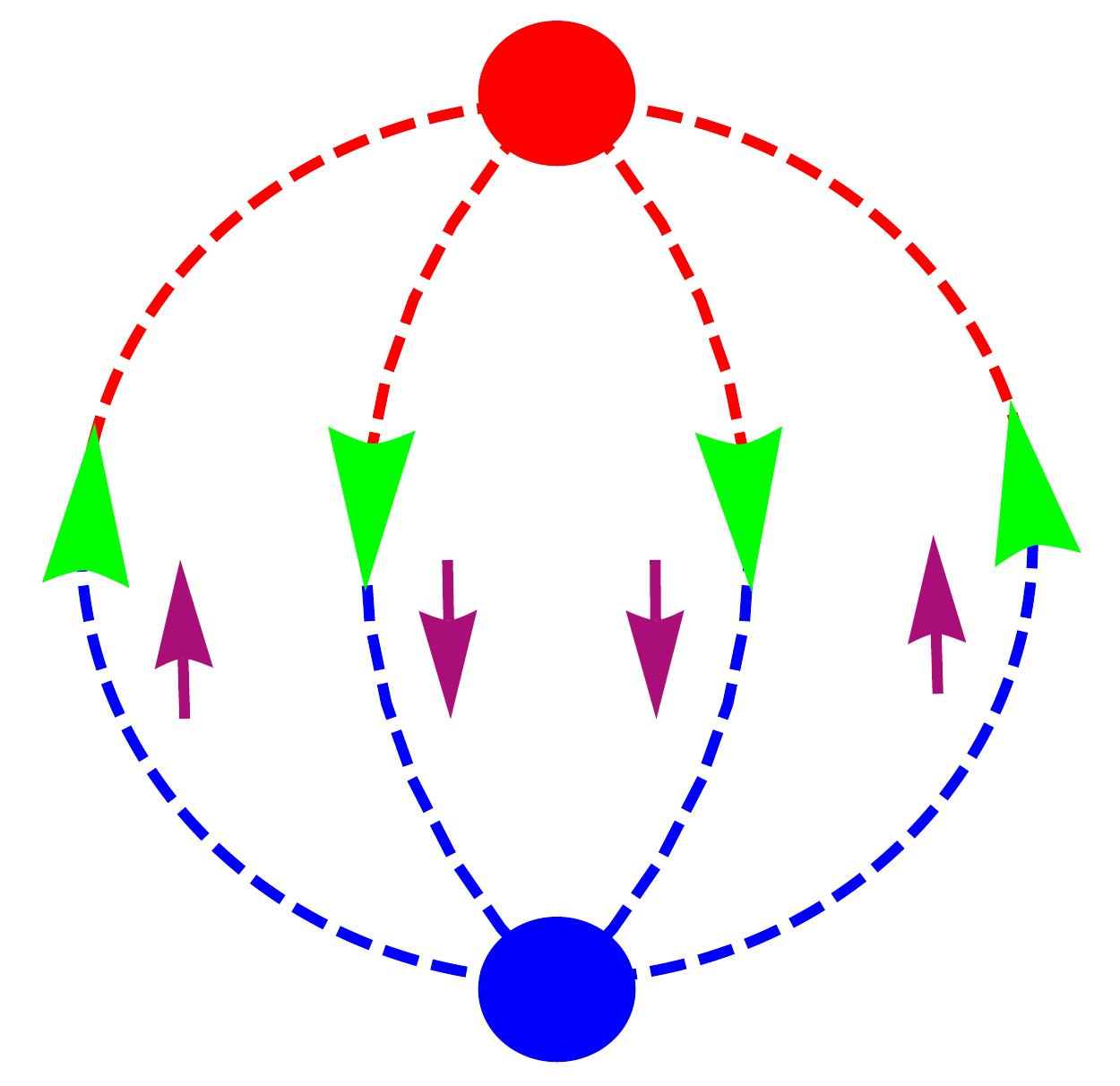}}_{\rm \text{type-I3 }}
\right]
\end{split}
\end{equation}
\caption{Feynman diagrams representing the second order interaction corrections to the CGF for 2SK model.}\label{LRFIG}
\end{figure}

We follow the similar procedure, as for the calculation of scattering contribution, to get the interaction correction to CGF. Substituting the interaction (inelastic) part of Hamiltonian, $H_{\rm in}$, into Eq.~\eqref{cdf1} and applying Wick's theorem, we obtain the Feynman diagrams accounting for the interaction effect in 2SK effect. These diagrams are shown in Fig.~\ref{LRFIG}. We allocate these interaction correction diagrams into three topologically different classes, namely type-I1, type-I2 and type-I3 as shown in lower panel of Fig.~\ref{codex1}. We introduce the notation, $\ln \chi_{\rm Ij}(\lambda)\;(\rm{j}=1, 2, 3)$, to represent the interaction correction to CGF corresponding to the digram of type-Ij. The real time GFs given in Eq.~\eqref{twogfs} pave the way for systematic calculation of $\ln \chi_{\rm Ij}(\lambda)$. As detailed in Appendix~\ref{inelastic_app}, we write the type-I1 and type-I3 diagrammatic contribution to CGF as,
\begin{align}\label{t1}
\ln\chi_{\rm I1/I3}&{=}{\pm}\frac{\Phi^2\mathcal{T}V}{24\pi}
\Big[\frac{V^2{+}4(\pi T)^2}{\sinh(V/2T)}
\sum_{x=\pm}\left(e^{-i\lambda x}{-}1\right)e^{xV/2T}\nonumber\\&
\pm 2\;\frac{V^2{+}(\pi T)^2}{\sinh(V/T)}\sum_{x=\pm}\left(e^{-2i\lambda x}-1\right)e^{xV/T}\Big].
\end{align}
Furthermore, the type-I2 diagram produces the interaction correction to CGF as,
\begin{equation}\label{t12}
\ln\chi_{\rm I2}=\frac{\Phi^2\mathcal{T}V}{12\pi}
\frac{V^2{+}({\pi} T)^2}{\sinh(V/T)}\sum_{x{=}{\pm}}\left(e^{-2i\lambda x}{-}1\right)e^{xV/T}.
\end{equation}
Substituting Eqs.~\eqref{t1} and~\eqref{t12} into Eq.~\eqref{mgfx} we get the $n$-th order cumulant of charge current, $\mathcal{C}^{\rm Ij}_n$,  corresponding to the diagram type-Ij. Of particular interest, the zero temperature results are 
\begin{align}
\mathcal{C}^{\rm I1/I3}_n&=\pm \frac{V^3}{12\pi}(-1)^n\left[1\pm2^{n+1}\right]\Phi^2.\label{c13}\\
\mathcal{C}^{\rm I2}_n&=\frac{V^3}{12\pi}(-1)^n2^{n+1}\Phi^2.\label{c2}
\end{align}
\setlength\belowcaptionskip{-3ex}
\setlength\belowcaptionskip{-3ex}
\setlength\belowcaptionskip{-5ex}
\begin{figure}
\begin{center}
\includegraphics[scale=0.5]{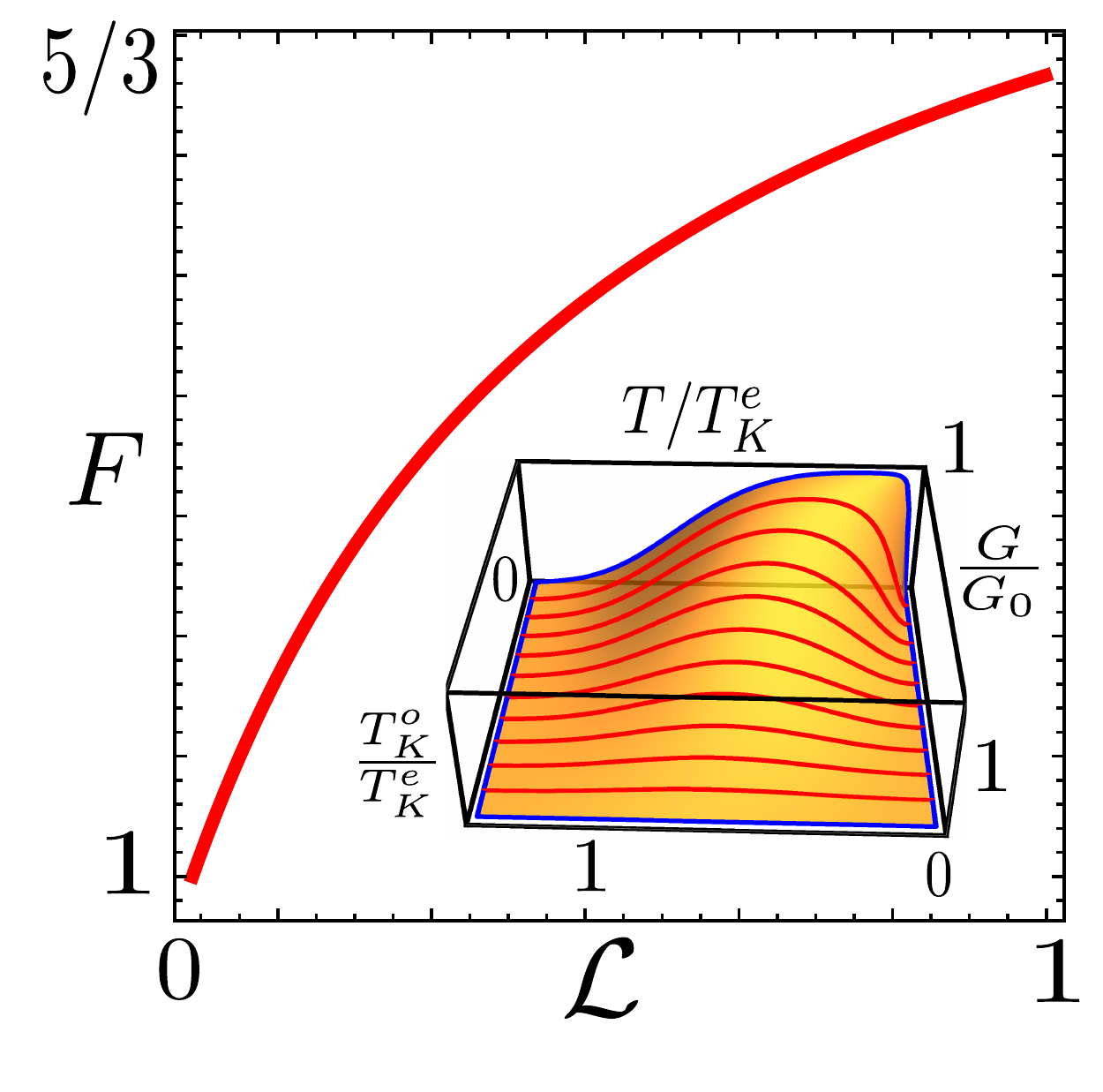}
\caption{The evolution of Fano factor ($F$) as a function of channel asymmetry parameter ($\mathcal{L}$) for a generic 2SK effect. Inset: The non-monotonic conductance behavior, the major hallmark of 2SK effect (see text for detail).}\label{evoln}
\end{center}
\end{figure}
\vspace*{-5mm}
\section{Results and discussion}\label{rd}
\vspace*{-2mm}
Collecting all the interaction contributions as detailed in Fig.~\ref{LRFIG}, and the scattering contribution given in Eq.~\eqref{m1}, we get the $n$-th cumulant of charge current at $T=0$ as
\begin{align}
\mathcal{C}_n&=({-}1)^{n}\frac{V^3}{6\pi}(\alpha_e-\alpha_o)^2\left[1+2^n\mathcal{L}\right],\label{totc1}
\end{align}
with 
\begin{equation}\label{forl}
\mathcal{L}\equiv 1+9\mathcal{Z},\;\;\mathcal{Z}=\frac{\left(\Phi{-}2/3\alpha_e\right)\left(\Phi{-}2/3\alpha_o\right)}{
\left(\alpha_e-\alpha_o\right)^2}.
\end{equation}
The parameter $\mathcal{Z}$ signifies the lack of universality away from the symmetry point, $\alpha_e=\alpha_o\;\text{and}\;\Phi=2/3\;\alpha_a$ of 2SK Hamiltonian. Besides, it has been predicted that the parameter $\mathcal{Z}$ is bounded such that $-1/9\leq\mathcal{Z}\leq 0$ \cite{KMDK_2018}. For the sake of simplicity, we introduce the new parameter $\mathcal{L}$ ($\equiv 1+9\mathcal{Z}$) in such a way that $0\leq\mathcal{L}\leq 1$. The minimum of $\mathcal{L}$ corresponds to the exact symmetry between two channels at resonance. The case of infinite asymmetry between even and odd channel, $T^o_K/T^e_K \to 0$, is characterizes by the upper bound of $\mathcal{L}$. This particular point, where the odd channel is decoupled from the impurity, recovers the 1CK paradigm. We see, form Eq.~\eqref{totc1}, that the $n$-th cumulant of charge current exactly vanishes at the symmetry point due to the destructive interference between two resonance channel. Same result holds true even at finite temperature. However, the l'Hopital's rule permits us to have the finite value of normalize $n$-th cumulant, $\mathcal{C}_n/\mathcal{C}_1$. Then we define the measure of backscattering via the generalized Fano factor
\begin{equation}\label{Fano}
F\equiv |\,\mathcal{C}_2/\mathcal{C}_1|=\frac{1+4\mathcal{L}}{1+2\mathcal{L}}.
\end{equation}
Plugging in the parameter $\mathcal{L}$ into Eq.~\eqref{Fano}, we get the Fano factor bounded from upper and below in such a way that $1\leq F\leq 5/3$. The upper bound recoups the Fano factor of 1CK effect, the super-Poissonian charge transferred statistics~\cite{Sela_Oreg_Oppen_Koch_PRL_(97)_2006}. The regime of maximum interaction in 2SK effect results the lower bound of $F$. This minimum of $F$ ($=1$) represent the Poissonian regime of charge distribution. Therefore, a generic 2SK effect exhibits the crossover regime of charge transferred statistics, from Poissonian to super-Poissonian, depending upon the channel asymmetry. This \textit{monotonic} dependence of $F$ on the channel-asymmetry parameter $\mathcal{L}$ is shown in Fig.~\ref{evoln}. The \textit{non-monotonic} conductance of 2SK effect as a function of temperature, extracted form $\left.\mathcal{C}_1\right|_{T\neq 0, V\to 0}$, is shown in the inset of Fig.~\ref{evoln} (see Ref.~\cite{KMDK_2018} for detail description).

In 1CK schemes, the definition of generalized Fano factor follows from $F\equiv \left.\delta \mathcal{C}_2/\delta \mathcal{C}_1\right|_{T\to 0}$, where $\delta \mathcal{C}_{1/2}$ represents the corresponding quantity after subtracting the linear part (those terms $\propto V$). Nevertheless, the $n$-th cumulant of charge current in 2SK schemes, the Eq.~\eqref{totc1}, does not shows up the linear terms in $V$. This makes very straightforward extraction of $F$ in 2SK effect, since it does not require the proper subtraction of linear terms.

The differential conductance of 2SK effect as a function of $B$ (the Zeeman field), $T$ and $V$ is given, in terms of FL transport coefficients, as $G/G_0=c_B B^2+c_T(\pi T)^2+c_V V^2$, where $G_0$ is the unitary conductance. The transport coefficients bear the compact form: $c_T/c_B=\left(\mathcal{L}+2\right)/3$ and $c_V/c_B=\mathcal{L}+1/2$~\cite{KMDK_2018}. Thus, the measurement of $\mathcal{L}$ would suffices the study of transport behaviors of 2SK effect. The compelling monotonic dependence of $F$ on $\mathcal{L}$, as shown in Fig.~\ref{evoln}, could furnish an experimental way to extract $\mathcal{L}$ as follows. Given an experimental setup in 2SK scheme, the independent measurements of charge current and noise impart the Fano factor. Thus obtained Fano factor uniquely defines the corresponding asymmetry parameter $\mathcal{L}$ via Eq.~\eqref{evoln}. Following this way of measurements of transport coefficients could be less involved than measuring the response functions.
\vspace*{-2mm}
\section{Conclusion}\label{con}
\vspace*{-2mm}
We extended the method FCS from conventional 1CK schemes to multi-channel Kondo paradigm. The developed framework of FCS has been demonstrated considering an example of 2SK effect. We analyzed the charge transferred statistics in the strong-coupling regime of a 2SK model using non-equilibrium Keldysh formulation. We found that the arbitrary cumulant of charge current get nullified at the symmetry point of 2SK model due to the destructive interference between two conducting channel. We studied the destructiveness/constructiveness of interference in terms of channel asymmetry parameter, $\mathcal{L}$. The $n$-th order normalized cumulant of charge current, $\mathcal{C}_n/\mathcal{C}_1$ took a compact function  of $\mathcal{L}$, only. A bounded value of Fano factor, $1\leq F\leq 5/3$, has been discovered. Studying the observed monotonic growth of $F$ as a function of $\mathcal{L}$, we uncovered the cross-over regimes of charge transfered statistics in 2SK effect, from Poissonian to super-Poissonian. We proposed a novel way of obtaining the FL transport coefficients of 2SK effect by the independent measurements of charge current and noise. The developed formalism imparts all the transport informations of 1CK effect as well. All the calculations have been performed at finite temperature, one can easily study the effect of temperature on an arbitrary cumulant of charge current~\footnote{One can study the evolution of Fano factor as a function of $V/T$ and the asymmetry parameter $\mathcal{L}$ similar to that of Ref.~\cite{Mora_Leyronas_Regnault_PRL_(100)_2008} devoted to 1CK effect.}.

\noindent \textit{Acknowledgments}$-$ We have benefited from illuminating discussions with Jan von Delft and Christophe Mora at the early stage of this project. We are thankful to Michele Fabrizio, Leonid Glazman, Yuval Oreg, Felix von Oppen, Achim Rosch and Eran Sela for fruitful discussions.
\appendix
\vspace*{-5mm}
\section{Scattering corrections to the CGF}\label{elastic_app}
\vspace*{-2mm}
The scattering correction to the MGF of 2SK model reads
\begin{equation}\label{dil1}
\chi_{\rm el}(\lambda)=\Big\langle T_C \exp\left[-i\int_C \mathcal{H}^{\lambda}_{\rm el}(t)dt\right]\Big\rangle_0.
\end{equation}
Where the scattering Hamiltonian is given in Eq.~\eqref{halpha}. The logarithm of Eq.~\eqref{dil1}, $\ln \chi_{\rm el}(\lambda)$, imparts the corresponding CGF. The second order expansion of $\ln \chi_{\rm el}(\lambda)$ in $\mathcal{H}^{\lambda}_{\rm el}$ followed by the use of Wick's theorem results in four different Feynman diagrams as shown in Fig.~\ref{codex1} (upper panel). The first and second diagrams are composed of only the channel-diagonal GFs. Owing to their similar geometry, we classified them as type-E1 diagram. Note that, in our convention, the two diagrams are geometrically similar if they contain equal number of channel-diagonal GFs (if present) and equal number of mixed-GFs (if present). The type-E2 diagrams shown in Fig.~\ref{codex1}, nonetheless, consist of only mixed-GFs. The interference between two channels, due to the scattering effects, is accounted for by these type-E2 diagrams. The contribution of type-E1 diagrams to the CGF is proportional to $\alpha^2_a$. Similarly, the CGF contribution of type-E2 diagrams is proportional to $\alpha_a\alpha_{\bar{a}}$. Therefore, the overall scattering contribution to CGF is written as

\begin{equation}\label{aama0}
\ln \chi_{\rm el}=\sum_a\left(\ln \chi_{\alpha^2_a}+\ln \chi_{\alpha_a\alpha_{\bar{a}}}\right).
\end{equation}

Topologically the type-E1 and type-E2 diagrams are quite distinct. The type-E1 diagrams has been already appeared in several previous works~\cite{Gogolin_Komnik_PRL(2006), Gogolin_Komnik_PRB(2006),tschmidt_2, tschmidt_1,ahes}, however, the type-E2 diagrams has not been considered yet. For completeness, we present the mathematical details of diagrammatic contribution to CGF for both diagrams. For type-E1 diagrams we write
\begin{align}\label{aama1}
&\ln\chi_{\alpha^2_a}=-\frac{1}{2}\frac{\alpha^2_a}{(2\pi\nu)^2}\sum_{kk'\sigma}\sum_{pp'\sigma'}\left(\varepsilon_k+\varepsilon_{k'}\right)
\left(\varepsilon_p+\varepsilon_{p'}\right)\times\nonumber\\
&\int_{\mathcal{C}}\!\!\! dt_1dt_2\Big< T_\mathcal{C}b^{\dagger}_{ak\sigma}(t_1)b_{ak'\sigma}(t_1)b^{\dagger}_{ap\sigma'}(t_2)b_{ap'\sigma'}(t_2)\Big>.
\end{align}
Here, we introduced the set of momentum indices ($k, k', p, p'$), spin indices ($\sigma, \sigma'$) and time indices ($t_1, t_2$). Equation.~\eqref{aama1} imparts the non-zero contribution only if $k=p'$, $k'=p$ and $\sigma=\sigma'$. Therefore, use of the method of Keldysh disentanglement to the Eq.~\eqref{aama1} results in
\begin{align}\label{aama2}
\ln\chi_{\alpha^2_a}=&-\frac{1}{2}\frac{\alpha^2_a\mathcal{T}}{(2\pi\nu)^2}\sum_{kk'\sigma}\left(\varepsilon_k+\varepsilon_{k'}\right)^2\times\nonumber\\
&\sum_{\eta_1\eta_2}\eta_1\eta_2\int_{\mathcal{C}}
dt\;\mathcal{G}^{\eta_1\eta_2}_{aa,k'\sigma}(t) \mathcal{G}^{\eta_2\eta_1}_{aa,k\sigma}(-t),
\end{align}
where, $\mathcal{T}$ is the measurement time (see Section~\ref{fcs}), $\eta_{i}$ ($i=1,2$) are the Keldysh branch (forward and backward) indices such that $\eta_i{=}\pm 1$. The channel-diagonal GFs, $\mathcal{G}^{\eta_1\eta_2}_{aa}$, are defined in Eqs.~\eqref{matrixgf},~\eqref{man1} and~\eqref{man2}. These GFs acquire the special property; $\nu\mathcal{G}^{\eta_1\eta_2}_{aa,k\sigma}(\varepsilon)=\mathcal{G}_{aa, \sigma}^{\eta_1\eta_2}(\varepsilon)\delta(\varepsilon-\varepsilon_k)$, where $\delta(\varepsilon-\varepsilon_k)$ stands for the Kronecker delta symbol. Then, Eq.~\eqref{aama2} reads
\begin{align}\label{aama3}
\ln\chi_{\alpha^2_a}{=}&{-}\frac{1}{2}\frac{\alpha^2_a\mathcal{T}}{(\pi\nu)^2}
\sum_{\eta_1\eta_2,\sigma}\eta_1\eta_2\int_{\mathcal{C}}
\frac{d\varepsilon}{2\pi}\;\varepsilon^2\;\mathcal{G}^{\eta_1\eta_2}_{aa,\sigma}(\varepsilon) \mathcal{G}^{\eta_2\eta_1}_{aa\sigma}(\varepsilon).
\end{align}
Summing Eq.~\eqref{aama3} over $\eta_{1/2}$ and then plugging in the various Keldysh GFs from Eqs.~\eqref{matrixgf},~\eqref{man1} and~\eqref{man2} leads
\begin{align}\label{aama4}
\ln\chi_{\alpha^2_a}=&\frac{\alpha^2_a\mathcal{T}}{2\pi}\sum_{\sigma}\mathcal{J}_{\rm el, \sigma},
\end{align}
with the compact form of the integral characterizing the scattering effects 
\begin{equation}\label{baba1}
\mathcal{J}_{\rm el}{=}\int^{\infty}_{-\infty}\!\!\!\varepsilon^2 d\varepsilon \left[f_S(1{-}f_D)(e^{-i\lambda}{-}1){+}f_D(1{-}f_S)(e^{i\lambda}{-}1)\right].
\end{equation}
We called this integral as the ``elastic integral", the mathematical details of it's computation will be discussed in Appendix~\ref{elint}.

Repeating all above calculations for type-E2 diagrams with same notations we get the quite similar result
\begin{align}\label{aama5}
\ln\chi_{\alpha_a\alpha_{\bar{a}}}=&-\frac{\alpha_a\alpha_{\bar{a}}\mathcal{T}}{2\pi}\sum_{\sigma}\mathcal{J}_{\rm el, \sigma}.
\end{align}
Though the origin of these two elastic diagrams and corresponding topology is quits distinct, it turns out that they produce equal but opposite contribution to the CGF. Plugging in the Eqs.~\eqref{aama4} and~\eqref{aama5} into Eq.~\eqref{aama0} imparts the contribution of scattering effects to the CGF, which is written as
\begin{equation}\label{aama6}
\ln\chi_{\rm el}=\frac{\left(\alpha_e-\alpha_0\right)^2\mathcal{T}}{2\pi}\sum_{\sigma}\mathcal{J}_{\rm el, \sigma}.
\end{equation}
Substituting the value of elastic integral from Eq.~\eqref{bishnud1} into Eq.~\eqref{aama6}, we get the final expression for the scattering contribution to the CGF in 2SK model, which is the Eq.~\eqref{cgf1} in the main text of this paper.
\vspace*{-5mm}
\section{Interaction corrections to the CGF}\label{inelastic_app}
\vspace*{-4mm}
In the same way and using the same notations as for the scattering contribution, we get several Feynman diagrams contributing to the CGF of 2SK model as shown in Fig.~\ref{LRFIG}. These diagrams are classified into three different group, type-Ij (j=1, 2, 3) diagram, based upon the numbers of channel-diagonal GFs and mixed-GFs present in a particular diagram (see the lower panel of Fig.~\ref{codex1}). For instance, the diagram with all (four) channel-diagonal GFs has been classified as type-I1 diagram, those with two channel-diagonal and remaining two mixed-GFs as the type-I2 and diagrams with all mixed-GFs has been termed as type-I3. As have been detailed in the Fig.~\ref{LRFIG}, we expressed the interaction contribution to the CGF in terms of these three diagrams. The two diagrams belonging to the same group might have different weight factor. The numbers of close fermion loops, the product of Pauli matrices and the renormalization factor in the Hamiltonian determines the weight factor corresponding to a particular diagram (for detail see Ref.~\cite{KMDK_2018}).

A single diagram of type-I1 (with CGF contribution proportional to $\phi^2_{e/o}$) completely characterizes the FCS of 1CK schemes~\cite{Gogolin_Komnik_PRL(2006), Gogolin_Komnik_PRB(2006),tschmidt_2, tschmidt_1,ahes}. However, the type-I2 and type-I3 diagrams are generic feature of multi-channel, multi-stage screening effects. These diagrams has not been studied yet. In this Appendix~\ref{inelastic_app} we provide the mathematical details of computing CGF contribution of type-Ij diagram.

The type-I1 diagram shown in lower panel of Fig.~\ref{codex1} produces the CGF contribution proportional to $\Phi^2$. Following the standard technique of Feynman diagram calculation, we cast the CGF contribution of type-I1 diagram into the form

\begin{align}\label{dd2}
&\ln\chi_{\rm I1}=-\frac{1}{2}\left(\frac{\Phi}{\pi\nu}\right)^2 \int_{\mathcal{C}} dt_1dt_2 \Big[\mathcal{G}_{ee, \sigma}(t_1-t_2)\times\nonumber\\
&\mathcal{G}_{ee, \sigma}(t_2-t_1)\mathcal{G}_{oo, \sigma}(t_1-t_2)\mathcal{G}_{oo, \sigma}(t_2-t_1)\Big].
\end{align}
In Eq.~\eqref{dd2} we did not consider the spin summation for being more general (implying that the spin index $\sigma$ is either down or up). Using the technique of Keldysh disentanglement we rewrite the Eq.~\eqref{dd2} as
\begin{align}\label{dd2}
\ln\chi_{\rm I1}&{=}{-}\frac{\mathcal{T}}{2}\left(\frac{\Phi}{\pi\nu}\right)^2\sum_{\eta_1\eta_2}\eta_1\eta_2 {\times} \int_{\mathcal{C}} dt \Big[\mathcal{G}^{\eta_1\eta_2}_{ee, \sigma}(t)\times\nonumber\\
&\mathcal{G}^{\eta_2\eta_1}_{ee, \sigma}(-t)\mathcal{G}^{\eta_1\eta_2}_{oo, \sigma}(t)\mathcal{G}^{\eta_2\eta_1}_{oo, \sigma}(-t)\Big].
\end{align}
Summing Eq.~\eqref{dd2} over Keldysh indices $\eta_{1/2}$ and using the expressions of GFs in Eq.~\eqref{twogfs}, we get
\begin{equation}\label{dd4}
\ln\chi_{\rm I1}=(\pi\Phi T^2)^2\mathcal{T}\int^{\infty+i\gamma}_{-\infty+i\gamma} \frac{\cos^4\left(\frac{\lambda}{2}+\frac{V}{2}t\right)}{\sinh^4(\pi T t)}dt.
\end{equation}
Coming from Eq.~\eqref{dd2} to Eq.~\eqref{dd4}, we retain only the $\lambda$-dependent terms. The integral involving into the Eq.~\eqref{dd4} is computed in Appendix~\ref{ine_a_b}. Then we write the CGF contribution of type-I1 diagram into the form 
\begin{align}\label{dd5}
&\ln\chi_{\rm I1}=\frac{\Phi^2\mathcal{T}}{96\pi}\Big[\frac{8V\left(V^2+(\pi T)^2\right)}{\sinh(V/T)}\sum_{x{=}\pm}e^{2ix\lambda}e^{-xV/T}\nonumber\\
&\frac{4V\left(V^2+4(\pi T)^2\right)}{\sinh(V/2T)}\sum_{x{=}\pm}e^{ix\lambda}e^{-xV/2T}\Big]
{+}\frac{\pi\Phi^2 T^3\mathcal{T}}{2}.
\end{align}
For proper renormalization of PDF we subtract the $\lambda = 0$ contribution of Eq.~\eqref{dd5} from the same Eq.~\eqref{dd5}, which results the final expression of CGF contributed by a diagram of type-I1
\begin{align}\label{dd6}
\ln\chi_{\rm I1}&=\frac{\Phi^2\mathcal{T}V}{24\pi}\Big[2\cdot\frac{V^2{+}(\pi T)^2}{\sinh(V/T)}\sum_{x{=}\pm}\left(e^{2ix\lambda}{-}1\right)e^{{-}xV/T}\nonumber\\
&+\frac{V^2{+}4(\pi T)^2}{\sinh(V/2T)}\sum_{x{=}\pm}\left(e^{ix\lambda}{-}1\right)e^{{-}xV/2T}\Big].
\end{align}

In the same way and using the same notations as for the type-I1 diagram, the CGF contribution of type-I2 diagram as shown in Fig.~\ref{codex1} reads
\begin{equation}\label{dd8}
\ln\chi_{\rm I2}{=}{-}(\pi\Phi T^2)^2\mathcal{T}\!\!\!\int^{\infty+i\gamma}_{-\infty+i\gamma} \!\!\frac{\sin^2\left(\frac{\lambda}{2}{+}\frac{V}{2}t\right)\cos^2\left(\frac{\lambda}{2}{+}\frac{V}{2}t\right)}{\sinh^4(\pi T t)}dt.
\end{equation}
Using the integral given in Appendix~\ref{ine_a_b} followed by the proper renormalization of PDF, the Eq.~\eqref{dd8} results in
\begin{equation}\label{dd9}
\ln\chi_{\rm I2}=\frac{\Phi^2\mathcal{T}V}{12\pi}\frac{V^2+(\pi T)^2}{\sinh(V/T)}\sum_{x=\pm}\left(e^{2ix\lambda}-1\right)e^{-xV/T}.
\end{equation}

Similarly, the CGF contribution of a type-I3 diagram as shown in Fig.~\ref{codex1} is given by
\begin{align}\label{baa1}
\ln\chi_{\rm I3}&=(\pi\Phi T^2)^2\mathcal{T}\int^{\infty+i\gamma}_{-\infty+i\gamma} \frac{\sin^4\left(\frac{\lambda}{2}{+}\frac{V}{2}t\right)}{\sinh^4(\pi T t)}dt.
\end{align}
The simplification of Eq.~\eqref{baa1} upon proper renormalization of PDF imparts
\begin{align}\label{baa3}
\ln\chi_{\rm I3}&=-\frac{\Phi^2\mathcal{T}V}{24\pi}\Big[\frac{V^2{+}4(\pi T)^2}{\sinh(V/2T)}\sum_{x{=}\pm}\left(e^{ix\lambda}{-}1\right)e^{{-}xV/2T}\nonumber\\
&-2\cdot\frac{V^2{+}(\pi T)^2}{\sinh(V/T)}\sum_{x{=}\pm}\left(e^{2ix\lambda}{-}1\right)e^{{-}xV/T}\Big].
\end{align}
\vspace*{-10mm}
\section{Elastic integral}\label{elint}
\vspace*{-3mm}
This section contain the detail of the calculation of the elastic integral Eq.~\eqref{baba1} using the properties of Fourier transform (FT). First we factorized the elastic integral as,
\begin{align}\label{elastic_c}
\mathcal{J}_{\rm el}=(e^{-i\lambda}-1)\mathcal{J}^{\rm 1}_{\rm el}+(e^{i\lambda}-1) \mathcal{J}^{\rm 2}_{\rm el}.
\end{align}
Here, we introduced the short hand notations;
\begin{equation}\label{baba3}
\mathcal{J}^{\rm 1}_{\rm el}{=}\int^{\infty}_{-\infty}\varepsilon^2 d\varepsilon f_S(1{-}f_D),\;\mathcal{J}^{\rm 2}_{\rm el}{=}\int^{\infty}_{-\infty}\varepsilon^2 d\varepsilon f_D(1{-}f_S).
\end{equation}
At $T=0$ we have, $f_{S/D}(\varepsilon)=\Theta(\mu_{S/D}-\varepsilon)$ and $1-f_{S/D}(\varepsilon)=\Theta(\varepsilon-\mu_{S/D})$. Thus the zero temperature limit of Eq.~\eqref{elastic_c} is quite trivial
\begin{equation}\label{aama7}
\left.\mathcal{J}_{\rm el}\right|_{T=0}=(e^{-i\lambda}-1)\int^{\mu_L}_{\mu_R}\varepsilon^2d\varepsilon=(e^{-i\lambda}-1)\frac{V^3}{12}.
\end{equation}
However, the Fermi distribution functions of the source and drain, and their FT at finite temperature are 
\begin{equation}\label{fermi_f}
f_{S/D}(\varepsilon){=}\frac{e^{-(\varepsilon-\mu_{S/D})/2T}}{2\cosh\left[(\varepsilon{-}\mu_{S/D})/2T\right]};f_{S/D}(t){=}\frac{iT}{2}\frac{e^{-it\mu_{S/D}}}{\sinh(\pi T t)}.
\end{equation}
For the sake of simplicity we define another function $h_{L/R}(\varepsilon)$ and its FT as
\begin{equation}\label{h}
h_{S/D}(\varepsilon)=e^{-\frac{\varepsilon}{T}}f_{S/D}(\varepsilon)
;\;h_{S/D}(t){=}{-}\frac{iT}{2}\frac{e^{-it\mu_{S/D}{-}\frac{\mu_{S/D}}{T}}}{\sinh(\pi T t)}.
\end{equation}
The function $h_{S/D}(\varepsilon)$ in Eq.~\eqref{h} imparts the way to convert the product of Fermi functions into weighted sum. For instance,
\begin{equation}\label{hjk}
f_S(\varepsilon)f_D(\varepsilon)=\frac{h_S(\varepsilon)-h_D(\varepsilon)}{e^{-\frac{\mu_D}{T}}-e^{-\frac{\mu_S}{T}}}=\frac{h_S(\varepsilon)-h_D(\varepsilon)}{2\sinh(V/2T)}.
\end{equation}
Then, $\mathcal{J}^{\rm 1}_{\rm el}$ and $\mathcal{J}^{\rm 2}_{\rm el}$ simplifies to 
\begin{align}\label{z1}
\mathcal{J}^{\rm 1/2}_{\rm el}&{=}{\int^{\infty}_{{-}\infty}}\varepsilon^2 d\varepsilon f_{S/D}(\varepsilon){-}\frac{\int^{\infty}_{-\infty}\varepsilon^2 d\varepsilon \left[h_L(\varepsilon)-h_R(\varepsilon)\right]}{2\sinh(V/2T)}.
\end{align}
Having defined the FT of the functions $f_{S/D}$ and $h_{S/D}$, we performed the integration of Eq.~\eqref{z1} by using the property of FT \cite{KMDK_2018},
\begin{equation}\label{ft}
\int_{-\infty}^{\infty}  \varepsilon^n  y(\varepsilon) d\varepsilon= \frac{2\pi}{(-i)^n}
\left. \partial_t^n \left[y(t)\right]\right|_{_{t=0}}.
\end{equation}
Here, $y(\varepsilon)$ is an arbitrary function with FT $y(t)$ and $\partial_t^n$ represents the $n$-th order differentiation with respect to $t$. Substitution of Eq.~\eqref{ft} for $n{=}2$ into Eq.~\eqref{z1} and then using the definitions of $f_{S/D}(t)$ and $h_{S/D}(t)$ defined in Eqs.~\eqref{fermi_f} and~\eqref{h} leads the result
\begin{equation}\label{z4}
\mathcal{J}^{\rm 1/2}_{\rm el}=\pm\frac{V}{2}\left[\frac{(\pi T)^2}{3}+\frac{V^2}{12}\right]\times\left[1\pm\coth\left(\frac{V}{2T}\right)\right].
\end{equation}
Plugging in $\mathcal{J}^{\rm 1/2}_{\rm el}$, from Eq.~\eqref{z4}, into Eq.~\eqref{elastic_c} we obtain the final expression for the elastic integral
\begin{equation}\label{bishnud1}
\mathcal{J}_{\rm el}{=}\frac{V}{24}\frac{V^2{+}4(\pi T)^2}{\sinh(V/2T)}
\left[(e^{-i\lambda}{-}1)e^{V/2T}{+}(e^{i\lambda}{-}1)e^{-V/2T}\right].
\end{equation}
This is the central equation governing the CGF contribution of scattering effects in 2SK model.
\vspace*{-4mm}
\section{Inelastic Integral}\label{ine_a_b}
\vspace*{-2mm}

For the computation of integrals in Eqs.~\eqref{dd4},~\eqref{dd8} and~\eqref{baa1}, we expand their numerators in powers of $e^{\pm i(\lambda/2+Vt/2)}$. Each term gives an integral of the form
\begin{equation}\label{mi1}
\mathcal{I}_{\pm}=\int^{\infty+i\gamma}_{-\infty+i\gamma} \frac{e^{\pm i\mathcal{A} t}}{\sinh^4(\pi T t)}dt,\;\;\;\mathcal{A}>0.
\end{equation}
The singularity of the integral in Eq.~\eqref{mi1} is removed by shifting the time contour by $i\gamma $ in the complex plane such that $\gamma D \gg 1$, $\gamma T \ll 1$ and $\gamma \mathcal{A} \ll 1$. Here, $D$ is the band cutoff. 
The poles of the integrand of $\mathcal{I}_{\pm}$ are given by the solution of $\sinh(\pi T t)=0$ for $t$. Which leads,
\begin{equation}\nonumber
\pi T t=\pm im\pi \Rightarrow t=\pm \frac{im}{T}, \quad m=0, \pm 1, \pm 2, \pm 3...
\end{equation}
With the choice of the rectangular contour shifted by $i/T$ in the complex plane which includes the pole of the integrand at $t=0$, the standard method of complex integration results in
\begin{equation}\label{6a}
\begin{split}
\mathcal{I}_+\left(1-e^{\mathcal{A}/T}\right)&=
-2\pi i\times  \left. \text{\text{Res}}\right|_{t=0},
\end{split}
\end{equation}
where $\left.\text{Res}\right|_{t=0}$ stands for Cauchy residue of the integrand in Eq.~\eqref{mi1} at $t=0$. Plugging in the residue into Eq.~\eqref{6a} results in
\begin{equation}\label{10a}
\mathcal{I}_+=-\frac{2\pi\left(\mathcal{A}^3+4\mathcal{A}(\pi T)^2\right)}{6(\pi T)^4}\times \frac{1}{1-e^{\frac{\mathcal{A}}{T}}}.
\end{equation} 
Similarly we computed $\mathcal{I}_-$. These integrals $\mathcal{I}_{\pm}$ are sufficient for the computation of all the inelastic diagrams shown in Fig.~\ref{LRFIG}.

\begin{thebibliography}{66}%
\makeatletter
\providecommand \@ifxundefined [1]{%
 \@ifx{#1\undefined}
}%
\providecommand \@ifnum [1]{%
 \ifnum #1\expandafter \@firstoftwo
 \else \expandafter \@secondoftwo
 \fi
}%
\providecommand \@ifx [1]{%
 \ifx #1\expandafter \@firstoftwo
 \else \expandafter \@secondoftwo
 \fi
}%
\providecommand \natexlab [1]{#1}%
\providecommand \enquote  [1]{``#1''}%
\providecommand \bibnamefont  [1]{#1}%
\providecommand \bibfnamefont [1]{#1}%
\providecommand \citenamefont [1]{#1}%
\providecommand \href@noop [0]{\@secondoftwo}%
\providecommand \href [0]{\begingroup \@sanitize@url \@href}%
\providecommand \@href[1]{\@@startlink{#1}\@@href}%
\providecommand \@@href[1]{\endgroup#1\@@endlink}%
\providecommand \@sanitize@url [0]{\catcode `\\12\catcode `\$12\catcode
  `\&12\catcode `\#12\catcode `\^12\catcode `\_12\catcode `\%12\relax}%
\providecommand \@@startlink[1]{}%
\providecommand \@@endlink[0]{}%
\providecommand \url  [0]{\begingroup\@sanitize@url \@url }%
\providecommand \@url [1]{\endgroup\@href {#1}{\urlprefix }}%
\providecommand \urlprefix  [0]{URL }%
\providecommand \Eprint [0]{\href }%
\providecommand \doibase [0]{http://dx.doi.org/}%
\providecommand \selectlanguage [0]{\@gobble}%
\providecommand \bibinfo  [0]{\@secondoftwo}%
\providecommand \bibfield  [0]{\@secondoftwo}%
\providecommand \translation [1]{[#1]}%
\providecommand \BibitemOpen [0]{}%
\providecommand \bibitemStop [0]{}%
\providecommand \bibitemNoStop [0]{.\EOS\space}%
\providecommand \EOS [0]{\spacefactor3000\relax}%
\providecommand \BibitemShut  [1]{\csname bibitem#1\endcsname}%
\let\auto@bib@innerbib\@empty
\bibitem [{\citenamefont {Blanter}\ and\ \citenamefont
  {Buttiker}(2000)}]{Butt_Blt}%
  \BibitemOpen
  \bibfield  {author} {\bibinfo {author} {\bibfnamefont {Y.}~\bibnamefont
  {Blanter}}\ and\ \bibinfo {author} {\bibfnamefont {M.}~\bibnamefont
  {Buttiker}},\ }\href
  {http://www.sciencedirect.com/science/article/pii/S0370157399001234}
  {\bibfield  {journal} {\bibinfo  {journal} {Physics Reports}\ }\textbf
  {\bibinfo {volume} {336}},\ \bibinfo {pages} {1 } (\bibinfo {year}
  {2000})}\BibitemShut {NoStop}%
\bibitem [{\citenamefont {Blanter}\ and\ \citenamefont
  {Nazarov}(2009)}]{Blanter_Nazarov}%
  \BibitemOpen
  \bibfield  {author} {\bibinfo {author} {\bibfnamefont {Y.~M.}\ \bibnamefont
  {Blanter}}\ and\ \bibinfo {author} {\bibfnamefont {Y.~V.}\ \bibnamefont
  {Nazarov}},\ }\href@noop {} {\emph {\bibinfo {title} {Quantum Transport:
  Introduction to Nanoscience}}}\ (\bibinfo  {publisher} {Cambridge University
  Press, Cambridge, England},\ \bibinfo {year} {2009})\BibitemShut {NoStop}%
\bibitem [{\citenamefont {Levitov}\ and\ \citenamefont
  {Lesovik}(1993)}]{Levitov_JETP}%
  \BibitemOpen
  \bibfield  {author} {\bibinfo {author} {\bibfnamefont {L.~S.}\ \bibnamefont
  {Levitov}}\ and\ \bibinfo {author} {\bibfnamefont {G.~B.}\ \bibnamefont
  {Lesovik}},\ }\href
  {http://www.jetpletters.ac.ru/ps/1186/article_17907.shtml} {\bibfield
  {journal} {\bibinfo  {journal} {JETP Lett.}\ }\textbf {\bibinfo {volume}
  {58}},\ \bibinfo {pages} {230} (\bibinfo {year} {1993})}\BibitemShut
  {NoStop}%
\bibitem [{\citenamefont {Lee}\ \emph {et~al.}(1995)\citenamefont {Lee},
  \citenamefont {Levitov},\ and\ \citenamefont {Yakovets}}]{Levitov_2}%
  \BibitemOpen
  \bibfield  {author} {\bibinfo {author} {\bibfnamefont {H.}~\bibnamefont
  {Lee}}, \bibinfo {author} {\bibfnamefont {L.~S.}\ \bibnamefont {Levitov}}, \
  and\ \bibinfo {author} {\bibfnamefont {A.~Y.}\ \bibnamefont {Yakovets}},\
  }\href {\doibase 10.1103/PhysRevB.51.4079} {\bibfield  {journal} {\bibinfo
  {journal} {Phys. Rev. B}\ }\textbf {\bibinfo {volume} {51}},\ \bibinfo
  {pages} {4079} (\bibinfo {year} {1995})}\BibitemShut {NoStop}%
\bibitem [{\citenamefont {Levitov}\ \emph {et~al.}(1996)\citenamefont
  {Levitov}, \citenamefont {Lee},\ and\ \citenamefont {Lesovik}}]{Levitov_1}%
  \BibitemOpen
  \bibfield  {author} {\bibinfo {author} {\bibfnamefont {L.~S.}\ \bibnamefont
  {Levitov}}, \bibinfo {author} {\bibfnamefont {H.}~\bibnamefont {Lee}}, \ and\
  \bibinfo {author} {\bibfnamefont {G.~B.}\ \bibnamefont {Lesovik}},\ }\href
  {https://doi.org/10.1063/1.531672} {\bibfield  {journal} {\bibinfo  {journal}
  {Journal of Mathematical Physics}\ }\textbf {\bibinfo {volume} {37}},\
  \bibinfo {pages} {4845} (\bibinfo {year} {1996})}\BibitemShut {NoStop}%
\bibitem [{\citenamefont {Levitov}(2003)}]{Levitov}%
  \BibitemOpen
  \bibfield  {author} {\bibinfo {author} {\bibfnamefont {L.~S.}\ \bibnamefont
  {Levitov}},\ }\href@noop {} {\emph {\bibinfo {title} {Quantum Noise in
  Mesoscopic Systems}}},\ edited by\ \bibinfo {editor} {\bibfnamefont {Y.~V.}\
  \bibnamefont {Nazarov}}\ (\bibinfo  {publisher} {Kluwer, Dordrecht},\
  \bibinfo {year} {2003})\BibitemShut {NoStop}%
\bibitem [{\citenamefont {Bagrets}\ and\ \citenamefont
  {Nazarov}(2003)}]{Nazarov_1}%
  \BibitemOpen
  \bibfield  {author} {\bibinfo {author} {\bibfnamefont {D.~A.}\ \bibnamefont
  {Bagrets}}\ and\ \bibinfo {author} {\bibfnamefont {Y.~V.}\ \bibnamefont
  {Nazarov}},\ }\href {\doibase 10.1103/PhysRevB.67.085316} {\bibfield
  {journal} {\bibinfo  {journal} {Phys. Rev. B}\ }\textbf {\bibinfo {volume}
  {67}},\ \bibinfo {pages} {085316} (\bibinfo {year} {2003})}\BibitemShut
  {NoStop}%
\bibitem [{\citenamefont {Levitov}\ and\ \citenamefont
  {Reznikov}(2004)}]{Levitov_PRB(2004)}%
  \BibitemOpen
  \bibfield  {author} {\bibinfo {author} {\bibfnamefont {L.~S.}\ \bibnamefont
  {Levitov}}\ and\ \bibinfo {author} {\bibfnamefont {M.}~\bibnamefont
  {Reznikov}},\ }\href
  {https://journals.aps.org/prb/abstract/10.1103/PhysRevB.70.115305} {\bibfield
   {journal} {\bibinfo  {journal} {Phys. Rev. B}\ }\textbf {\bibinfo {volume}
  {70}},\ \bibinfo {pages} {115305} (\bibinfo {year} {2004})}\BibitemShut
  {NoStop}%
\bibitem [{\citenamefont {Gustavsson}\ \emph {et~al.}(2006)\citenamefont
  {Gustavsson}, \citenamefont {Leturcq}, \citenamefont
  {Simovi\ifmmode~\check{c}\else \v{c}\fi{}}, \citenamefont {Schleser},
  \citenamefont {Ihn}, \citenamefont {Studerus}, \citenamefont {Ensslin},
  \citenamefont {Driscoll},\ and\ \citenamefont {Gossard}}]{Gustavsson}%
  \BibitemOpen
  \bibfield  {author} {\bibinfo {author} {\bibfnamefont {S.}~\bibnamefont
  {Gustavsson}}, \bibinfo {author} {\bibfnamefont {R.}~\bibnamefont {Leturcq}},
  \bibinfo {author} {\bibfnamefont {B.}~\bibnamefont
  {Simovi\ifmmode~\check{c}\else \v{c}\fi{}}}, \bibinfo {author} {\bibfnamefont
  {R.}~\bibnamefont {Schleser}}, \bibinfo {author} {\bibfnamefont
  {T.}~\bibnamefont {Ihn}}, \bibinfo {author} {\bibfnamefont {P.}~\bibnamefont
  {Studerus}}, \bibinfo {author} {\bibfnamefont {K.}~\bibnamefont {Ensslin}},
  \bibinfo {author} {\bibfnamefont {D.~C.}\ \bibnamefont {Driscoll}}, \ and\
  \bibinfo {author} {\bibfnamefont {A.~C.}\ \bibnamefont {Gossard}},\ }\href
  {\doibase 10.1103/PhysRevLett.96.076605} {\bibfield  {journal} {\bibinfo
  {journal} {Phys. Rev. Lett.}\ }\textbf {\bibinfo {volume} {96}},\ \bibinfo
  {pages} {076605} (\bibinfo {year} {2006})}\BibitemShut {NoStop}%
\bibitem [{\citenamefont {Belzig}(2005)}]{Belzig}%
  \BibitemOpen
  \bibfield  {author} {\bibinfo {author} {\bibfnamefont {W.}~\bibnamefont
  {Belzig}},\ }\href {\doibase 10.1103/PhysRevB.71.161301} {\bibfield
  {journal} {\bibinfo  {journal} {Phys. Rev. B}\ }\textbf {\bibinfo {volume}
  {71}},\ \bibinfo {pages} {161301} (\bibinfo {year} {2005})}\BibitemShut
  {NoStop}%
\bibitem [{\citenamefont {Clerk}\ \emph {et~al.}(2010)\citenamefont {Clerk},
  \citenamefont {Devoret}, \citenamefont {Girvin}, \citenamefont {Marquardt},\
  and\ \citenamefont {Schoelkopf}}]{clerk}%
  \BibitemOpen
  \bibfield  {author} {\bibinfo {author} {\bibfnamefont {A.~A.}\ \bibnamefont
  {Clerk}}, \bibinfo {author} {\bibfnamefont {M.~H.}\ \bibnamefont {Devoret}},
  \bibinfo {author} {\bibfnamefont {S.~M.}\ \bibnamefont {Girvin}}, \bibinfo
  {author} {\bibfnamefont {F.}~\bibnamefont {Marquardt}}, \ and\ \bibinfo
  {author} {\bibfnamefont {R.~J.}\ \bibnamefont {Schoelkopf}},\ }\href
  {\doibase 10.1103/RevModPhys.82.1155} {\bibfield  {journal} {\bibinfo
  {journal} {Rev. Mod. Phys.}\ }\textbf {\bibinfo {volume} {82}},\ \bibinfo
  {pages} {1155} (\bibinfo {year} {2010})}\BibitemShut {NoStop}%
\bibitem [{\citenamefont {Reulet}\ \emph {et~al.}(2003)\citenamefont {Reulet},
  \citenamefont {Senzier},\ and\ \citenamefont {Prober}}]{m1}%
  \BibitemOpen
  \bibfield  {author} {\bibinfo {author} {\bibfnamefont {B.}~\bibnamefont
  {Reulet}}, \bibinfo {author} {\bibfnamefont {J.}~\bibnamefont {Senzier}}, \
  and\ \bibinfo {author} {\bibfnamefont {D.~E.}\ \bibnamefont {Prober}},\
  }\href {\doibase 10.1103/PhysRevLett.91.196601} {\bibfield  {journal}
  {\bibinfo  {journal} {Phys. Rev. Lett.}\ }\textbf {\bibinfo {volume} {91}},\
  \bibinfo {pages} {196601} (\bibinfo {year} {2003})}\BibitemShut {NoStop}%
\bibitem [{\citenamefont {Bomze}\ \emph {et~al.}(2005)\citenamefont {Bomze},
  \citenamefont {Gershon}, \citenamefont {Shovkun}, \citenamefont {Levitov},\
  and\ \citenamefont {Reznikov}}]{m2}%
  \BibitemOpen
  \bibfield  {author} {\bibinfo {author} {\bibfnamefont {Y.}~\bibnamefont
  {Bomze}}, \bibinfo {author} {\bibfnamefont {G.}~\bibnamefont {Gershon}},
  \bibinfo {author} {\bibfnamefont {D.}~\bibnamefont {Shovkun}}, \bibinfo
  {author} {\bibfnamefont {L.~S.}\ \bibnamefont {Levitov}}, \ and\ \bibinfo
  {author} {\bibfnamefont {M.}~\bibnamefont {Reznikov}},\ }\href {\doibase
  10.1103/PhysRevLett.95.176601} {\bibfield  {journal} {\bibinfo  {journal}
  {Phys. Rev. Lett.}\ }\textbf {\bibinfo {volume} {95}},\ \bibinfo {pages}
  {176601} (\bibinfo {year} {2005})}\BibitemShut {NoStop}%
\bibitem [{\citenamefont {Fujisawa}\ \emph {et~al.}(2006)\citenamefont
  {Fujisawa}, \citenamefont {Hayashi}, \citenamefont {Tomita},\ and\
  \citenamefont {Hirayama}}]{Fujisawa1634}%
  \BibitemOpen
  \bibfield  {author} {\bibinfo {author} {\bibfnamefont {T.}~\bibnamefont
  {Fujisawa}}, \bibinfo {author} {\bibfnamefont {T.}~\bibnamefont {Hayashi}},
  \bibinfo {author} {\bibfnamefont {R.}~\bibnamefont {Tomita}}, \ and\ \bibinfo
  {author} {\bibfnamefont {Y.}~\bibnamefont {Hirayama}},\ }\href {\doibase
  10.1126/science.1126788} {\bibfield  {journal} {\bibinfo  {journal}
  {Science}\ }\textbf {\bibinfo {volume} {312}},\ \bibinfo {pages} {1634}
  (\bibinfo {year} {2006})}\BibitemShut {NoStop}%
\bibitem [{\citenamefont {Fricke}\ \emph {et~al.}(2007)\citenamefont {Fricke},
  \citenamefont {Hohls}, \citenamefont {Wegscheider},\ and\ \citenamefont
  {Haug}}]{m4}%
  \BibitemOpen
  \bibfield  {author} {\bibinfo {author} {\bibfnamefont {C.}~\bibnamefont
  {Fricke}}, \bibinfo {author} {\bibfnamefont {F.}~\bibnamefont {Hohls}},
  \bibinfo {author} {\bibfnamefont {W.}~\bibnamefont {Wegscheider}}, \ and\
  \bibinfo {author} {\bibfnamefont {R.~J.}\ \bibnamefont {Haug}},\ }\href
  {\doibase 10.1103/PhysRevB.76.155307} {\bibfield  {journal} {\bibinfo
  {journal} {Phys. Rev. B}\ }\textbf {\bibinfo {volume} {76}},\ \bibinfo
  {pages} {155307} (\bibinfo {year} {2007})}\BibitemShut {NoStop}%
\bibitem [{\citenamefont {Timofeev}\ \emph {et~al.}(2007)\citenamefont
  {Timofeev}, \citenamefont {Meschke}, \citenamefont {Peltonen}, \citenamefont
  {Heikkil\"a},\ and\ \citenamefont {Pekola}}]{m5}%
  \BibitemOpen
  \bibfield  {author} {\bibinfo {author} {\bibfnamefont {A.~V.}\ \bibnamefont
  {Timofeev}}, \bibinfo {author} {\bibfnamefont {M.}~\bibnamefont {Meschke}},
  \bibinfo {author} {\bibfnamefont {J.~T.}\ \bibnamefont {Peltonen}}, \bibinfo
  {author} {\bibfnamefont {T.~T.}\ \bibnamefont {Heikkil\"a}}, \ and\ \bibinfo
  {author} {\bibfnamefont {J.~P.}\ \bibnamefont {Pekola}},\ }\href {\doibase
  10.1103/PhysRevLett.98.207001} {\bibfield  {journal} {\bibinfo  {journal}
  {Phys. Rev. Lett.}\ }\textbf {\bibinfo {volume} {98}},\ \bibinfo {pages}
  {207001} (\bibinfo {year} {2007})}\BibitemShut {NoStop}%
\bibitem [{\citenamefont {Gershon}\ \emph {et~al.}(2008)\citenamefont
  {Gershon}, \citenamefont {Bomze}, \citenamefont {Sukhorukov},\ and\
  \citenamefont {Reznikov}}]{m6}%
  \BibitemOpen
  \bibfield  {author} {\bibinfo {author} {\bibfnamefont {G.}~\bibnamefont
  {Gershon}}, \bibinfo {author} {\bibfnamefont {Y.}~\bibnamefont {Bomze}},
  \bibinfo {author} {\bibfnamefont {E.~V.}\ \bibnamefont {Sukhorukov}}, \ and\
  \bibinfo {author} {\bibfnamefont {M.}~\bibnamefont {Reznikov}},\ }\href
  {\doibase 10.1103/PhysRevLett.101.016803} {\bibfield  {journal} {\bibinfo
  {journal} {Phys. Rev. Lett.}\ }\textbf {\bibinfo {volume} {101}},\ \bibinfo
  {pages} {016803} (\bibinfo {year} {2008})}\BibitemShut {NoStop}%
\bibitem [{\citenamefont {Gabelli}\ and\ \citenamefont {Reulet}(2009)}]{m7}%
  \BibitemOpen
  \bibfield  {author} {\bibinfo {author} {\bibfnamefont {J.}~\bibnamefont
  {Gabelli}}\ and\ \bibinfo {author} {\bibfnamefont {B.}~\bibnamefont
  {Reulet}},\ }\href {\doibase 10.1103/PhysRevB.80.161203} {\bibfield
  {journal} {\bibinfo  {journal} {Phys. Rev. B}\ }\textbf {\bibinfo {volume}
  {80}},\ \bibinfo {pages} {161203} (\bibinfo {year} {2009})}\BibitemShut
  {NoStop}%
\bibitem [{\citenamefont {Flindt}\ \emph {et~al.}(2009)\citenamefont {Flindt},
  \citenamefont {Fricke}, \citenamefont {Hohls}, \citenamefont {Novotn{\'y}},
  \citenamefont {Neto{\v c}n{\'y}}, \citenamefont {Brandes},\ and\
  \citenamefont {Haug}}]{m8}%
  \BibitemOpen
  \bibfield  {author} {\bibinfo {author} {\bibfnamefont {C.}~\bibnamefont
  {Flindt}}, \bibinfo {author} {\bibfnamefont {C.}~\bibnamefont {Fricke}},
  \bibinfo {author} {\bibfnamefont {F.}~\bibnamefont {Hohls}}, \bibinfo
  {author} {\bibfnamefont {T.}~\bibnamefont {Novotn{\'y}}}, \bibinfo {author}
  {\bibfnamefont {K.}~\bibnamefont {Neto{\v c}n{\'y}}}, \bibinfo {author}
  {\bibfnamefont {T.}~\bibnamefont {Brandes}}, \ and\ \bibinfo {author}
  {\bibfnamefont {R.~J.}\ \bibnamefont {Haug}},\ }\href {\doibase
  10.1073/pnas.0901002106} {\bibfield  {journal} {\bibinfo  {journal}
  {Proceedings of the National Academy of Sciences}\ }\textbf {\bibinfo
  {volume} {106}},\ \bibinfo {pages} {10116} (\bibinfo {year}
  {2009})}\BibitemShut {NoStop}%
\bibitem [{\citenamefont {Muzykantskii}\ and\ \citenamefont
  {Khmelnitskii}(1994)}]{sita1}%
  \BibitemOpen
  \bibfield  {author} {\bibinfo {author} {\bibfnamefont {B.~A.}\ \bibnamefont
  {Muzykantskii}}\ and\ \bibinfo {author} {\bibfnamefont {D.~E.}\ \bibnamefont
  {Khmelnitskii}},\ }\href {\doibase 10.1103/PhysRevB.50.3982} {\bibfield
  {journal} {\bibinfo  {journal} {Phys. Rev. B}\ }\textbf {\bibinfo {volume}
  {50}},\ \bibinfo {pages} {3982} (\bibinfo {year} {1994})}\BibitemShut
  {NoStop}%
\bibitem [{\citenamefont {Belzig}\ and\ \citenamefont {Nazarov}(2001)}]{sita2}%
  \BibitemOpen
  \bibfield  {author} {\bibinfo {author} {\bibfnamefont {W.}~\bibnamefont
  {Belzig}}\ and\ \bibinfo {author} {\bibfnamefont {Y.~V.}\ \bibnamefont
  {Nazarov}},\ }\href {\doibase 10.1103/PhysRevLett.87.067006} {\bibfield
  {journal} {\bibinfo  {journal} {Phys. Rev. Lett.}\ }\textbf {\bibinfo
  {volume} {87}},\ \bibinfo {pages} {067006} (\bibinfo {year}
  {2001})}\BibitemShut {NoStop}%
\bibitem [{\citenamefont {Pilgram}\ \emph {et~al.}(2003)\citenamefont
  {Pilgram}, \citenamefont {Jordan}, \citenamefont {Sukhorukov},\ and\
  \citenamefont {B\"uttiker}}]{sita4}%
  \BibitemOpen
  \bibfield  {author} {\bibinfo {author} {\bibfnamefont {S.}~\bibnamefont
  {Pilgram}}, \bibinfo {author} {\bibfnamefont {A.~N.}\ \bibnamefont {Jordan}},
  \bibinfo {author} {\bibfnamefont {E.~V.}\ \bibnamefont {Sukhorukov}}, \ and\
  \bibinfo {author} {\bibfnamefont {M.}~\bibnamefont {B\"uttiker}},\ }\href
  {\doibase 10.1103/PhysRevLett.90.206801} {\bibfield  {journal} {\bibinfo
  {journal} {Phys. Rev. Lett.}\ }\textbf {\bibinfo {volume} {90}},\ \bibinfo
  {pages} {206801} (\bibinfo {year} {2003})}\BibitemShut {NoStop}%
\bibitem [{\citenamefont {Taddei}\ and\ \citenamefont {Fazio}(2002)}]{sita5}%
  \BibitemOpen
  \bibfield  {author} {\bibinfo {author} {\bibfnamefont {F.}~\bibnamefont
  {Taddei}}\ and\ \bibinfo {author} {\bibfnamefont {R.}~\bibnamefont {Fazio}},\
  }\href {\doibase 10.1103/PhysRevB.65.075317} {\bibfield  {journal} {\bibinfo
  {journal} {Phys. Rev. B}\ }\textbf {\bibinfo {volume} {65}},\ \bibinfo
  {pages} {075317} (\bibinfo {year} {2002})}\BibitemShut {NoStop}%
\bibitem [{\citenamefont {Di~Lorenzo}\ and\ \citenamefont
  {Nazarov}(2004)}]{sita6}%
  \BibitemOpen
  \bibfield  {author} {\bibinfo {author} {\bibfnamefont {A.}~\bibnamefont
  {Di~Lorenzo}}\ and\ \bibinfo {author} {\bibfnamefont {Y.~V.}\ \bibnamefont
  {Nazarov}},\ }\href {\doibase 10.1103/PhysRevLett.93.046601} {\bibfield
  {journal} {\bibinfo  {journal} {Phys. Rev. Lett.}\ }\textbf {\bibinfo
  {volume} {93}},\ \bibinfo {pages} {046601} (\bibinfo {year}
  {2004})}\BibitemShut {NoStop}%
\bibitem [{\citenamefont {Braggio}\ \emph {et~al.}(2006)\citenamefont
  {Braggio}, \citenamefont {K\"onig},\ and\ \citenamefont {Fazio}}]{sita7}%
  \BibitemOpen
  \bibfield  {author} {\bibinfo {author} {\bibfnamefont {A.}~\bibnamefont
  {Braggio}}, \bibinfo {author} {\bibfnamefont {J.}~\bibnamefont {K\"onig}}, \
  and\ \bibinfo {author} {\bibfnamefont {R.}~\bibnamefont {Fazio}},\ }\href
  {\doibase 10.1103/PhysRevLett.96.026805} {\bibfield  {journal} {\bibinfo
  {journal} {Phys. Rev. Lett.}\ }\textbf {\bibinfo {volume} {96}},\ \bibinfo
  {pages} {026805} (\bibinfo {year} {2006})}\BibitemShut {NoStop}%
\bibitem [{\citenamefont {Pistolesi}(2004)}]{sita8}%
  \BibitemOpen
  \bibfield  {author} {\bibinfo {author} {\bibfnamefont {F.}~\bibnamefont
  {Pistolesi}},\ }\href {\doibase 10.1103/PhysRevB.69.245409} {\bibfield
  {journal} {\bibinfo  {journal} {Phys. Rev. B}\ }\textbf {\bibinfo {volume}
  {69}},\ \bibinfo {pages} {245409} (\bibinfo {year} {2004})}\BibitemShut
  {NoStop}%
\bibitem [{\citenamefont {Bennett}\ and\ \citenamefont {Clerk}(2008)}]{sita9}%
  \BibitemOpen
  \bibfield  {author} {\bibinfo {author} {\bibfnamefont {S.~D.}\ \bibnamefont
  {Bennett}}\ and\ \bibinfo {author} {\bibfnamefont {A.~A.}\ \bibnamefont
  {Clerk}},\ }\href {\doibase 10.1103/PhysRevB.78.165328} {\bibfield  {journal}
  {\bibinfo  {journal} {Phys. Rev. B}\ }\textbf {\bibinfo {volume} {78}},\
  \bibinfo {pages} {165328} (\bibinfo {year} {2008})}\BibitemShut {NoStop}%
\bibitem [{\citenamefont {Kambly}\ \emph {et~al.}(2011)\citenamefont {Kambly},
  \citenamefont {Flindt},\ and\ \citenamefont {B\"uttiker}}]{gora}%
  \BibitemOpen
  \bibfield  {author} {\bibinfo {author} {\bibfnamefont {D.}~\bibnamefont
  {Kambly}}, \bibinfo {author} {\bibfnamefont {C.}~\bibnamefont {Flindt}}, \
  and\ \bibinfo {author} {\bibfnamefont {M.}~\bibnamefont {B\"uttiker}},\
  }\href {\doibase 10.1103/PhysRevB.83.075432} {\bibfield  {journal} {\bibinfo
  {journal} {Phys. Rev. B}\ }\textbf {\bibinfo {volume} {83}},\ \bibinfo
  {pages} {075432} (\bibinfo {year} {2011})}\BibitemShut {NoStop}%
\bibitem [{\citenamefont {Hewson}(1993)}]{Hewson}%
  \BibitemOpen
  \bibfield  {author} {\bibinfo {author} {\bibfnamefont {A.}~\bibnamefont
  {Hewson}},\ }\href@noop {} {\emph {\bibinfo {title} {The Kondo Problem to
  Heavy Fermions}}}\ (\bibinfo  {publisher} {Cambridge University Press,
  Cambridge, England},\ \bibinfo {year} {1993})\BibitemShut {NoStop}%
\bibitem [{\citenamefont {Coleman}(2015)}]{Coleman_book}%
  \BibitemOpen
  \bibfield  {author} {\bibinfo {author} {\bibfnamefont {P.}~\bibnamefont
  {Coleman}},\ }\href {https://doi.org/10.1017/CBO9781139020916} {\emph
  {\bibinfo {title} {Introduction to Many-Body Physics}}}\ (\bibinfo
  {publisher} {Cambridge University Press, Cambridge},\ \bibinfo {year}
  {2015})\BibitemShut {NoStop}%
\bibitem [{\citenamefont {Cox}\ and\ \citenamefont
  {Zawadowski}(1998)}]{Cox_Adv_Phys(47)_1998}%
  \BibitemOpen
  \bibfield  {author} {\bibinfo {author} {\bibfnamefont {D.~L.}\ \bibnamefont
  {Cox}}\ and\ \bibinfo {author} {\bibfnamefont {A.}~\bibnamefont
  {Zawadowski}},\ }\href {http://dx.doi.org/10.1080/000187398243500} {\bibfield
   {journal} {\bibinfo  {journal} {Adv. Phys.}\ }\textbf {\bibinfo {volume}
  {47}},\ \bibinfo {pages} {599} (\bibinfo {year} {1998})}\BibitemShut
  {NoStop}%
\bibitem [{\citenamefont {Kondo}(1964)}]{dk1}%
  \BibitemOpen
  \bibfield  {author} {\bibinfo {author} {\bibfnamefont {J.}~\bibnamefont
  {Kondo}},\ }\href {http://dx.doi.org/10.1143/PTP.32.37} {\bibfield  {journal}
  {\bibinfo  {journal} {Progress of Theoretical Physics}\ }\textbf {\bibinfo
  {volume} {32}},\ \bibinfo {pages} {37} (\bibinfo {year} {1964})}\BibitemShut
  {NoStop}%
\bibitem [{\citenamefont {Nozi{\'e}res}(1974)}]{Nozieres}%
  \BibitemOpen
  \bibfield  {author} {\bibinfo {author} {\bibfnamefont {P.}~\bibnamefont
  {Nozi{\'e}res}},\ }\href {https://doi.org/10.1007/BF00654541} {\bibfield
  {journal} {\bibinfo  {journal} {J. Low Temp. Phys.}\ }\textbf {\bibinfo
  {volume} {17}},\ \bibinfo {pages} {31} (\bibinfo {year} {1974})}\BibitemShut
  {NoStop}%
\bibitem [{\citenamefont {Nozieres}\ and\ \citenamefont
  {Blandin}(1980)}]{Nozieres_Blandin_JPhys_1980}%
  \BibitemOpen
  \bibfield  {author} {\bibinfo {author} {\bibfnamefont {P.}~\bibnamefont
  {Nozieres}}\ and\ \bibinfo {author} {\bibfnamefont {A.}~\bibnamefont
  {Blandin}},\ }\href {https://doi.org/10.1051/jphys:01980004103019300}
  {\bibfield  {journal} {\bibinfo  {journal} {J. Phys}\ }\textbf {\bibinfo
  {volume} {41}},\ \bibinfo {pages} {193} (\bibinfo {year} {1980})}\BibitemShut
  {NoStop}%
\bibitem [{\citenamefont {Affleck}\ and\ \citenamefont
  {Ludwig}(1993)}]{Affleck_Lud_PRB(48)_1993}%
  \BibitemOpen
  \bibfield  {author} {\bibinfo {author} {\bibfnamefont {I.}~\bibnamefont
  {Affleck}}\ and\ \bibinfo {author} {\bibfnamefont {A.~W.~W.}\ \bibnamefont
  {Ludwig}},\ }\href
  {https://journals.aps.org/prb/abstract/10.1103/PhysRevB.48.7297} {\bibfield
  {journal} {\bibinfo  {journal} {Phys. Rev. B}\ }\textbf {\bibinfo {volume}
  {48}},\ \bibinfo {pages} {7297} (\bibinfo {year} {1993})}\BibitemShut
  {NoStop}%
\bibitem [{\citenamefont {Bickers}\ \emph {et~al.}(1985)\citenamefont
  {Bickers}, \citenamefont {Cox},\ and\ \citenamefont
  {Wilkins}}]{Wilkins_PRL(54)_1985}%
  \BibitemOpen
  \bibfield  {author} {\bibinfo {author} {\bibfnamefont {N.~E.}\ \bibnamefont
  {Bickers}}, \bibinfo {author} {\bibfnamefont {D.~L.}\ \bibnamefont {Cox}}, \
  and\ \bibinfo {author} {\bibfnamefont {J.~W.}\ \bibnamefont {Wilkins}},\
  }\href {\doibase 10.1103/PhysRevLett.54.230} {\bibfield  {journal} {\bibinfo
  {journal} {Phys. Rev. Lett.}\ }\textbf {\bibinfo {volume} {54}},\ \bibinfo
  {pages} {230} (\bibinfo {year} {1985})}\BibitemShut {NoStop}%
\bibitem [{\citenamefont {Goldhaber-Gordon}\ \emph {et~al.}(1998)\citenamefont
  {Goldhaber-Gordon}, \citenamefont {Shtrikman}, \citenamefont {Mahalu},
  \citenamefont {Abusch-Magder}, \citenamefont {Meirav},\ and\ \citenamefont
  {Kastner}}]{Goldhaber_nat(391)_1998}%
  \BibitemOpen
  \bibfield  {author} {\bibinfo {author} {\bibfnamefont {D.}~\bibnamefont
  {Goldhaber-Gordon}}, \bibinfo {author} {\bibfnamefont {H.}~\bibnamefont
  {Shtrikman}}, \bibinfo {author} {\bibfnamefont {D.}~\bibnamefont {Mahalu}},
  \bibinfo {author} {\bibfnamefont {D.}~\bibnamefont {Abusch-Magder}}, \bibinfo
  {author} {\bibfnamefont {U.}~\bibnamefont {Meirav}}, \ and\ \bibinfo {author}
  {\bibfnamefont {M.~A.}\ \bibnamefont {Kastner}},\ }\href
  {http://dx.doi.org/10.1038/34373} {\bibfield  {journal} {\bibinfo  {journal}
  {Nature}\ }\textbf {\bibinfo {volume} {391}},\ \bibinfo {pages} {156}
  (\bibinfo {year} {1998})}\BibitemShut {NoStop}%
\bibitem [{\citenamefont {Cronenwett}\ \emph {et~al.}(1998)\citenamefont
  {Cronenwett}, \citenamefont {Oosterkamp},\ and\ \citenamefont
  {Kouwenhoven}}]{Cronewett_SCI(281)_1998}%
  \BibitemOpen
  \bibfield  {author} {\bibinfo {author} {\bibfnamefont {S.~M.}\ \bibnamefont
  {Cronenwett}}, \bibinfo {author} {\bibfnamefont {T.~H.}\ \bibnamefont
  {Oosterkamp}}, \ and\ \bibinfo {author} {\bibfnamefont {L.~P.}\ \bibnamefont
  {Kouwenhoven}},\ }\href {\doibase 10.1126/science.281.5376.540} {\bibfield
  {journal} {\bibinfo  {journal} {Science}\ }\textbf {\bibinfo {volume}
  {281}},\ \bibinfo {pages} {540} (\bibinfo {year} {1998})}\BibitemShut
  {NoStop}%
\bibitem [{\citenamefont {Nyg{\aa}ard}\ \emph {et~al.}(2000)\citenamefont
  {Nyg{\aa}ard}, \citenamefont {Cobden},\ and\ \citenamefont
  {Lindelof}}]{Jesper_NAT(408)_2000}%
  \BibitemOpen
  \bibfield  {author} {\bibinfo {author} {\bibfnamefont {J.}~\bibnamefont
  {Nyg{\aa}ard}}, \bibinfo {author} {\bibfnamefont {D.~H.}\ \bibnamefont
  {Cobden}}, \ and\ \bibinfo {author} {\bibfnamefont {P.~E.}\ \bibnamefont
  {Lindelof}},\ }\href {http://dx.doi.org/10.1038/35042545} {\bibfield
  {journal} {\bibinfo  {journal} {Nature}\ }\textbf {\bibinfo {volume} {408}},\
  \bibinfo {pages} {342} (\bibinfo {year} {2000})}\BibitemShut {NoStop}%
\bibitem [{\citenamefont {Kouwenhoven}\ and\ \citenamefont
  {Glazman}(2001)}]{revival}%
  \BibitemOpen
  \bibfield  {author} {\bibinfo {author} {\bibfnamefont {L.}~\bibnamefont
  {Kouwenhoven}}\ and\ \bibinfo {author} {\bibfnamefont {L.}~\bibnamefont
  {Glazman}},\ }\href {http://stacks.iop.org/2058-7058/14/i=1/a=28} {\bibfield
  {journal} {\bibinfo  {journal} {Physics World}\ }\textbf {\bibinfo {volume}
  {14}},\ \bibinfo {pages} {33} (\bibinfo {year} {2001})}\BibitemShut {NoStop}%
\bibitem [{\citenamefont {Jezouin}\ \emph {et~al.}(2013)\citenamefont
  {Jezouin}, \citenamefont {Parmentier}, \citenamefont {Anthore}, \citenamefont
  {Gennser}, \citenamefont {Cavanna}, \citenamefont {Jin},\ and\ \citenamefont
  {Pierre}}]{Pierre_SCI(342)_2013}%
  \BibitemOpen
  \bibfield  {author} {\bibinfo {author} {\bibfnamefont {S.}~\bibnamefont
  {Jezouin}}, \bibinfo {author} {\bibfnamefont {F.~D.}\ \bibnamefont
  {Parmentier}}, \bibinfo {author} {\bibfnamefont {A.}~\bibnamefont {Anthore}},
  \bibinfo {author} {\bibfnamefont {U.}~\bibnamefont {Gennser}}, \bibinfo
  {author} {\bibfnamefont {A.}~\bibnamefont {Cavanna}}, \bibinfo {author}
  {\bibfnamefont {Y.}~\bibnamefont {Jin}}, \ and\ \bibinfo {author}
  {\bibfnamefont {F.}~\bibnamefont {Pierre}},\ }\href {\doibase
  10.1126/science.1241912} {\bibfield  {journal} {\bibinfo  {journal}
  {Science}\ }\textbf {\bibinfo {volume} {342}},\ \bibinfo {pages} {601}
  (\bibinfo {year} {2013})}\BibitemShut {NoStop}%
\bibitem [{\citenamefont {Keller}\ \emph {et~al.}(2014)\citenamefont {Keller},
  \citenamefont {Amasha}, \citenamefont {Weymann}, \citenamefont {Moca},
  \citenamefont {Rau}, \citenamefont {Katine}, \citenamefont {Shtrikman},
  \citenamefont {Zar{\'a}nd},\ and\ \citenamefont
  {Goldhaber-Gordon}}]{Keller_Goldhaber_nat_phys(10)_2014}%
  \BibitemOpen
  \bibfield  {author} {\bibinfo {author} {\bibfnamefont {A.~J.}\ \bibnamefont
  {Keller}}, \bibinfo {author} {\bibfnamefont {S.}~\bibnamefont {Amasha}},
  \bibinfo {author} {\bibfnamefont {I.}~\bibnamefont {Weymann}}, \bibinfo
  {author} {\bibfnamefont {C.~P.}\ \bibnamefont {Moca}}, \bibinfo {author}
  {\bibfnamefont {I.~G.}\ \bibnamefont {Rau}}, \bibinfo {author} {\bibfnamefont
  {J.~A.}\ \bibnamefont {Katine}}, \bibinfo {author} {\bibfnamefont
  {H.}~\bibnamefont {Shtrikman}}, \bibinfo {author} {\bibfnamefont
  {G.}~\bibnamefont {Zar{\'a}nd}}, \ and\ \bibinfo {author} {\bibfnamefont
  {D.}~\bibnamefont {Goldhaber-Gordon}},\ }\href
  {http://dx.doi.org/10.1038/nphys2844} {\bibfield  {journal} {\bibinfo
  {journal} {Nature Physics}\ }\textbf {\bibinfo {volume} {10}},\ \bibinfo
  {pages} {145} (\bibinfo {year} {2014})}\BibitemShut {NoStop}%
\bibitem [{\citenamefont {Iftikhar}\ \emph {et~al.}(2015)\citenamefont
  {Iftikhar}, \citenamefont {Jezouin}, \citenamefont {Anthore}, \citenamefont
  {Gennser}, \citenamefont {Parmentier}, \citenamefont {Cavanna},\ and\
  \citenamefont {Pierre}}]{Pierre_NAT(526)_2015}%
  \BibitemOpen
  \bibfield  {author} {\bibinfo {author} {\bibfnamefont {Z.}~\bibnamefont
  {Iftikhar}}, \bibinfo {author} {\bibfnamefont {S.}~\bibnamefont {Jezouin}},
  \bibinfo {author} {\bibfnamefont {A.}~\bibnamefont {Anthore}}, \bibinfo
  {author} {\bibfnamefont {U.}~\bibnamefont {Gennser}}, \bibinfo {author}
  {\bibfnamefont {F.~D.}\ \bibnamefont {Parmentier}}, \bibinfo {author}
  {\bibfnamefont {A.}~\bibnamefont {Cavanna}}, \ and\ \bibinfo {author}
  {\bibfnamefont {F.}~\bibnamefont {Pierre}},\ }\href
  {http://dx.doi.org/10.1038/nature15384} {\bibfield  {journal} {\bibinfo
  {journal} {Nature}\ }\textbf {\bibinfo {volume} {526}},\ \bibinfo {pages}
  {233} (\bibinfo {year} {2015})}\BibitemShut {NoStop}%
\bibitem [{\citenamefont {Jezouin}\ \emph {et~al.}(2016)\citenamefont
  {Jezouin}, \citenamefont {Iftikhar}, \citenamefont {Anthore}, \citenamefont
  {Parmentier}, \citenamefont {Gennser}, \citenamefont {Cavanna}, \citenamefont
  {Ouerghi}, \citenamefont {Levkivskyi}, \citenamefont {Idrisov}, \citenamefont
  {Sukhorukov}, \citenamefont {Glazman},\ and\ \citenamefont
  {Pierre}}]{Pierre_NAT(536)_2016}%
  \BibitemOpen
  \bibfield  {author} {\bibinfo {author} {\bibfnamefont {S.}~\bibnamefont
  {Jezouin}}, \bibinfo {author} {\bibfnamefont {Z.}~\bibnamefont {Iftikhar}},
  \bibinfo {author} {\bibfnamefont {A.}~\bibnamefont {Anthore}}, \bibinfo
  {author} {\bibfnamefont {F.~D.}\ \bibnamefont {Parmentier}}, \bibinfo
  {author} {\bibfnamefont {U.}~\bibnamefont {Gennser}}, \bibinfo {author}
  {\bibfnamefont {A.}~\bibnamefont {Cavanna}}, \bibinfo {author} {\bibfnamefont
  {A.}~\bibnamefont {Ouerghi}}, \bibinfo {author} {\bibfnamefont {I.~P.}\
  \bibnamefont {Levkivskyi}}, \bibinfo {author} {\bibfnamefont
  {E.}~\bibnamefont {Idrisov}}, \bibinfo {author} {\bibfnamefont {E.~V.}\
  \bibnamefont {Sukhorukov}}, \bibinfo {author} {\bibfnamefont {L.~I.}\
  \bibnamefont {Glazman}}, \ and\ \bibinfo {author} {\bibfnamefont
  {F.}~\bibnamefont {Pierre}},\ }\href {http://dx.doi.org/10.1038/nature19072}
  {\bibfield  {journal} {\bibinfo  {journal} {Nature}\ }\textbf {\bibinfo
  {volume} {536}},\ \bibinfo {pages} {60} (\bibinfo {year} {2016})}\BibitemShut
  {NoStop}%
\bibitem [{\citenamefont {Pustilnik}\ and\ \citenamefont
  {Glazman}(2004)}]{GP_Review_2005}%
  \BibitemOpen
  \bibfield  {author} {\bibinfo {author} {\bibfnamefont {M.}~\bibnamefont
  {Pustilnik}}\ and\ \bibinfo {author} {\bibfnamefont {L.}~\bibnamefont
  {Glazman}},\ }\href {http://stacks.iop.org/0953-8984/16/i=16/a=R01}
  {\bibfield  {journal} {\bibinfo  {journal} {J. Phys.: Condens. Matter}\
  }\textbf {\bibinfo {volume} {16}},\ \bibinfo {pages} {R513} (\bibinfo {year}
  {2004})}\BibitemShut {NoStop}%
\bibitem [{\citenamefont {Mora}\ \emph {et~al.}(2008)\citenamefont {Mora},
  \citenamefont {Leyronas},\ and\ \citenamefont
  {Regnault}}]{Mora_Leyronas_Regnault_PRL_(100)_2008}%
  \BibitemOpen
  \bibfield  {author} {\bibinfo {author} {\bibfnamefont {C.}~\bibnamefont
  {Mora}}, \bibinfo {author} {\bibfnamefont {X.}~\bibnamefont {Leyronas}}, \
  and\ \bibinfo {author} {\bibfnamefont {N.}~\bibnamefont {Regnault}},\ }\href
  {https://journals.aps.org/prl/abstract/10.1103/PhysRevLett.100.036604}
  {\bibfield  {journal} {\bibinfo  {journal} {Phys. Rev. Lett.}\ }\textbf
  {\bibinfo {volume} {100}},\ \bibinfo {pages} {036604} (\bibinfo {year}
  {2008})}\BibitemShut {NoStop}%
\bibitem [{\citenamefont {Mora}\ \emph {et~al.}(2009)\citenamefont {Mora},
  \citenamefont {Vitushinsky}, \citenamefont {Leyronas}, \citenamefont
  {Clerk},\ and\ \citenamefont {Hur}}]{Mora_Clerk_Hur_PRB(80)_2009}%
  \BibitemOpen
  \bibfield  {author} {\bibinfo {author} {\bibfnamefont {C.}~\bibnamefont
  {Mora}}, \bibinfo {author} {\bibfnamefont {P.}~\bibnamefont {Vitushinsky}},
  \bibinfo {author} {\bibfnamefont {X.}~\bibnamefont {Leyronas}}, \bibinfo
  {author} {\bibfnamefont {A.~A.}\ \bibnamefont {Clerk}}, \ and\ \bibinfo
  {author} {\bibfnamefont {K.~L.}\ \bibnamefont {Hur}},\ }\href
  {https://journals.aps.org/prb/abstract/10.1103/PhysRevB.80.155322} {\bibfield
   {journal} {\bibinfo  {journal} {Phys. Rev. B}\ }\textbf {\bibinfo {volume}
  {80}},\ \bibinfo {pages} {155322} (\bibinfo {year} {2009})}\BibitemShut
  {NoStop}%
\bibitem [{\citenamefont {Hanl}\ \emph {et~al.}(2014)\citenamefont {Hanl},
  \citenamefont {Weichselbaum}, \citenamefont {von Delft},\ and\ \citenamefont
  {Kiselev}}]{HWDK_PRB_(89)_2014}%
  \BibitemOpen
  \bibfield  {author} {\bibinfo {author} {\bibfnamefont {M.}~\bibnamefont
  {Hanl}}, \bibinfo {author} {\bibfnamefont {A.}~\bibnamefont {Weichselbaum}},
  \bibinfo {author} {\bibfnamefont {J.}~\bibnamefont {von Delft}}, \ and\
  \bibinfo {author} {\bibfnamefont {M.}~\bibnamefont {Kiselev}},\ }\href
  {https://journals.aps.org/prb/abstract/10.1103/PhysRevB.89.195131} {\bibfield
   {journal} {\bibinfo  {journal} {Phys. Rev. B}\ }\textbf {\bibinfo {volume}
  {89}},\ \bibinfo {pages} {195131} (\bibinfo {year} {2014})}\BibitemShut
  {NoStop}%
\bibitem [{\citenamefont {Gogolin}\ and\ \citenamefont
  {Komnik}(2006{\natexlab{a}})}]{Gogolin_Komnik_PRL(2006)}%
  \BibitemOpen
  \bibfield  {author} {\bibinfo {author} {\bibfnamefont {A.~O.}\ \bibnamefont
  {Gogolin}}\ and\ \bibinfo {author} {\bibfnamefont {A.}~\bibnamefont
  {Komnik}},\ }\href
  {https://journals.aps.org/prl/abstract/10.1103/PhysRevLett.97.016602}
  {\bibfield  {journal} {\bibinfo  {journal} {Phys. Rev. Lett.}\ }\textbf
  {\bibinfo {volume} {97}},\ \bibinfo {pages} {016602} (\bibinfo {year}
  {2006}{\natexlab{a}})}\BibitemShut {NoStop}%
\bibitem [{\citenamefont {Gogolin}\ and\ \citenamefont
  {Komnik}(2006{\natexlab{b}})}]{Gogolin_Komnik_PRB(2006)}%
  \BibitemOpen
  \bibfield  {author} {\bibinfo {author} {\bibfnamefont {A.~O.}\ \bibnamefont
  {Gogolin}}\ and\ \bibinfo {author} {\bibfnamefont {A.}~\bibnamefont
  {Komnik}},\ }\href
  {https://journals.aps.org/prb/abstract/10.1103/PhysRevB.73.195301} {\bibfield
   {journal} {\bibinfo  {journal} {Phys. Rev. B}\ }\textbf {\bibinfo {volume}
  {73}},\ \bibinfo {pages} {195301} (\bibinfo {year}
  {2006}{\natexlab{b}})}\BibitemShut {NoStop}%
\bibitem [{\citenamefont {Golub}(2006)}]{golub}%
  \BibitemOpen
  \bibfield  {author} {\bibinfo {author} {\bibfnamefont {A.}~\bibnamefont
  {Golub}},\ }\href {\doibase 10.1103/PhysRevB.73.233310} {\bibfield  {journal}
  {\bibinfo  {journal} {Phys. Rev. B}\ }\textbf {\bibinfo {volume} {73}},\
  \bibinfo {pages} {233310} (\bibinfo {year} {2006})}\BibitemShut {NoStop}%
\bibitem [{\citenamefont {Schmidt}\ \emph
  {et~al.}(2007{\natexlab{a}})\citenamefont {Schmidt}, \citenamefont
  {Gogolin},\ and\ \citenamefont {Komnik}}]{tschmidt_2}%
  \BibitemOpen
  \bibfield  {author} {\bibinfo {author} {\bibfnamefont {T.~L.}\ \bibnamefont
  {Schmidt}}, \bibinfo {author} {\bibfnamefont {A.~O.}\ \bibnamefont
  {Gogolin}}, \ and\ \bibinfo {author} {\bibfnamefont {A.}~\bibnamefont
  {Komnik}},\ }\href {\doibase 10.1103/PhysRevB.75.235105} {\bibfield
  {journal} {\bibinfo  {journal} {Phys. Rev. B}\ }\textbf {\bibinfo {volume}
  {75}},\ \bibinfo {pages} {235105} (\bibinfo {year}
  {2007}{\natexlab{a}})}\BibitemShut {NoStop}%
\bibitem [{\citenamefont {Schmidt}\ \emph
  {et~al.}(2007{\natexlab{b}})\citenamefont {Schmidt}, \citenamefont {Komnik},\
  and\ \citenamefont {Gogolin}}]{tschmidt_1}%
  \BibitemOpen
  \bibfield  {author} {\bibinfo {author} {\bibfnamefont {T.~L.}\ \bibnamefont
  {Schmidt}}, \bibinfo {author} {\bibfnamefont {A.}~\bibnamefont {Komnik}}, \
  and\ \bibinfo {author} {\bibfnamefont {A.~O.}\ \bibnamefont {Gogolin}},\
  }\href {\doibase 10.1103/PhysRevB.76.241307} {\bibfield  {journal} {\bibinfo
  {journal} {Phys. Rev. B}\ }\textbf {\bibinfo {volume} {76}},\ \bibinfo
  {pages} {241307} (\bibinfo {year} {2007}{\natexlab{b}})}\BibitemShut
  {NoStop}%
\bibitem [{\citenamefont {Sakano}\ \emph {et~al.}(2012)\citenamefont {Sakano},
  \citenamefont {Nishikawa}, \citenamefont {Oguri}, \citenamefont {Hewson},\
  and\ \citenamefont {Tarucha}}]{ahes}%
  \BibitemOpen
  \bibfield  {author} {\bibinfo {author} {\bibfnamefont {R.}~\bibnamefont
  {Sakano}}, \bibinfo {author} {\bibfnamefont {Y.}~\bibnamefont {Nishikawa}},
  \bibinfo {author} {\bibfnamefont {A.}~\bibnamefont {Oguri}}, \bibinfo
  {author} {\bibfnamefont {A.~C.}\ \bibnamefont {Hewson}}, \ and\ \bibinfo
  {author} {\bibfnamefont {S.}~\bibnamefont {Tarucha}},\ }\href {\doibase
  10.1103/PhysRevLett.108.266401} {\bibfield  {journal} {\bibinfo  {journal}
  {Phys. Rev. Lett.}\ }\textbf {\bibinfo {volume} {108}},\ \bibinfo {pages}
  {266401} (\bibinfo {year} {2012})}\BibitemShut {NoStop}%
\bibitem [{\citenamefont {Pustilnik}\ and\ \citenamefont
  {Glazman}(2001)}]{Glazman_PRL_2001}%
  \BibitemOpen
  \bibfield  {author} {\bibinfo {author} {\bibfnamefont {M.}~\bibnamefont
  {Pustilnik}}\ and\ \bibinfo {author} {\bibfnamefont {L.~I.}\ \bibnamefont
  {Glazman}},\ }\href
  {https://journals.aps.org/prl/abstract/10.1103/PhysRevLett.87.216601}
  {\bibfield  {journal} {\bibinfo  {journal} {Phys. Rev. Lett.}\ }\textbf
  {\bibinfo {volume} {87}},\ \bibinfo {pages} {216601} (\bibinfo {year}
  {2001})}\BibitemShut {NoStop}%
\bibitem [{\citenamefont {Sasaki}\ \emph {et~al.}(2000)\citenamefont {Sasaki},
  \citenamefont {De~Franceschi}, \citenamefont {Elzerman}, \citenamefont
  {van~der Wiel}, \citenamefont {Eto}, \citenamefont {Tarucha},\ and\
  \citenamefont {Kouwenhoven}}]{intspin}%
  \BibitemOpen
  \bibfield  {author} {\bibinfo {author} {\bibfnamefont {S.}~\bibnamefont
  {Sasaki}}, \bibinfo {author} {\bibfnamefont {S.}~\bibnamefont
  {De~Franceschi}}, \bibinfo {author} {\bibfnamefont {J.~M.}\ \bibnamefont
  {Elzerman}}, \bibinfo {author} {\bibfnamefont {W.~G.}\ \bibnamefont {van~der
  Wiel}}, \bibinfo {author} {\bibfnamefont {M.}~\bibnamefont {Eto}}, \bibinfo
  {author} {\bibfnamefont {S.}~\bibnamefont {Tarucha}}, \ and\ \bibinfo
  {author} {\bibfnamefont {L.~P.}\ \bibnamefont {Kouwenhoven}},\ }\href
  {\doibase doi:10.1038/35015509} {\bibfield  {journal} {\bibinfo  {journal}
  {Nature}\ }\textbf {\bibinfo {volume} {405}},\ \bibinfo {pages} {764}
  (\bibinfo {year} {2000})}\BibitemShut {NoStop}%
\bibitem [{\citenamefont {Eto}\ and\ \citenamefont {Nazarov}(2000)}]{KMaddy}%
  \BibitemOpen
  \bibfield  {author} {\bibinfo {author} {\bibfnamefont {M.}~\bibnamefont
  {Eto}}\ and\ \bibinfo {author} {\bibfnamefont {Y.~V.}\ \bibnamefont
  {Nazarov}},\ }\href {\doibase 10.1103/PhysRevLett.85.1306} {\bibfield
  {journal} {\bibinfo  {journal} {Phys. Rev. Lett.}\ }\textbf {\bibinfo
  {volume} {85}},\ \bibinfo {pages} {1306} (\bibinfo {year}
  {2000})}\BibitemShut {NoStop}%
\bibitem [{\citenamefont {Quay}\ \emph {et~al.}(2007)\citenamefont {Quay},
  \citenamefont {Cumings}, \citenamefont {Gamble}, \citenamefont {Picciotto},
  \citenamefont {Kataura},\ and\ \citenamefont {Goldhaber-Gordon}}]{lla}%
  \BibitemOpen
  \bibfield  {author} {\bibinfo {author} {\bibfnamefont {C.~H.~L.}\
  \bibnamefont {Quay}}, \bibinfo {author} {\bibfnamefont {J.}~\bibnamefont
  {Cumings}}, \bibinfo {author} {\bibfnamefont {S.~J.}\ \bibnamefont {Gamble}},
  \bibinfo {author} {\bibfnamefont {R.~d.}\ \bibnamefont {Picciotto}}, \bibinfo
  {author} {\bibfnamefont {H.}~\bibnamefont {Kataura}}, \ and\ \bibinfo
  {author} {\bibfnamefont {D.}~\bibnamefont {Goldhaber-Gordon}},\ }\href
  {\doibase 10.1103/PhysRevB.76.245311} {\bibfield  {journal} {\bibinfo
  {journal} {Phys. Rev. B}\ }\textbf {\bibinfo {volume} {76}},\ \bibinfo
  {pages} {245311} (\bibinfo {year} {2007})}\BibitemShut {NoStop}%
\bibitem [{\citenamefont {Mitchell}\ \emph {et~al.}(2017)\citenamefont
  {Mitchell}, \citenamefont {Pedersen}, \citenamefont {Hedegard},\ and\
  \citenamefont {Paaske}}]{bdaama}%
  \BibitemOpen
  \bibfield  {author} {\bibinfo {author} {\bibfnamefont {A.~K.}\ \bibnamefont
  {Mitchell}}, \bibinfo {author} {\bibfnamefont {K.~G.~L.}\ \bibnamefont
  {Pedersen}}, \bibinfo {author} {\bibfnamefont {P.}~\bibnamefont {Hedegard}},
  \ and\ \bibinfo {author} {\bibfnamefont {J.}~\bibnamefont {Paaske}},\ }\href
  {http://dx.doi.org/10.1038/ncomms15210} {\bibfield  {journal} {\bibinfo
  {journal} {Nature Communications}\ }\textbf {\bibinfo {volume} {8}},\
  \bibinfo {pages} {15210} (\bibinfo {year} {2017})}\BibitemShut {NoStop}%
\bibitem [{\citenamefont {Posazhennikova}\ and\ \citenamefont
  {Coleman}(2005)}]{Coleman_PRL(94)_2005}%
  \BibitemOpen
  \bibfield  {author} {\bibinfo {author} {\bibfnamefont {A.}~\bibnamefont
  {Posazhennikova}}\ and\ \bibinfo {author} {\bibfnamefont {P.}~\bibnamefont
  {Coleman}},\ }\href
  {https://journals.aps.org/prl/abstract/10.1103/PhysRevLett.94.036802}
  {\bibfield  {journal} {\bibinfo  {journal} {Phys. Rev. Lett.}\ }\textbf
  {\bibinfo {volume} {94}},\ \bibinfo {pages} {036802} (\bibinfo {year}
  {2005})}\BibitemShut {NoStop}%
\bibitem [{\citenamefont {Karki}\ \emph {et~al.}(2018)\citenamefont {Karki},
  \citenamefont {Mora}, \citenamefont {von Delft},\ and\ \citenamefont
  {Kiselev}}]{KMDK_2018}%
  \BibitemOpen
  \bibfield  {author} {\bibinfo {author} {\bibfnamefont {D.~B.}\ \bibnamefont
  {Karki}}, \bibinfo {author} {\bibfnamefont {C.}~\bibnamefont {Mora}},
  \bibinfo {author} {\bibfnamefont {J.}~\bibnamefont {von Delft}}, \ and\
  \bibinfo {author} {\bibfnamefont {M.~N.}\ \bibnamefont {Kiselev}},\ }\href
  {\doibase 10.1103/PhysRevB.97.195403} {\bibfield  {journal} {\bibinfo
  {journal} {Phys. Rev. B}\ }\textbf {\bibinfo {volume} {97}},\ \bibinfo
  {pages} {195403} (\bibinfo {year} {2018})}\BibitemShut {NoStop}%
\bibitem [{\citenamefont {Posazhennikova}\ \emph {et~al.}(2007)\citenamefont
  {Posazhennikova}, \citenamefont {Bayani},\ and\ \citenamefont
  {Coleman}}]{Coleman_PRB(75)_2007}%
  \BibitemOpen
  \bibfield  {author} {\bibinfo {author} {\bibfnamefont {A.}~\bibnamefont
  {Posazhennikova}}, \bibinfo {author} {\bibfnamefont {B.}~\bibnamefont
  {Bayani}}, \ and\ \bibinfo {author} {\bibfnamefont {P.}~\bibnamefont
  {Coleman}},\ }\href
  {https://journals.aps.org/prb/abstract/10.1103/PhysRevB.75.245329} {\bibfield
   {journal} {\bibinfo  {journal} {Phys. Rev. B}\ }\textbf {\bibinfo {volume}
  {75}},\ \bibinfo {pages} {245329} (\bibinfo {year} {2007})}\BibitemShut
  {NoStop}%
\bibitem [{\citenamefont {Sela}\ \emph {et~al.}(2006)\citenamefont {Sela},
  \citenamefont {Oreg}, \citenamefont {von Oppen},\ and\ \citenamefont
  {Koch}}]{Sela_Oreg_Oppen_Koch_PRL_(97)_2006}%
  \BibitemOpen
  \bibfield  {author} {\bibinfo {author} {\bibfnamefont {E.}~\bibnamefont
  {Sela}}, \bibinfo {author} {\bibfnamefont {Y.}~\bibnamefont {Oreg}}, \bibinfo
  {author} {\bibfnamefont {F.}~\bibnamefont {von Oppen}}, \ and\ \bibinfo
  {author} {\bibfnamefont {J.}~\bibnamefont {Koch}},\ }\href
  {https://journals.aps.org/prl/abstract/10.1103/PhysRevLett.97.086601}
  {\bibfield  {journal} {\bibinfo  {journal} {Phys. Rev. Lett.}\ }\textbf
  {\bibinfo {volume} {97}},\ \bibinfo {pages} {086601} (\bibinfo {year}
  {2006})}\BibitemShut {NoStop}%
\bibitem [{\citenamefont {Schrieffer}\ and\ \citenamefont
  {Wolf}(1966)}]{Schrieffer_Wolf_PR(149)_1966}%
  \BibitemOpen
  \bibfield  {author} {\bibinfo {author} {\bibfnamefont {J.~R.}\ \bibnamefont
  {Schrieffer}}\ and\ \bibinfo {author} {\bibfnamefont {P.}~\bibnamefont
  {Wolf}},\ }\href
  {https://journals.aps.org/pr/abstract/10.1103/PhysRev.149.491} {\bibfield
  {journal} {\bibinfo  {journal} {Phys. Rev.}\ }\textbf {\bibinfo {volume}
  {149}},\ \bibinfo {pages} {491} (\bibinfo {year} {1966})}\BibitemShut
  {NoStop}%
\bibitem [{\citenamefont {Karki}\ and\ \citenamefont
  {Kiselev}(2017)}]{Karki_MK_TE}%
  \BibitemOpen
  \bibfield  {author} {\bibinfo {author} {\bibfnamefont {D.~B.}\ \bibnamefont
  {Karki}}\ and\ \bibinfo {author} {\bibfnamefont {M.~N.}\ \bibnamefont
  {Kiselev}},\ }\href
  {https://journals.aps.org/prb/abstract/10.1103/PhysRevB.96.121403} {\bibfield
   {journal} {\bibinfo  {journal} {Phys. Rev. B}\ }\textbf {\bibinfo {volume}
  {96}},\ \bibinfo {pages} {121403(R)} (\bibinfo {year} {2017})}\BibitemShut
  {NoStop}%
\bibitem [{Note1()}]{Note1}%
  \BibitemOpen
  \bibinfo {note} {One can study the evolution of Fano factor as a function of
  $V/T$ and the asymmetry parameter $\protect \mathcal {L}$ similar to that of
  Ref.~\cite {Mora_Leyronas_Regnault_PRL_(100)_2008} devoted to 1CK
  effect.}\BibitemShut {Stop}%
\end{thebibliography}
\end{document}